\documentclass[draftcls,onecolumn]{IEEEtran}
\usepackage{graphicx,amsmath,amssymb,array,stfloats,cuted,midfloat,psfrag}
\usepackage[noadjust]{cite}
\newtheorem{theorem}{\textbf{Theorem}}[section]
\newtheorem{definition}{\textbf{Definition}}[section]

\begin{document}
\title{Multiple Description Quantization Via Gram-Schmidt Orthogonalization}
\author{Jun Chen,~\IEEEmembership{Student Member,~IEEE}, Chao Tian,~\IEEEmembership{Student Member,~IEEE}, Toby Berger,
~\IEEEmembership{Fellow,~IEEE},
Sheila~S.~Hemami,~\IEEEmembership{Senior Member,~IEEE}
\thanks{Jun Chen and Toby Berger are supported in part by NSF Grant
CCR-033 0059 and a grant from the National Academies Keck Futures
Initiative (NAKFI). This work has been presented in part at the
39th Annual Conference on Information Sciences and Systems in
March 2005.}}
\date{}
\markboth{} {Chen \MakeLowercase{\textit{et al.}}} \maketitle

\begin{abstract}
The multiple description (MD) problem has received considerable
attention as a model of information transmission over unreliable
channels. A general framework for designing efficient multiple
description quantization schemes is proposed in this paper. We
provide a systematic treatment of the El Gamal-Cover (EGC)
achievable MD rate-distortion region, and show that any point in
the EGC region can be achieved via a successive quantization
scheme along with quantization splitting. For the quadratic
Gaussian case, the proposed scheme has an intrinsic connection
with the Gram-Schmidt orthogonalization, which implies that the
whole Gaussian MD rate-distortion region is achievable with a
sequential dithered lattice-based quantization scheme as the
dimension of the (optimal) lattice quantizers becomes large.
Moreover, this scheme is shown to be universal for all i.i.d.
smooth sources with performance no worse than that for an i.i.d.
Gaussian source with the same variance and asymptotically optimal
at high resolution. A class of low-complexity MD scalar quantizers
in the proposed general framework also is constructed and is
illustrated geometrically; the performance is analyzed in the high
resolution regime, which exhibits a noticeable improvement over
the existing MD scalar quantization schemes.
\end{abstract}
\begin{keywords}
Gram-Schmidt orthogonalization, lattice quantization, MMSE,
multiple description, quantization splitting.
\end{keywords}

\IEEEpeerreviewmaketitle

\normalsize
\section{Introduction}
In the multiple description problem the total available bit rate
is split between two channels and either channel may be subject to
failure. It is desired to allocate rate and coded representations
between the two channels, such that if one channel fails, an
adequate reconstruction of the source is possible, but if both
channels are available, an improved reconstruction over the
single-channel reception results. The formal definition of the MD
problem is as follows (also see Fig. 1).

Let $\{X(t)\}_{t=1}^\infty$ be an i.i.d. random process with
$X(t)\sim p(x)$ for all $t$. Let
$d(\cdot,\cdot):\mathcal{X}\times\mathcal{X}\rightarrow
[0,d_{\max}]$ be a distortion measure.
\begin{definition}
The quintuple $(R_1, R_2, D_1, D_2, D_3)$ is called achievable if
for all $\varepsilon>0$, there exist, for $n$ sufficiently large,
encoding functions:
\begin{equation*}
  f_{i}^{(n)}:
  \mathcal{X}^{n}\rightarrow\mathcal{C}_i^{(n)} \quad
  \log|\mathcal{C}_i^{(n)}|\leq n(R_i+\varepsilon) \quad i=1,2,
\end{equation*}
and decoding functions:
\begin{equation*}
g^{(n)}_{i}:\ \mathcal{C}_{i}^{(n)}\rightarrow \mathcal{X}^n \quad
i=1,2
\end{equation*}
\begin{equation*}
  g^{(n)}_{3}:\ \mathcal{C}_{1}^{(n)}\times
\mathcal{C}_{2}^{(n)}\rightarrow \mathcal{X}^n
\end{equation*}
such that for $\mathbf{\hat
X}_i=g^{(n)}_{i}(f^{(n)}_{i}(\mathbf{X})), i=1,2$, and for
$\mathbf{\hat
X}_3=g^{(n)}_{3}(f^{(n)}_{1}(\mathbf{X}),f^{(n)}_{2}(\mathbf{X}))$,
\begin{equation*}
\frac{1}{n}\mathbb{E}\sum\limits_{t=1}^n d(X(t), \hat
X_i(t))<D_i+\varepsilon \quad i=1,2,3.
\end{equation*}
The MD rate-distortion region, denoted by $\mathcal{Q}$, is the
set of all achievable quintuples.
\end{definition}

In this paper the encoding functions $f^{(n)}_1$ and $f^{(n)}_2$
are referred to as encoder 1 and encoder 2, respectively.
Similarly, decoding functions $g^{(n)}_1$, $g^{(n)}_2$ and
$g^{(n)}_3$ are referred to as decoder 1, decoder 2, and decoder
3, respectively. It should be emphasized that in a real system,
encoders 1 and 2 are just two different encoding functions of a
single encoder while decoders 1, 2 and 3 are different decoding
functions of a single decoder. Alternatively,  in the MD
literature decoders 1 and 2 are sometimes referred to as the side
decoders because of their positions in Fig. 1, while decoder 3 is
referred to as the central decoder.

\begin{figure}[hbt]
\centering
\begin{psfrags}
\psfrag{x}[r]{$\mathbf{X}$}%
\psfrag{xhat1}[l]{$\mathbf{\hat{X}}_1$}%
\psfrag{xhat2}[l]{$\mathbf{\hat{X}}_2$}%
\psfrag{xhat3}[l]{$\mathbf{\hat{X}}_3$}%
\psfrag{en1}[c]{Encoder $1$}%
\psfrag{en2}[c]{Encoder $2$}%
\psfrag{r1}[c]{$R_1$}%
\psfrag{r2}[c]{$R_2$}%
\psfrag{de1}[c]{Decoder $1$}%
\psfrag{de2}[c]{Decoder $2$}%
\psfrag{de3}[c]{Decoder $3$}%
\includegraphics[scale=1]{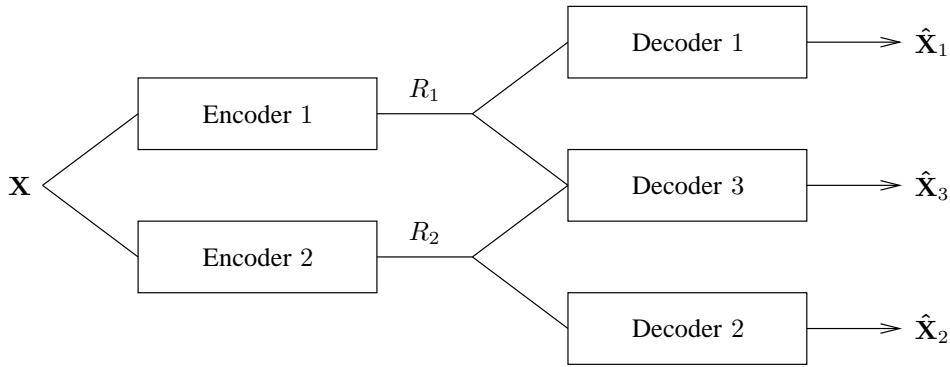}
\end{psfrags}
\caption{Encoder and decoder diagram for multiple descriptions.}
\end{figure}

Early contributions to the MD problem can be found in
\cite{Witsenhausen, WWZ, Ozarow, WW}. The first general result was
El Gamal and Cover's achievable region.

\begin{definition}[EGC region]
For random variables $U_1,U_2$ and $U_3$ jointly distributed with
the generic source variable $X$ via conditional distribution
$p(u_1,u_2,u_3|x)$, let
\begin{eqnarray*}
\mathcal{R}(U_1,U_2,U_3)=\left\{(R_1,R_2):R_1+R_2\geq I(X; U_1,
U_2, U_3)+I(U_1;U_2), R_i\geq I(X;U_i), i=1,2\right\}.
\end{eqnarray*}
Let
\begin{eqnarray*}
\mathcal{Q}(U_1,U_2,U_3)=\left\{(R_1,R_2,D_1,D_2,D_3):(R_1,R_2)\in\mathcal{R}(U_1,U_2,U_3),
\exists  \hat X_i=g_i(U_i) \mbox{ with } \mathbb{E}d(X,\hat
X_i)\leq D_i, i=1,2,3 \right\}.
\end{eqnarray*}
The EGC region\footnote{The form of the EGC region here is
slightly different from the one given in \cite{ElGamal}, but it is
straightforward to show they are equivalent. $g_3(U_3)$ can be also replaced
by a function of $(U_1,U_2,U_3)$, say
$\widetilde{g}(U_1,U_2,U_3)$, but the resulting
$\mathcal{Q}_{EGC}$ is still the same because for any
$(U_1,U_2,U_3)$ jointly distributed with $X$, there exist
$(U_1,U_2,\widetilde{U}_3)$ with $\widetilde{U}_3=(U_1,U_2,U_3)$
such that $\widetilde{g}_3(U_1,U_2,U_3)=g_3(\widetilde{U}_3)$.} is
then defined as
\begin{eqnarray*}
\mathcal{Q}_{EGC}=\mbox{conv}\left(\bigcup\limits_{p(u_1,u_2,u_3|x)}\mathcal{Q}(U_1,U_2,U_3)\right),
\end{eqnarray*}
where $\mbox{conv}(\mathcal{S})$ denotes the convex hull
of $\mathcal{S}$ for any set $\mathcal{S}$ in the Euclidean space.
\end{definition}

It was proved in \cite{ElGamal} that
$\mathcal{Q}_{EGC}\subseteq\mathcal{Q}$. Ozarow \cite{Ozarow}
showed that $\mathcal{Q}_{EGC}=\mathcal{Q}$ for the quadratic
Gaussian source. Ahlswede \cite{Ahlswede} showed that the EGC
region is also tight for the ``no excess sum-rate" case. Zhang and
Berger \cite{ZhangBerger} constructed a counterexample for which
$\mathcal{Q}_{EGC}\subsetneqq\mathcal{Q}$. Further results can be
found in \cite{ref27,ref29,ref30, Zamir, Fu, Fengthesis, Luis}.
The MD problem has also been generalized to the $n$-channel case
\cite{ref33, Pradhan}, but even the quadratic Gaussian case is far
from being completely understood. The extension of the MD problem
to the distributed source coding scenario has been considered in
\cite{Ishwar,ChenRobust}, where the problem is again widely open.

The first constructive method to generate multiple descriptions is the
 multiple description scalar quantization (MDSQ), which was proposed by Vaishampayan
\cite{Vaishampayan1,Vaishampayan2}. The key component of this
method is the index assignment, which maps an index to an index
pair as the two descriptions. However, the design of the index
assignment turns out to be a difficult problem. Since optimal
solution cannot be found efficiently, Vaishampayan
\cite{Vaishampayan1} provided several heuristic methods to
construct balanced index assignments which are not optimal but
likely to perform well. The analysis of this class of balanced
quantizers reveals that asymptotically (in rate) it is 3.07 dB
away from the rate-distortion bound \cite{Vai:98} in terms of
central and side distortion product, when a uniform central
quantizer is used; this granular distortion gap can be reduced by
0.4 dB when the central quantizer cells are better optimized
\cite{Cha:041}. The design of balanced index assignment was
recently more thoroughly addressed in \cite{Bal:04} from an
algorithm perspective, and the index assignments for more than two
description appeared in \cite{Ber:02}. Other methods have also
been proposed to optimize the index assignments
\cite{Gor:03,Kou:03}.

The framework of MDSQ was later extended to multiple description
lattice vector quantization (MDLVQ) for balanced descriptions in
\cite{Vai:01} and for the asymmetric case in \cite{Dig:02}. The
design relies heavily on the choice of lattice/sublattice
structure to facilitate the construction of index assignments. The
analysis on these quantizers shows that the constructions are
high-resolution optimal in asymptotically high dimensions;
however, in lower dimension, optimization of the code-cells can
also improve the high-resolution performance
\cite{Goy:02}\cite{Cha:042}. The major difficulty in constructing both MDSQ and MDLVQ is to
find good index assignments, and thus it would simplify the overall design
significantly if the component of index assignment can be
eliminated altogether.

Frank-Dayan and Zamir \cite{Zamir2} proposed a class of MD schemes
which use entropy-coded dithered lattice quantizers (ECDQs). The
system consists of two independently dithered lattice quantizers
as the two side quantizers, with a possible third dithered lattice
quantizer to provide refinement information for the central
decoder. It was found that even with the quadratic Gaussian
source, this system is only optimal in asymptotically high
dimensions for the degenerate cases such as successive refinement
and the ``no excess marginal-rate'' case, but not optimal in
general. The difficulty lies in generating dependent quantization
errors of two side quantizers to simulate the Gaussian multiple
description test channel. Several possible improvements were
provided in \cite{Zamir2}, but the problem remains unsolved.

The method of MD coding using correlating transforms was first
proposed by Orchard, Wang, Vaishampayan, and Reibman
\cite{Orchard, Wang}, and this technique has then been further
developed in \cite{Pradhan1} and \cite{Goyal}. However, the
transform-based approach is mainly designed for vector sources,
and it is most suitable when the redundancy between the
descriptions is kept relatively low.

In this paper we provide a systematic treatment of the El
Gamal-Cover (EGC) achievable MD rate-distortion region and show it
can be decomposed into a simplified-EGC (SEGC) region and an
superimposed refinement operation. Furthermore, any point in the
SEGC region can be achieved via a successive quantization scheme
along with quantization splitting. For the quadratic Gaussian
case, the MD rate-distortion region is the same as the SEGC
region, and the proposed scheme has an intrinsic connection with
the Gram-Schmidt orthogonalization method. Thus we use
single-description ECDQs, with independent subtractive dithers as
building blocks for this MD coding scheme, by which the difficulty
of generating dependent quantization errors is circumvented.
Analytical expressions for the rate-distortion performance of this
system are then derived for general sources, and compared to the
optimal rate regions at both high and low lattice dimensions.

The proposed scheme is conceptually different
from those in \cite{Zamir2}, and it can achieve the whole Gaussian
MD rate-distortion region as the dimension of the (optimal)
lattice quantizers becomes large, unlike the method proposed in
\cite{Zamir2}. From a construction perspective, the new MD coding
system can be realized by 2-3 conventional lattice quantizers
along with some linear operations, and thus it is considerably
simpler than MDSQ and MDLVQ by removing the index assignment and
the reliance on the lattice/sublattice structure. Though the
proposed coding scheme suggests many possible implementations of
practical quantization methods, the focus of this article is on the
information theoretic framework; thus instead of providing
detailed designs of quantizers, a geometric interpretation of the
scalar MD quantization scheme is given as an illustration to
connect the information theoretic description of coding scheme and
its practical counterpart.

The remainder of this paper is divided into 6 sections. In Section
II, ECDQ and the Gram-Schmidt orthogonalization method are breifly
reviewed and a connection between the successive quantization
scheme and the Gram-Schmidt orthogonalization method is
established. In Section III we present a systematic treatment of
the EGC region and show the sufficiency of a successive
quantization scheme along with quantization splitting. In Section
IV the quadratic Gaussian case is considered in more depth. In
Section V the proposed scheme based on ECDQ is shown to be
universal for all i.i.d. smooth sources with performance no worse
than that for an i.i.d. Gaussian source with the same variance and
asymptotically optimal at high resolution. A geometric
interpretation of the scalar MD quantization scheme in our
framework is given in Section VI. Some further extensions are
suggested in Section VII, which also serves as the conclusion.
Throughout, we use boldfaced letters to indicate ($n$-dimensional)
vectors, capital letters for random objects, and small letters for
their realizations. For example, we let
$\mathbf{X}=(X(1),\cdots,X(n))^T$ and
$\mathbf{x}=(x(1),\cdots,x(n))^T$.

\section{Entropy-Coded Dithered Quantization and Gram-Schmidt Orthogonalization}
In this section, we first give a brief review of ECDQ, and then
explain the difficulty of applying ECDQ directly to the MD
problem. As a method to resolve this difficulty, the Gram-Schmidt
orthogonalization is introduced and a connection between the
sequential (dithered) quantization and the Gram-Schmidt
orthogonalization is established. The purpose of this section is
two-fold: The first is to review related results on ECDQ and the
Gram-Schmidt orthogonalization and show their connection, while
the second is to explicate the intuition that motivated this work.

\subsection{Review of Entropy-Coded Dithered Quantization}

Some basic definitions and properties of ECDQ from \cite{Zamir2}
are quoted below. More detailed discussion and derivation can be
found in \cite{Feder1,Feder,Feder3,Zamir1}.

\begin{figure}[hbt]
\centering
\begin{psfrags}
\psfrag{x}[c]{$\mathbf{X}$}%
\psfrag{y}[c]{$\mathbf{Y}$}%
\psfrag{z}[c]{$\mathbf{Z}$}%
\psfrag{-z}[c]{$-\mathbf{Z}$}%
\psfrag{q}[c]{$Q_n(\cdot)$}%
\psfrag{e}[c]{E}%
\psfrag{d}[c]{D}%
\psfrag{r}[c]{$R$}%
\psfrag{n}[c]{$\mathbf{N}=-\mathbf{Z}$}%
\psfrag{sim}[c]{$\sim$}%
\psfrag{ecdq}[c]{ECDQ}%
\includegraphics[scale=1]{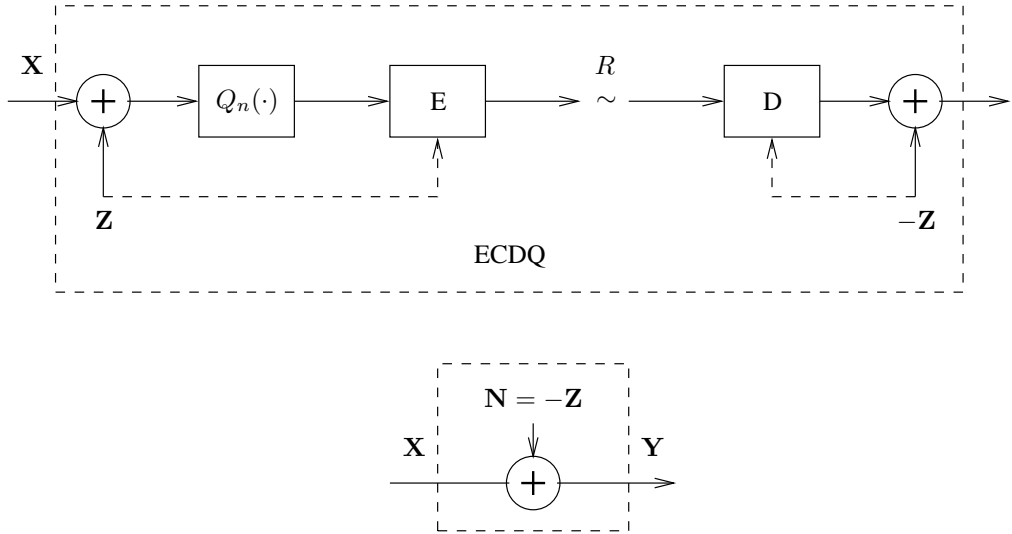}
\end{psfrags}
\caption{ECDQ and its equivalent additive-noise
channel.\label{ECDQ}}
\end{figure}

An $n$-dimensional lattice quantizer is formed from a lattice
$\mathbf{L}_n$. The quantizer $Q_n(\cdot)$ maps each vector
$\mathbf{x}\in\mathcal{R}^n$ into the lattice point
$\mathbf{l}_i\in \mathbf{L}_n$ that is nearest to $\mathbf{x}$.
The region of all $n$-vectors mapped into a lattice point
$\mathbf{l}_i\in \mathbf{L}_n$ is the Voronoi region
\begin{eqnarray*}
V(\mathbf{l}_i)=\left\{\mathbf{x}\in\mathcal{R}^n:||\mathbf{x}-\mathbf{l}_i||\leq
||\mathbf{x}-\mathbf{l}_j||, \forall j\neq i\right\}.
\end{eqnarray*}
The dither $\mathbf{Z}$ is an $n$-dimensional random vector,
independent of the source, and uniformly distributed over the
basic cell $V_0$ of the lattice which is the Voronoi region of the
lattice point $\mathbf{0}$. The dither vector is assumed to be
available to both the encoder and the decoder. The normalized
second moment $G_n$ of the lattice characterizes the second moment
of the dither vector
\begin{eqnarray*}
\frac{1}{n}\mathbb{E}||\mathbf{Z}||^2=G_nV^{2/n},
\end{eqnarray*}
where $V$ denotes the volume of $V_0$. Both the entropy encoder
and the decoder are conditioned on the dither sample $\mathbf{Z}$;
furthermore, the entropy coder is assumed to be ideal. The lattice
quantizer with dither represents the source vector $\mathbf{X}$ by
the vector $\mathbf{W}=Q_n(\mathbf{X}+\mathbf{Z})-\mathbf{Z}$. The
resulting properties of the ECDQ are as follows.
\begin{enumerate}
\item The quantization error vector $\mathbf{W}-\mathbf{X}$ is independent of $\mathbf{X}$ and is
distributed as $-\mathbf{Z}$. In particular, the mean-squared
quantization error is given by the second moment of the dither,
independently of the source distribution, i.e.,
\begin{eqnarray*}
\frac{1}{n}\mathbb{E}||\mathbf{W}-\mathbf{X}||^2=\frac{1}{n}\mathbb{E}||\mathbf{Z}||^2=G_nV^{2/n}.
\end{eqnarray*}

\item The coding rate of the ECDQ is equal to the mutual
information between the input and output of an additive noise
channel $\mathbf{Y}=\mathbf{X}+\mathbf{N}$, where $\mathbf{N}$,
the channel's noise, has the same probability density function as
$-\mathbf{Z}$ (see Fig. \ref{ECDQ}) ,
\begin{eqnarray*}
H(Q_n(\mathbf{X}+\mathbf{Z})|\mathbf{Z})=I(\mathbf{X};\mathbf{Y})=h(\mathbf{Y})-h(\mathbf{N}).
\end{eqnarray*}

\item For optimal lattice quantizers, i.e., lattice quantizers with
the minimal normalized second moment $G_n$, the autocorrelation of
the quantizer noise is ``white" , i.e.,
$\mathbb{E}\mathbf{Z}\mathbf{Z}^T=\sigma^2I_n $ where $I_n$ is the
$n\times n$ identity matrix, $\sigma^2=G^{opt}_nV^{2/n}$ is the
second moment of the lattice, and
\begin{eqnarray*}
G^{opt}_n=\min\limits_{Q_n(\cdot)}\frac{\int_{V_0}||\mathbf{x}||^2\mbox{d}\mathbf{x}}{nV^{1+\frac{2}{n}}}
\end{eqnarray*}
is the minimal normalized second moment of an $n$-dimensional
lattice.
\end{enumerate}

Consider the following problem to motivate the general result.
Suppose a quantization system is needed with input $X_1$ and
outputs $(X_2,\cdots,X_M)$ such that the quantization errors
$X_i-X_1$, $i=2,\cdots,M$, are correlated with each other in a
certain predetermined way, but are uncorrelated with $X_1$.
Seemingly, $M-1$ quantizers may be used, each with $X_1$ as the
input and $X_i$ as the output for some $i$, $i=2,\cdots,M$. By
property 1) of ECDQ, if dithers are introduced, the quantization
errors are uncorrelated (actually independent) of the input of the
quantizer. However, it is difficult to make the quantization
errors of these $M-1$ quantizers correlated in the desired manner.
One may expect it to be possible to correlate the quantization
errors by simply correlating the dithers of different quantizers,
but this turns out to be not true as pointed out in \cite{Zamir2}.
Next, we present a solution to this problem by exploiting the
relationship between the Gram-Schmidt orthogonalization and
sequential (dithered) quantization.

\subsection{Gram-Schmidt Orthogonalization}
In order to facilitate the treatment, the problem is reformulated in
an equivalent form: Given $X^M_1$ with an arbitrary covariance
matrix, construct a quantization system with $\widetilde{X}_1$ as
the input and $(\widetilde{X}_2,\cdots,\widetilde{X}_M)$ as the
outputs such that the covariance matrices of $X^M_1$ and
$\widetilde{X}^M_1$ are the same.

Let $\mathcal{H}_s$ denote the set of all finite-variance,
zero-mean, real scalar random variables. It is well known
\cite{Kailath, Guess} that $\mathcal{H}_s$ becomes a Hilbert space
under the inner product mapping
\begin{eqnarray*}
\langle X,Y\rangle=\mathbb{E}(XY):
\mathcal{H}_s\times\mathcal{H}_s\rightarrow\mathcal{R}.
\end{eqnarray*}
The norm induced by this inner product is
\begin{eqnarray*}
\|X\|^2=\langle X,X\rangle=\mathbb{E}X^2.
\end{eqnarray*}
For $X^M_1=(X_1,\cdots,X_M)^T$ with $X_i\in\mathcal{H}_s$,
$i=1,\cdots,M$, the Gram-Schmidt orthogonalization can be used to
construct an orthogonal basis $B^M_1=(B_1,\cdots,B_M)^T$ for
$X^M_1$. Specifically, the Gram-Schmidt orthogonalization proceeds
as follows:
\begin{eqnarray*}
B_1&=&X_1,\\
B_i&=&X_i-\sum\limits_{j=1}^{i-1}\frac{\langle X_i,B_j\rangle}{\|
B_j\|^2}B_j\nonumber\\
&=&X_i-\sum\limits_{j=1}^{i-1}\frac{\mathbb{E}(X_iB_j)}{\mathbb{E}B^2_j}B_j,\quad
i=2,\cdots,M.
\end{eqnarray*}
Note: $\frac{\mathbb{E}(X_iB_j)}{\mathbb{E}B^2_j}$ can assume any
real number if $B_j=0$. Alternatively, $B^M_1$ can also be
computed using the method of linear estimation. Let $K_{X^m_1}$
denote the covariance matrix of $(X_1,\cdots,X_m)^T$ and let
$K_{X_mX_1^{m-1}}=\mathbb{E}[X_m(X_1,\cdots,X_{m-1})^T]$, then
\begin{eqnarray}
B_1&=&X_1,\label{GS1}\\
B_i&=&X_i-K_{i-1}X^{i-1}_{1},\quad i=2,\cdots,M. \label{GS2}
\end{eqnarray}
Here $K_{i-1}\in\mathcal{R}^{1\times (i-1)}$ is a row vector
satisfying $K_{i-1}K_{X^{i-1}_1}=K_{X_iX^{i-1}_1}$. When
$K_{X^{i-1}_1}$ is invertible, $K_{i-1}$ is uniquely given by
$K_{X_iX^{i-1}_1}K^{-1}_{X^{i-1}_1}$. The product
$K_{i-1}X^{i-1}_1$ is the \textit{linear MMSE estimate} of $X_i$
given $X^{i-1}_1$, and $\mathbb{E}B^2_i$ is its corresponding
\textit{linear MMSE estimation error}.

The Gram-Schmidt orthogonalization is closely related to the
$LDL^T$ factorization. That is, if all leading minors of
$K_{X^M_1}$ are nonzero, then there exists a unique factorization
such that $K_{X^M_1}=LDL^T$, where $D$ is diagonal, and $L$ is
lower triangular with unit diagonal. Specifically,
$D=\mbox{diag}\left\{\| B_1\|^2,\cdots,\| B_M\|^2\right\}$ and
\begin{eqnarray*}
L&=&\begin{pmatrix}
  1 &   &   &   &   \\
  \frac{\langle X_2,B_1\rangle}{\|B_1\|^2} & 1 &   &  &   \\
  \frac{\langle X_3,B_1\rangle}{\|B_1\|^2} & \frac{\langle X_3,B_2\rangle}{\|B_2\|^2} & 1 &   &   \\
  \vdots & \vdots & \vdots & \ddots &   \\
  \frac{\langle X_L,B_1\rangle}{\|B_1\|^2} & \frac{\langle X_L,B_2\rangle}{\|B_2\|^2} & \frac{\langle X_L,B_3\rangle}{\|B_3\|^2} & \cdots & 1
\end{pmatrix}.
\end{eqnarray*}
$B^M_1=L^{-1}X^M_1$ is sometimes referred to as the
\textit{innovation process} \cite{Kailath}.

In the special case in which $X^M_1$ are jointly Gaussian, the
elements of $B^M_1$ are given by
\begin{eqnarray*}
B_1&=&X_1,\\
B_i&=&X_i-\mathbb{E}(X_i|X^{i-1}_1)\\
&=&X_i-\sum\limits_{j=1}^{i-1}\mathbb{E}(X_i|B_j),\quad
i=2,\cdots,M,
\end{eqnarray*}
and $B^M_1$ are zero-mean, independent and jointly Gaussian.
Moreover, since $X^i_1$ is a deterministic function of $B^i_1$, it
follows that $B^M_{i+1}$ is independent of $X^i_1$, for
$i=1,\cdots,M-1$. Note: For $i=2,\cdots,M$,
$\mathbb{E}(X_i|X^{i-1}_1)$ (or
$\sum_{j=1}^{i-1}\mathbb{E}(X_i|B_j)$) is a \textit{sufficient
statistic}\footnote{Actually, $\mathbb{E}(X_i|X^{i-1}_1)$ (or
$\sum_{j=1}^{i-1}\mathbb{E}(X_i|B_j)$) is a \textit{minimal}
sufficient statistic; i.e., $\mathbb{E}(X_i|X^{i-1}_1)$ (or
$\sum_{j=1}^{i-1}\mathbb{E}(X_i|B_j)$) is a function of every
other sufficient statistic $f(X^{i-1}_1)$ (or $f(B^{i-1}_1)$).}
for estimation of $X_i$ from $X^{i-1}_1$ (or $B^{i-1}_1$);
$\mathbb{E}(X_i|X^{i-1}_1)$ (or
$\sum_{j=1}^{i-1}\mathbb{E}(X_i|B_j)$) also is the \textit{MMSE
estimate} of $X_i$ given $X^{i-1}_1$ (or $B^{i-1}_1$) and
$\mathbb{E}B^2_i$ is the \textit{MMSE estimation error}.

We now show that one can construct a sequential quantization system
with $X_1$ as the input to generate a zero-mean random vector
$\widetilde{X}^M_1=(\widetilde{X}_1,\widetilde{X}_2,\cdots,\widetilde{X}_M)^T$
whose covariance matrix is also $K_{X^M_1}$. Let $X^M_1$ be a zero-mean random vector with covariance
matrix $K_{X^M_1}$. By (\ref{GS1}) and (\ref{GS2}), it is true that
\begin{eqnarray}
X_1&=&B_1,\label{XS1}\\
X_i&=&K_{i-1}X^{i-1}_{1}+B_i,\quad i=2,\cdots,M. \label{XS2}
\end{eqnarray}
 Assume that
$B_i\neq 0$ for $i=2,\cdots,M$. Let $Q_{i,1}(\cdot)$ be a scalar
lattice quantizer with step size
$\Delta_i=\sqrt{12\mathbb{E}B^2_{i+1}}$, $i=1,2,\cdots,M-1$. Let
the dither $Z_i\sim\mathcal{U}(-\Delta_i/2,\Delta_i/2)$ be a
random variable uniformly distributed over the basic cell of
$Q_{i,1}$, $i=1,2,\cdots,M-1$. Note: the second subscript $n$ of
$Q_{i,n}$ denotes the dimension of the lattice quantizer. In this
case $n=1$, so it is a scalar quantizer.

Suppose $(X_1,Z_1,\cdots,Z_{M-1})$ are independent. Define
\begin{eqnarray*}
\widetilde{X}_1&=&X_1,\\
\widetilde{X}_i&=&Q_{i-1,1}\left(K_{i-1}\widetilde{X}^{i-1}_{1}+Z_{i-1}\right)-Z_{i-1},\quad
i=2,\cdots,M.
\end{eqnarray*}
By property 2) of the ECDQ, we have
\begin{eqnarray}
\widetilde{X}_1&=&X_1,\label{XQS1}\\
\widetilde{X}_i&=&K_{i-1}\widetilde{X}^{i-1}_{1}+N_i,\quad
i=2,\cdots,M, \label{XQS2}
\end{eqnarray}
where $N_i\sim\mathcal{U}(\Delta_i/2,\Delta_i/2)$ with
$\mathbb{E}N^2_i=\mathbb{E}B^2_{i+1}$, $i=1,\cdots,M-1$, and
$(X_1,N_1,\cdots,N_M)$ are independent. By comparing (\ref{XS1}),
(\ref{XS2}) and (\ref{XQS1}), (\ref{XQS2}), it is straightforward
to verify that $X^M_1$ and $\widetilde{X}^M_1$ have the same
covariance matrix.

Since $\mathbb{E}B^2_i$ $(i=2,\cdots,M)$ are not necessarily the
same, it follows that the quantizers $Q_{i,1}(\cdot)$
$(i=1,\cdots,M-1)$ are different in general. But by incorporating
linear pre- and post-filters \cite{Feder3}, all these quantizers
can be made identical. Specifically, given a scalar lattice
quantizer $Q_1(\cdot)$ with step size $\Delta$, let the dither
$Z'_i\sim\mathcal{U}(-\Delta/2,\Delta/2)$ be a random variable
uniformly distributed over the basic cell of $Q_1$,
$i=1,2,\cdots,M-1$. Suppose $(X_1,Z'_1,\cdots,Z'_{M-1})$ are
independent. Define
\begin{eqnarray*}
\overline{X}_1&=&X_1,\\
\overline{X}_i&=&a_{i-1}\left[Q_1\left(\frac{1}{a_{i-1}}K_{i-1}\overline{X}^{i-1}_{1}+Z'_{i-1}\right)-Z'_{i-1}\right],\quad
i=2,\cdots,M,
\end{eqnarray*}
where $a_{i}=\pm\sqrt{\frac{12\mathbb{E}B^2_{i+1}}{\Delta^2}}$,
$i=1,2,\cdots,M-1$. By property 2) of the ECDQ, it is again straightforward to
verify that $X^M_1$ and $\overline{X}^M_1$ have the same
covariance matrix. Essentially by introducing the prefilter
$\frac{1}{a_{i}}$ and the postfilter $a_{i}$, the quantizer $Q_1(\cdot)$ is converted
to the quantizer $Q_{i,1}(\cdot)$ for which
\begin{eqnarray*}
Q_{i,1}(x)=a_{i}Q_1(\frac{x}{a_i}).
\end{eqnarray*}
This is referred to as the \textit{shaping} \cite{Feder} of the
quantizer $Q_1(\cdot)$ by $a_i$. In the case where $\Delta^2=12$,
we have $\mathbb{E}(Z'_i)^2=1$, $i=1,\cdots,M-1$, and the
constructed sequential (dithered) quantization system can be
regarded as a simulation of Gram-Schmidt orthonormalization.

If $B_i=0$ for some $i$, then
$\widetilde{X}_i=K_{i-1}\widetilde{X}^{i-1}_{1}$ (or
$\overline{X}_i=K_{i-1}\overline{X}^{i-1}_{1}$) and therefore no
quantization operation is needed to generate $\widetilde{X}_i$ (or
$\overline{X}_i$ ) from $\widetilde{X}^{i-1}_{1}$ (or
$\overline{X}^{i-1}_{1}$).

The generalization of the correspondence between the Gram-Schmidt
orthogonalization and the sequential (dithered) quantization to
the vector case is straightforward; see Appendix I.

\section{Successive Quantization and Quantization Splitting}
In this section, an information-theoretic analysis of the EGC
region is provided. Two coding schemes, namely successive
quantization and quantization splitting, are subsequently
introduced. Together with Gram-Schmidt orthogonalization, they are
the main components of the quantization schemes that will be
presented in the next two sections.

\subsection{An information theoretic analysis of the EGC region}
Rewrite $\mathcal{R}(U_1,U_2,U_3)$ in the following form:
\begin{eqnarray*}
\mathcal{R}(U_1,U_2,U_3)=\left\{(R_1,R_2):R_1+R_2\geq I(X; U_1,
U_2)+I(U_1;U_2)+I(X;U_3|U_1,U_2), R_i\geq I(X;U_i), i=1,2\right\}.
\end{eqnarray*}
Without loss of generality, assume that $X\rightarrow
U_3\rightarrow (U_1,U_2)$ form a Markov chain since otherwise
$U_3$ can be replaced by $\widetilde{U}_3=(U_1,U_2,U_3)$ without
affecting the rate and distortion constraints. Therefore $U_3$ can
be viewed as a fine description of $X$ and $(U_1,U_2)$ as coarse
descriptions of $X$. The term $I(X, U_3|U_1,U_2)$ is the rate used
for the superimposed refinement from the pair of coarse
descriptions $(U_1,U_2)$ to the fine description $U_3$; in
general, this refinement rate is split between the two channels.
Since description refinement schemes have been studied extensively
in the multiresolution or layered source coding scenario and are
well-understood, this operation can be separated from other parts
of the EGC scheme.
\begin{definition}[SEGC region]
For random variables $U_1$ and $U_2$ jointly distributed with the
generic source variable $X$ via conditional distribution
$p(u_1,u_2|x)$, let
\begin{eqnarray*}
\mathcal{R}(U_1,U_2)=\left\{(R_1,R_2):R_1+R_2\geq I(X; U_1,
U_2)+I(U_1;U_2), R_i\geq I(X;U_i), i=1,2\right\}.
\end{eqnarray*}
Let
\begin{eqnarray*}
\mathcal{Q}(U_1,U_2)=\left\{(R_1,R_2,D_1,D_2,D_3):(R_1,R_2)\in\mathcal{R}(U_1,U_2),
\exists  \hat X_1=g_1(U_1), \hat X_2=g_2(U_2), \hat
X_3=g_3(U_1,U_2)\right.\\\left. \mbox{ with } \mathbb{E}d(X,\hat
X_i)\leq D_i, i=1,2,3 \right\}.
\end{eqnarray*}
The SEGC region is defined as
\begin{eqnarray*}
\mathcal{Q}_{SEGC}=\mbox{conv}\left(\bigcup\limits_{p(u_1,u_2|x)}\mathcal{Q}(U_1,U_2)\right).
\end{eqnarray*}
\end{definition}

The SEGC region first appeared in \cite{Witsenhausen} and was
attributed to El Gamal and Cover. It was shown in
\cite{ZhangBerger} that
$\mathcal{Q}_{SEGC}\subseteq\mathcal{Q}_{EGC}$.

\begin{figure}[hbt]
\centering
\begin{psfrags}
\psfrag{r1}[r]{$R_1$}%
\psfrag{r2}[r]{$R_2$}%
\psfrag{v1}[r]{$V_1$}%
\psfrag{v2}[c]{$V_2$}%
\includegraphics[scale=0.8]{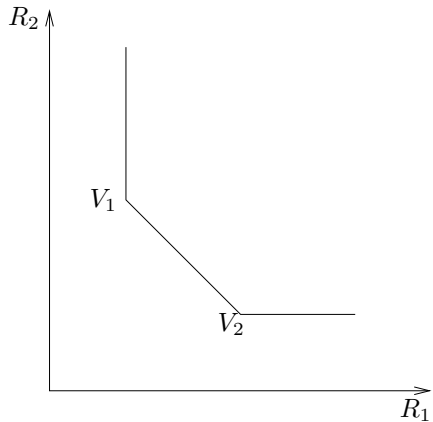}
\end{psfrags}
\caption{The shape of $\mathcal{R}(U_1,U_2)$.\label{rate region}}
\end{figure}

Using the identity
\begin{eqnarray*}
I(A;BC)=I(A;B)+I(A;C)+I(B;C|A)-I(B;C),
\end{eqnarray*}
$\mathcal{R}(U_1,U_2)$ can be written as
\begin{eqnarray*}
\mathcal{R}(U_1,U_2)=\left\{(R_1,R_2):R_1+R_2\geq I(X;
U_1)+I(X;U_2)+I(U_1;U_2|X), R_i\geq I(X;U_i), i=1,2\right\}.
\end{eqnarray*}
The typical shape of $\mathcal{R}(U_1,U_2)$ is shown in Fig.
\ref{rate region}.

It is noteworthy that $\mathcal{R}(U_1,U_2)$ resembles Marton's
achievable region \cite{Marton} for a two-user broadcast channel.
This is not surprising since the proof of the EGC theorem relies
heavily on the results in \cite{Meulen} which were originally for
a simplified proof of Marton's coding theorem for the discrete
memoryless broadcast channel. Since the corner points of Marton's
region can be achieved via a relatively simple coding scheme due
to Gel'fand and Pinsker \cite{Pinsker}, which for the Gaussian
case becomes Costa's dirty paper coding \cite{Costa}, it is
natural to conjecture that simple quantization schemes may exist
for the corner points of $\mathcal{R}(U_1,U_2)$. This conjecture
turns out to be correct as will be shown below.

Since $I(U_1;U_2|X)\geq 0$, the sum-rate constraint in
$\mathcal{R}(U_1,U_2)$ is always effective. Thus
\begin{eqnarray*}
\left\{(R_1,R_2):R_1+R_2=I(X;U_1)+I(X;U_2)+I(U_1;U_2|X), R_i\geq
I(X;U_i),i=1,2\right\}
\end{eqnarray*}
will be called {\em the dominant face} of $\mathcal{R}(U_1,U_2)$.
Any rate pair inside $\mathcal{R}(U_1,U_2)$ is inferior to some
rate pair on the dominant face in terms of compression efficiency.
Hence, in searching for the optimal scheme, attention can be
restricted to rate pairs on the dominant face without loss of
generality. The dominant face of $\mathcal{R}(U_1,U_2)$ has two
vertices $V_1$ and $V_2$. Let $(R_1(V_i),R_2(V_i))$ denote the
coordinates of vertex $V_i$, $i=1,2$, then
\begin{enumerate}
\item[] $V_1$: $R_1(V_1)=I(X;U_1)$, $R_2(V_1)=I(X,U_1;U_2)$;
\item[] $V_2$: $R_1(V_2)=I(X,U_2;U_1)$, $R_2(V_2)=I(X;U_2)$.
\end{enumerate}
The expressions of these two vertices directly lead to the
following successive quantization scheme. By symmetry, we shall
only consider $V_1$.

\subsection{Successive Quantization Scheme}
The successive quantization scheme is given as follows:
\begin{enumerate}
\item Codebook Generation: Encoder 1 independently generates $2^{n[I(X;U_1)+\epsilon_1]}$ codewords
$\{\mathbf{U}_1(j)\}_{j=1}^{2^{n[I(X;U_1)+\epsilon_1]}}$ according
to the distribution $\prod p(u_1)$. Encoder 2 independently
generates $2^{n[I(X,U_1;U_2)+\epsilon_2]}$ codewords
$\{\mathbf{U}_2(k)\}_{k=1}^{2^{n[I(X,U_1;U_2)+\epsilon_2]}}$
according to the distribution $\prod p(u_2)$.

\item Encoding Procedure: Given $\mathbf{X}$, encoder 1 finds the codeword
$\mathbf{U}_1(j^*)$ such that $\mathbf{U}_1(j^*)$ is strongly
typical with $\mathbf{X}$. Then encoder 2 finds the codeword
$\mathbf{U}_2(k^*)$ such that $\mathbf{U}_2(k^*)$ is strongly
typical with $\mathbf{X}$ and $\mathbf{U}_1(j^*)$. Index $j^*$ is
transmitted through channel 1 and index $k^*$ is transmitted
through channel 2.

\item Reconstruction: Decoder $1$ reconstructs $\mathbf{\hat X}_1$ with $\hat
X_1(t)=g_1(U_1(j^*,t))$. Decoder $2$ reconstructs $\mathbf{\hat
X}_2$ with $\hat X_2(t)=g_2(U_2(k^*,t))$. Decoder $3$ reconstructs
$\mathbf{\hat X}_3$ with $\hat X_3(t)=g_3(U_1(j^*,t),U_2(k^*,t))$.
Here, $U_1(j^*,t)$ and $U_2(k^*,t)$ are the $t$-th entries of
$\mathbf{U}_1(j^*)$ and $\mathbf{U}_2(k^*)$, respectively,
$t=1,2,\cdots,n$.
\end{enumerate}
It can be shown rigorously that
\begin{eqnarray*}
\frac{1}{n}\sum\limits_{t=1}^n\mathbb{E}d(X_i(t),\hat
X_i(t))&\leq&
\mathbb{E}d(X, g_i(U_i))+\epsilon_{2+i},\quad i=1,2,\\
\frac{1}{n}\sum\limits_{t=1}^n\mathbb{E}d(X_3(t),\hat
X_3(t))&\leq& \mathbb{E}d(X, g_3(U_1,U_2))+\epsilon_5
\end{eqnarray*}
as $n$ goes to infinity and $\epsilon_i$ $(i=1,2,\cdots,5)$ can be
made arbitrarily close to zero. The proof is conventional and thus
is omitted.

For this scheme, encoder 1 does the encoding first and then
encoder 2 follows. The main complexity of this scheme resides in
encoder 2, since it needs to construct a codebook that covers the
$(\mathbf{X},\mathbf{U}_1)$-space instead of just the
$\mathbf{X}$-space. Observe that, if a function $f(X,U_1)=V$ can
be found
 such that $V$ is a s\textit{ufficient statistic} for estimation $U_2$ from
$(X,U_1)$, i.e., $(X,U_1)\rightarrow V\rightarrow U_2$ form a
Markov chain\footnote{Such a function $f(\cdot,\cdot)$ always
exists provided $|\mathcal{V}|\geq
|\mathcal{X}||\mathcal{U}_1|$.}, then
\begin{eqnarray*}
I(X,U_1;U_2)=I(V;U_2).
\end{eqnarray*}
The importance of this observation is that encoder 2 then only
needs to construct a codebook that covers the $\mathbf{V}$-space
instead of the $(\mathbf{X},\mathbf{U}_1)$-space. This is because
the Markov lemma \cite{Berger} implies that if $\mathbf{U}_2$ is
jointly typical with $\mathbf{V}$, then $\mathbf{U}_2$ is jointly
typical with $(\mathbf{X},\mathbf{U}_1)$ with high probability.
This observation turns out to be crucial for the quadratic Gaussian
case.

We point out that the successive coding structure associated with
the corner points of $\mathcal{R}(U_1,U_2)$ is not a special case
in network information theory. Besides its resemblance to the
successive Gel'fand-Pinsker coding structure associated with the
corner points of the Marton's region previously mentioned, other
noteworthy examples include the successive decoding structure
associated with the corner points of the Slepian-Wolf region
\cite{Slepian} (and more generally, the Berger-Tung region
\cite{Berger, Tung, Housewright}) and the corner points of the
capacity region of the memoryless multiaccess channel
\cite{AhlswedeMulti, Liao}.

\subsection{Successive Quantization Scheme with Quantization
Splitting}

A straightforward method to achieve an arbitrary rate pair on the
dominant face of $\mathcal{R}(U_1,U_2)$ is timesharing of coding
schemes that achieve the two vertices. However, such a scheme
requires four quantizers in general. Instead, the scheme based on
quantization splitting introduced below needs only three
quantizers. Before presenting it, we shall first prove the
following theorem.

\begin{theorem}
For any rate pair $(R_1,R_2)$ on the dominant face of
$\mathcal{R}(U_1,U_2)$, there exists a random variable $U'_2$ with
$(X,U_1)\rightarrow U_2\rightarrow U'_2$ such that
\begin{eqnarray*}
R_1&=&I(X,U'_2; U_1),\\
R_2&=&I(X;U'_2)+I(X,U_1;U_2|U'_2).
\end{eqnarray*}
Similarly, there exists a random variable $U'_1$ with
$(X,U_2)\rightarrow U_1\rightarrow U'_1$ such that
\begin{eqnarray*}
R_1&=&I(X,U'_1)+I(X,\hat X_2;U_1|U'_1),\\
R_2&=&I(X,U'_1; U_2).
\end{eqnarray*}
\end{theorem}

Before proceeding to prove this theorem, we make the following remarks.
\begin{itemize}
\item By the symmetry between the two forms, only the statement regarding the first form needs to be proved.
\item Since $(X,U_1)\rightarrow U_2\rightarrow U'_2$ form a Markov chain, if $U'_2$ is independent of $U_2$, then it must be independent of
$(X,U_1,U_2)$ altogether\footnote{This is because
$p(x,u_1,u_2|u'_2)=p(u_2|u'_2)p(x,u_1|u_2,u'_2)=p(u_2)p(x,u_1|u_2)=p(x,u_1,u_2)$.}.
Then in this case,
\begin{eqnarray*}
R_1=I(X;U_1),\quad R_2=I(X,U_1;U_2),
\end{eqnarray*}
which are the coordinates of $V_1$.
\item At the other extreme, letting $U'_2$ be $U_2$ gives
\begin{eqnarray*}
R_1=I(X,U_2;U_1),\quad R_2=I(X,U_2),
\end{eqnarray*}
which are the coordinates of $V_2$.
\end{itemize}

\begin{proof}
First construct a class of transition probabilities\footnote{There
are many ways to construct such a class of transition
probabilities. For example, we can let $p_0(u'_2|u_2)=p(u'_2)$,
$p_1(u'_2|u_2)=\delta(u_2,u'_2)$, and set
$p_{\epsilon}(u'_2|u_2)=(1-\epsilon)p_0(u'_2|u_2)+\epsilon
p_1(u'_2|u_2)$. Here $\delta(u_2,u'_2)=1$ if $u_2=u'_2$ and $=0$
otherwise.} $p_{\epsilon}(u'_2|u_2)$ indexed by $\epsilon$ such
that $I(U;U'_2)$ varies continuously from 0 to $H(U_2)$ as
$\epsilon$ changes from 0 to 1, with $(X,U_1)\rightarrow
U_2\rightarrow U'_2$ holding for all the members of this class. It
remains to show that
\begin{eqnarray*}
R_1+R_2=I(X,U_1)+I(X;U_2)+I(U_1;U_2|X).
\end{eqnarray*}
This is indeed true since
\begin{eqnarray*}
R_1+R_2&=&I(X,U'_2; U_1)+I(X;U'_2)+I(X,U_1;U_2|U'_2)\\
&=&I(X,U'_2; U_1)+I(X;U'_2)+I(X;U_2|U'_2)+I(U_1;U_2|X,U'_2)\\
&=&I(X, U_2, U'_2;U_1)+I(X;U_2,U'_2).
\end{eqnarray*}
By the construction $(X,U_1)\rightarrow U_2\rightarrow U'_2$, it follows that
\begin{eqnarray*}
&&I(X, U_2, U'_2;U_1)+I(X;U_2,U'_2)\\
&=&I(X, U_2;U_1)+I(X;U_2)\\
&=&I(X,U_1)+I(X;U_2)+I(U_1;U_2|X),
\end{eqnarray*}
which completes the proof.
\end{proof}

The successive quantization scheme with quantization splitting is given as follows:
\begin{enumerate}
\item Codebook Generation: Encoder 1 independently generates
$2^{n[I(X, U'_2;U_1)+\epsilon'_1]}$ codewords
$\{\mathbf{U}_1(i)\}_{i=1}^{2^{n[I(X,U'_2;U_1)+\epsilon'_1]}}$
according to the marginal distribution $\prod p(u_1)$. Encoder 2
independently generates $2^{n[I(X;U'_2)+\epsilon'_2]}$ codewords
$\{{\mathbf{U'}_2}(j)\}_{j=1}^{2^{n[I(X;U'_2)+\epsilon'_2]}}$
according to the marginal distribution $\prod p(u'_2)$. For each
codeword ${\mathbf{U'}_2}(j)$, encoder 2 independently generates
$2^{n[I(X,U_1 ;U_2|U'_2)+\epsilon'_3]}$ codewords
$\{{\mathbf{U}_2}(j,k)\}_{k=1}^{2^{n[I(X,U_1;U_2|U'_2)+\epsilon'_3]}}$
according to the conditional distribution $\prod\limits_{t}
p(u_2|U'_2(j,t))$. Here $U'_2(j,t)$ is the $t$-th entry of
$\mathbf{U'}_2(j)$

\item Encoding Procedure: Given $\mathbf{X}$, encoder 2 finds the codeword
${\mathbf{U'}_2}(j^*)$ such that ${\mathbf{U'}_2}(j^*)$ is
strongly typical with $\mathbf{X}$. Then encoder 1 finds the
codeword $\mathbf{U}_1(i^*)$ such that $\mathbf{U}_1(i^*)$ is
strongly typical with $\mathbf{X}$ and ${\mathbf{U'}_2}(j^*)$.
Finally, encoder 2 finds the codeword ${\mathbf{U}_2}(j^*,k^*)$
such that $\mathbf{U}_2(j^*,k^*)$ is strongly typical with
$\mathbf{X}$, $\mathbf{U}_1(i^*)$ and ${\mathbf{U'}_2}(j^*)$.
Index $i^*$ is transmitted through channel 1. Indices $j^*$ and
$k^*$ are transmitted through channel 2.

\item Reconstruction: Decoder $1$ reconstructs $\mathbf{\hat X}_1$ with $\hat
X_1(t)=g_1(U_1(i^*,t))$. Decoder $2$ reconstructs $\mathbf{\hat
X}_2$ with $\hat X_2(t)=g_2(U_2(j^*,k^*,t))$. Decoder $3$
reconstructs $\mathbf{\hat X}_3$ with $\hat
X_3(t)=g_3(U_1(i^*,t),U_2(j^*,k^*,t))$. Here $U_1(i^*,t)$ is the
$t$-th entry of $\mathbf{U}_1(i^*)$ and $U_2(j^*,k^*,t)$ is the
$t$-th entry of $\mathbf{U}_2(j^*,k^*)$, $t=1,2,\cdots,n$.
\end{enumerate}

Again, it can be shown rigorously that
\begin{eqnarray*}
\frac{1}{n}\sum\limits_{t=1}^n\mathbb{E}d(X_i(t),\hat
X_i(t))&\leq&
\mathbb{E}d(X, g_i(U_i))+\epsilon'_{3+i},\quad i=1,2,\\
\frac{1}{n}\sum\limits_{t=1}^n\mathbb{E}d(X_3(t),\hat
X_3(t))&\leq& \mathbb{E}d(X, g_3(U_1,U_2))+\epsilon'_6
\end{eqnarray*}
as $n$ goes to infinity and $\epsilon'_i$ $(i=1,2,\cdots,6)$ can
be made arbitrarily close to zero. The proof is standard, so the
details are omitted.

This approach is a natural generalization of the successive
quantization scheme for the vertices of $\mathcal{R}(U_1,U_2)$.
$U'_2$ can be viewed as a coarse description of $X$ and $U_2$ as a
fine description of $X$. The idea of introducing an auxiliary
coarse description to convert a joint coding scheme to a
successive coding scheme has been widely used in the distributed
source coding problems \cite{Rimoldi, Coleman, SuccessiveCEO}.
Similar ideas have also found application in multiaccess
communications \cite{Carleial, Urbanke, Grant, Rimoldi2}.

\section{The Gaussian Multiple Description Region}
In this section we apply the general results in the preceding
section to the quadratic Gaussian case\footnote{All our results
derived under the assumption of discrete memoryless source and
bounded distortion measure can be generalized to the quadratic
Gaussian case, using the technique in \cite{Gallager}.}. The
Gaussian MD rate-distortion region is first analyzed to show that
$\mathcal{Q}_{EGC}=\mathcal{Q}_{SEGC}$ in this case. Then, by
incorporating the Gram-Schmidt orthogonalization with successive
quantization and quantization splitting, a coding scheme that
achieves the whole Gaussian MD region is presented.

\subsection{An Analysis of the Gaussian MD Region}
Let $\{X^G(t)\}_{t=1}^{\infty}$ be an i.i.d. Gaussian process with
$X^G(t)\sim\mathcal{N}(0,\sigma^2_X)$ for all $t$. Let
$d(\cdot,\cdot)$ be the squared error distortion measure. For the
quadratic Gaussian case, the MD rate-distortion region was
characterized in \cite{ElGamal, Ozarow, Feng}. Namely,
$(R_1,R_2,D_1,D_2,D_3)\in\mathcal{Q}$ if and only if
\begin{eqnarray*}
R_i&\geq&\frac{1}{2}\log\frac{\sigma^2_X}{D_i},\quad i=1,2,\\
R_1+R_2&\geq&\frac{1}{2}\log\frac{\sigma^2_X}{D_3}+\frac{1}{2}\log\psi(D_1,D_2,D_3),
\end{eqnarray*}
where
\begin{eqnarray*}
\psi(D_1,D_2,D_3)=\left\{\begin{array}{ll}
  1, \hspace{1.95in} D_3<D_1+D_2-\sigma^2_X\\
  \frac{\sigma^2_XD_3}{D_1D_2}, \hspace{1.7in}D_3> \left(\frac{1}{D_1}+\frac{1}{D_2}-\frac{1}{\sigma^2_X}\right)^{-1}\\
  \frac{(\sigma^2_X-D_3)^2}
{(\sigma^2_X-D_3)^2-[\sqrt{(\sigma^2_X-D_1)(\sigma^2_X-D_2)}-\sqrt{(D_1-D_3)(D_2-D_3)}]^2},  \mbox{ o.w.} \\
\end{array}\right.&\\
\end{eqnarray*}

The case $D_3<D_1+D_2-\sigma^2_X$ and the case
$D_3>\left(1/{D_1}+1/{D_2}-1/{\sigma^2_X}\right)^{-1}$ are
degenerate. It is easy to verify that for any
$(R_1,R_2,D_1,D_2,D_3)\in\mathcal{Q}$ with
$D_3<D_1+D_2-\sigma^2_X$, there exist $D^*_1\leq D_1$, $D^*_2\leq
D_2$ such that $(R_1,R_2,D^*_1,D^*_2,D_3)\in\mathcal{Q}$ and
$D_3=D^*_1+D^*_2-\sigma^2_X$. Similarly, for any
$(R_1,R_2,D_1,D_2,D_3)\in\mathcal{Q}$ with
$D_3>\left(1/{D_1}+1/{D_2}-1/{\sigma^2_X}\right)^{-1}$, there
exist  $D^*_3=\left(1/{D_1}+1/{D_2}-1/{\sigma^2_X}\right)^{-1}<
D_3$ such that $(R_1,R_2,D_1,D_2,D^*_3)\in\mathcal{Q}$. Henceforth
we shall only consider the subregion when
$\left(1/{D_1}+1/{D_2}-1/{\sigma^2_X}\right)^{-1}\geq D_3\geq
D_1+D_2-\sigma^2_X$, for which $D_1,D_2$ and $D_3$ all are
effective.

Following the approach in \cite{ElGamal}, let
\begin{eqnarray}
U_1&=&X^G+T_0+T_1,\label{U_1}\\
U_2&=&X^G+T_0+T_2,\label{U_2}
\end{eqnarray}
where $(T_1,T_2)$, $T_0$, $X$ are zero-mean, jointly Gaussian and
independent, and $\mathbb{E}(T_1T_2)=-\sigma_{T_1}\sigma_{T_2}$.
Let $\hat X^G_i=\mathbb{E}(X^G|U_i)=\alpha_i U_i$ $(i=1,2)$, and
$\hat X^G_3=\mathbb{E}(X^G|U_1,U_2)=\beta_1 U_1+\beta_2 U_2$,
where
\begin{eqnarray*}
\alpha_i&=&\frac{\sigma^2_X}{\sigma^2_X+\sigma^2_{T_0}+\sigma^2_{T_i}},\quad i=1,2,\\
\beta_1&=&\frac{\sigma^2_X\sigma_{T_2}}{(\sigma_{T_1}+\sigma_{T_2})(\sigma^2_X+\sigma^2_{T_0})},\\
\beta_2&=&\frac{\sigma^2_X\sigma_{T_1}}{(\sigma_{T_1}+\sigma_{T_2})(\sigma^2_X+\sigma^2_{T_0})}.
\end{eqnarray*}
Set $\mathbb{E}(X^G-\hat X^G_i)^2=D_i$, $i=1,2,3$; then
\begin{eqnarray}
\sigma^2_{T_0}&=&\frac{D_3\sigma^2_X}{\sigma^2_X-D_3},\label{T_0}\\
\sigma^2_{T_i}&=&\frac{D_i\sigma^2_X}{\sigma^2_X-D_i}-\frac{D_3\sigma^2_X}{\sigma^2_X-D_3},\quad
i=1,2.\label{T_1T_2}
\end{eqnarray}

With these $\sigma^2_{T_i}$ $(i=0,1,2)$, it is straightforward to verify that
\begin{eqnarray*}
I(X^G;U_i)&=&\frac{1}{2}\log\frac{\sigma^2_X+\sigma^2_{T_0}+\sigma^2_{T_i}}{\sigma^2_{T_0}+\sigma^2_{T_i}}\\
&=&\frac{1}{2}\log\frac{\sigma^2_X}{D_i}\quad i=1,2,\\
I(X^G;U_1)+I(X^G;U_2)+I(U_1;U_2|X^G)&=&\frac{1}{2}\log\frac{\sigma^2_X+\sigma^2_{T_0}+\sigma^2_{T_1}}{\sigma^2_{T_0}+\sigma^2_{T_1}}+\frac{1}{2}\log\frac{\sigma^2_X+\sigma^2_{T_0}+\sigma^2_{T_2}}{\sigma^2_{T_0}+\sigma^2_{T_2}}\\
&&+\frac{1}{2}\log\frac{(\sigma^2_{T_0}+\sigma^2_{T_1})(\sigma^2_{T_0}+\sigma^2_{T_2})}{\sigma^2_{T_0}(\sigma_{T_1}+\sigma_{T_2})^2}
\\&=&\frac{1}{2}\log\frac{\sigma^2_X}{D_3}+\frac{1}{2}\log\psi(D_1,D_2,D_3).
\end{eqnarray*}
Therefore, we have
\begin{eqnarray}
\mathcal{R}^G(U_1,U_2)&\triangleq&\left\{(R_1,R_2):R_1+R_2\geq
I(X^G;U_1)+I(X^G;U_2)+I(U_1;U_2|X^G),R_i\geq I(X^G;U_i), i=1,2
\right\}\nonumber\\
&=&\left\{(R_1,R_2):R_1+R_2\geq\frac{1}{2}\log\frac{\sigma^2_X}{D_3}+\frac{1}{2}\log\psi(D_1,D_2,D_3),R_i\geq\frac{1}{2}\log\frac{\sigma^2_X}{D_i},
i=1,2\right\}. \label{R(U_1,U_2)}
\end{eqnarray}
Hence for the quadratic Gaussian case,
\begin{eqnarray*}
\mathcal{Q}=\mathcal{Q}_{EGC}=\mathcal{Q}_{SEGC}
\end{eqnarray*}
and there is no need to introduce $U_3$ (more precisely, $U_3$ can
be represented as a deterministic function of $U_1$ and $U_2$).

The coordinates of the vertices $V^G_1$ and $V^G_2$ of
$\mathcal{R}^G(U_1,U_2)$ can be computed as follows.
\begin{eqnarray}
R_1(V^G_1)&=&\frac{1}{2}\log\frac{\sigma^2_X+\sigma^2_{T_0}+\sigma^2_{T_1}}{\sigma^2_{T_0}+\sigma^2_{T_1}}\nonumber\\
&=&\frac{1}{2}\log\frac{\sigma^2_X}{D_1},\\
R_2(V^G_1)&=&\frac{1}{2}\log\frac{(\sigma^2_X+\sigma^2_{T_0}+\sigma^2_{T_2})(\sigma^2_{T_0}+\sigma^2_{T_1})}{\sigma^2_{T_0}(\sigma_{T_1}+\sigma_{T_2})^2}\nonumber\\
&=&\frac{1}{2}\log\frac{D_1}{D_3}+\frac{1}{2}\log\psi(D_1,D_2,D_3).
\end{eqnarray}
and
\begin{eqnarray}
R_1(V^G_2)&=&\frac{1}{2}\log\frac{(\sigma^2_X+\sigma^2_{T_0}+\sigma^2_{T_1})(\sigma^2_{T_0}+\sigma^2_{T_2})}{\sigma^2_{T_0}(\sigma_{T_1}+\sigma_{T_2})^2}\nonumber\\
&=&\frac{1}{2}\log\frac{D_2}{D_3}+\frac{1}{2}\log\psi(D_1,D_2,D_3),\\
R_2(V^G_2)&=&\frac{1}{2}\log\frac{\sigma^2_X+\sigma^2_{T_0}+\sigma^2_{T_2}}{\sigma^2_{T_0}+\sigma^2_{T_2}}\nonumber\\
&=&\frac{1}{2}\log\frac{\sigma^2_X}{D_2}.
\end{eqnarray}
Henceforth we shall assume that for fixed $(D_1,D_2,D_3)$,
$\sigma^2_{T_i}$ $(i=0,1,2)$ are uniquely determined by
(\ref{T_0}) and (\ref{T_1T_2}), and consequently
$\mathcal{R}^G(U_1,U_2)$ is given by (\ref{R(U_1,U_2)}). Since
only the optimal MD coding scheme is of interest, the sum-rate
$R_1+R_2$ should be minimized with respect to the distortion
constraints $(D_1,D_2,D_3)$, i.e., $(R_1,R_2)$ must be on the
dominant face of $\mathcal{R}^G(U_1,U_2)$. Thus for fixed
$(D_1,D_2,D_3)$,
\begin{eqnarray}
R_1+R_2=\frac{1}{2}\log\frac{\sigma^2_X}{D_3}+\frac{1}{2}\log\psi(D_1,D_2,D_3).
\label{R_1+R_2}
\end{eqnarray}

\subsection{Successive Quantization for Gaussian Source}
If we view $U_1, U_2$ as two different quantizations of $X^G$ and
let $U_1-X^G$ and $U_2-X^G$ be their corresponding quantization
errors, then it follows
\begin{eqnarray}
\mathbb{E}[(U_1-X^G)(U_2-X^G)]&=&\mathbb{E}[(T_0+T_1)(T_0+T_2)]\nonumber\\
&=&\sigma^2_{T_0}-\sigma_{T_1}\sigma_{T_2}\nonumber\\
&=&\frac{D_3\sigma^2_X}{\sigma^2_X-D_3}-\sqrt{\left(\frac{D_1\sigma^2_X}{\sigma^2_X-D_1}-\frac{D_3\sigma^2_X}{\sigma^2_X-D_3}\right)\left(\frac{D_2\sigma^2_X}{\sigma^2_X-D_2}-\frac{D_3\sigma^2_X}{\sigma^2_X-D_3}\right)},\label{errorcorrelaton}
\end{eqnarray}
which is non-zero unless
$D_3=\left(1/{D_1}+1/{D_2}-1/{\sigma^2_X}\right)^{-1}$. The
existence of correlation between the quantization errors is the
main difficulty in designing the optimal MD quantization schemes.
To circumvent this difficulty, $U_1$ and $U_2$ can be represented in a
different form by using the Gram-Schmidt orthogonalization. It
yields that
\begin{eqnarray*}
B_1&=&X^G,\\
B_2&=&U_1-E(U_1|X^G)=U_1-X^G,\\
B_3&=&U_2-E(U_2|X^G,U_1)=U_2-a_1X^G-a_2U_1,
\end{eqnarray*}
where
\begin{eqnarray}
\label{eqn:gaussiana1a2}
a_1&=&\frac{\sigma^2_{T_1}+\sigma_{T_1}\sigma_{T_2}}{\sigma^2_{T_0}+\sigma^2_{T_1}},\label{a_1}\\
a_2&=&\frac{\sigma^2_{T_0}-\sigma_{T_1}\sigma_{T_2}}{\sigma^2_{T_0}+\sigma^2_{T_1}}.\label{a_2}
\end{eqnarray}
It can be computed that
\begin{eqnarray}
\mathbb{E}B^2_2&=&\sigma^2_{T_0}+\sigma^2_{T_1}, \label{B_2}\\
\mathbb{E}B^2_3&=&\frac{\sigma^2_{T_0}(\sigma_{T_1}+\sigma_{T_2})^2}{\sigma^2_{T_0}+\sigma^2_{T_1}}.
\label{B_3}
\end{eqnarray}

\begin{figure}[hbt]
\centering
\begin{psfrags}
\psfrag{x}[c]{$\mathbf{X}^G$}%
\psfrag{xhat1}[l]{$\mathbf{\hat{X}}^G_1$}%
\psfrag{xhat2}[l]{$\mathbf{\hat{X}}^G_2$}%
\psfrag{xhat3}[l]{$\mathbf{\hat{X}}^G_3$}%
\psfrag{w1}[c]{$\mathbf{U}_1$}%
\psfrag{w2}[l]{$\mathbf{U}_2$}%
\psfrag{t1}[c]{$\mathbf{B}_2$}%
\psfrag{t2}[c]{$\mathbf{B}_3$}%
\psfrag{a1}[c]{$a_1$}%
\psfrag{a2}[l]{$a_2$}%
\psfrag{alpha1}[c]{$\alpha_1$}%
\psfrag{alpha2}[c]{$\alpha_2$}%
\psfrag{beta1}[r]{$\beta_1$}%
\psfrag{beta2}[r]{$\beta_2$}%
\psfrag{plus}[c]{$+$}%
\psfrag{times}[c]{$\times$}%
\includegraphics[scale=1]{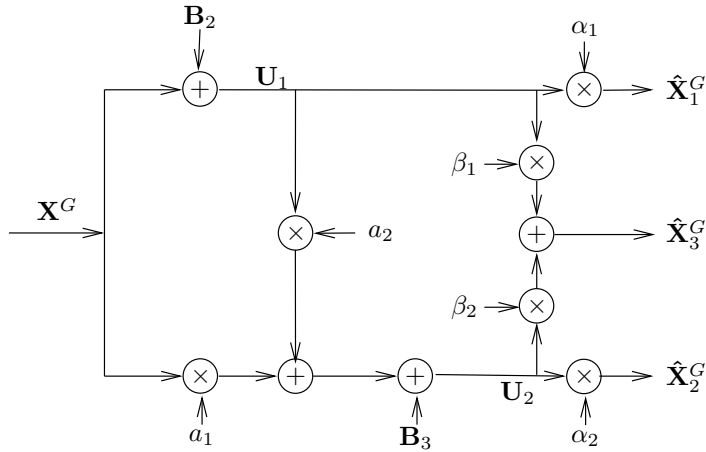}
\end{psfrags}
\caption{MD quantization scheme for $V^G_1$.\label{GVQ}}
\end{figure}

Now consider the quantization scheme for vertex $V^G_1$
of $\mathcal{R}^G(U_1,U_2)$ (see Fig. \ref{GVQ}). $R_1(V^G_1)$ is given by
\begin{eqnarray}
R_1(V^G_1)=I(X^G;U_1)=I(X^G;X^G+B_2).\label{R_1}
\end{eqnarray}
Since $U_2=\mathbb{E}(U_2|X^G,U_1)+B_3$, where $B_3$ is
independent of $(X^G,U_1)$, it follows that
$(X^G,U_1)\rightarrow\mathbb{E}(U_2|X^G,U_1)\rightarrow U_2$ form
a Markov chain. Clearly, $\mathbb{E}(U_2|X^G,U_1)\rightarrow
(X^G,U_1)\rightarrow U_2$ also form a Markov chain since
$\mathbb{E}(U_2|X^G,U_1)$ is a deterministic function of
$(X^G,U_1)$. These two Markov relationships imply that
\begin{eqnarray*}
I(X^G,U_1;U_2)=I(\mathbb{E}(U_2|X^G,U_1);U_2),
\end{eqnarray*}
and thus
\begin{eqnarray}
R_2(V^G_2)&=&I(X^G,U_1;U_2)\nonumber\\
&=&I(\mathbb{E}(U_2|X^G,U_1);U_2)\nonumber\\
&=&I(a_1X^G+a_2U_1;a_1X^G+a_2U_1+B_3).\label{R_2}
\end{eqnarray}
Although the above expressions are all of single letter type, it
does not mean that symbol by symbol operations can achieve the
optimal bound. Instead, when interpreting these information
theoretic results, one should think of a system that operates on
long blocks. Roughly speaking, (\ref{R_1}) and (\ref{R_2}) imply that
\begin{enumerate}
\item Encoder 1 is a quantizer of rate $R_1(V^G_1)$ whose input is $\mathbf{X}^G$ and output is
$\mathbf{U}_1$. The quantization error is
$\mathbf{B}_2=\mathbf{U}_1-\mathbf{X}^G$, which is a zero-mean
Gaussian vector with covariance matrix $\mathbb{E}B^2_2I_n$.
\item Encoder 2 is a quantizer of rate $R_2(V^G_1)$ with input $a_1\mathbf{X}^G+a_2\mathbf{U}_1$ and
output $\mathbf{U}_2$. The quantization error
$\mathbf{B}_3=\mathbf{U}_2-a_1\mathbf{X}^G-a_2\mathbf{U}_1$ is a
zero-mean Gaussian vector with covariance matrix
$\mathbb{E}B^2_3I_n$.
\end{enumerate}

Remarks:
\begin{enumerate}
\item $\mathbf{U}_1$ (or $\mathbf{U}_2$) is not a deterministic function of
$\mathbf{X}^G$ (or $a_1\mathbf{X}^G+a_2\mathbf{U}_1$), and for
classical quantizers the quantization noise is generally not
Gaussian. Thus strictly speaking, the ``noise-adding'' components
in Fig. \ref{GVQ} are not quantizers in the traditional sense. We
nevertheless refer to them as quantizers\footnote{This slight
abuse of the word ``quantizer'' can be justified in the context of
ECDQ (as we will show in the next section) since the quantization
noise of the optimal lattice quantizer is indeed asymptotically
Gaussian; furthermore, the quantization noise is indeed
independent of the input for ECDQ \cite{Feder}.} in this section
for simplicity.
\item $\mathbf{U}_1$ is revealed to decoder 1 and decoder 3, and
$\mathbf{U}_2$ is revealed to decoder 2 and decoder 3. Decoder $i$
approximates $\mathbf{X}$ by $\mathbf{\hat X}_i=\alpha_i
\mathbf{U}_i$, $i=1,2$. Decoder 3 approximates $\mathbf{X}$ by
$\mathbf{\hat X}_3=\beta_1\mathbf{U}_1+\beta_2 \mathbf{U}_2$. The
rates to reveal $\mathbf{U}_1$ and $\mathbf{U}_2$ are the rates of
description 1 and description 2, respectively.
\item From Fig. \ref{GVQ}, it is obvious that the MD quantization for $V^G_1$
is essentially the Gram-Schmidt orthogonalization of
$(\mathbf{X}^G,\mathbf{U}_1,\mathbf{U}_2)$. As previously shown in
Section II, the Gram-Schmidt orthogonalization can be simulated by
sequential (dithered) quantization. The formal description and
analysis of this quantization scheme in the context of multiple
descriptions for general sources will be given in Section V.
\end{enumerate}

\subsection{Successive Quantization with Quantization Splitting for Gaussian Source}

Now we study the quantization scheme for an arbitrary rate pair
$(R^G_1,R^G_2)$ on the dominant face of $\mathcal{R}^G(U_1,U_2)$.
Note that since the rate sum $R^G_1+R^G_2$ is given by
(\ref{R_1+R_2}), $(R^G_1,R^G_2)$ only has one degree of freedom.

Let $U'_2=X^G+T_0+T_2+T_3$, where $T_3$ is zero-mean, Gaussian and
independent of $(X^G,T_0,T_1,T_2)$. It is easy to verify that
$(X^G,U_1)\rightarrow U_2\rightarrow U'_2$ form a Markov chain.
Applying the Gram-Schmidt orthogonalization algorithm to
$(X^G,U'_2,U_1)$, we have
\begin{eqnarray*}
\widetilde{B}_1&=&X,\\
\widetilde{B}_2&=&U'_2-\mathbb{E}(U'_2|X^G)=U'_2-X^G,\\
\widetilde{B}_3&=&U_1-\mathbb{E}(U_1|X^G,U'_2)=U_1-b_1X^G-b_2U'_2,
\end{eqnarray*}
where
\begin{eqnarray}
b_1&=&\frac{\sigma^2_{T_2}+\sigma^2_{T_3}+\sigma_{T_1}\sigma_{T_2}}{\sigma^2_{T_0}+\sigma^2_{T_2}+\sigma^2_{T_3}},\label{b_1}\\
b_2&=&\frac{\sigma^2_{T_0}-\sigma_{T_1}\sigma_{T_2}}{\sigma^2_{T_0}+\sigma^2_{T_2}+\sigma^2_{T_3}}.\label{b_2}
\end{eqnarray}
The variances of $\widetilde{B}_2$ and $\widetilde{B}_3$ are
\begin{eqnarray}
\mathbb{E}{\widetilde{B}_2}^2&=&\sigma^2_{T_0}+\sigma^2_{T_2}+\sigma^2_{T_3},\label{EB_2}\\
\mathbb{E}{\widetilde{B}_3}^2&=&\frac{\sigma^2_{T_0}(\sigma_{T_1}+\sigma_{T_2})^2+\sigma^2_{T_3}(\sigma^2_{T_0}+\sigma^2_{T_1})}{\sigma^2_{T_0}+\sigma^2_{T_2}+\sigma^2_{T_3}}.\label{EB_3}
\end{eqnarray}
Since $U_1=\mathbb{E}(U_12|X^G,U'_2)+\widetilde{B}_3$, where
$\widetilde{B}_3$ is independent of $(X^G,U'_2)$, it follows that
$(X^G,U'_2)\rightarrow\mathbb{E}(U_1|X^G,U'_2)\rightarrow U_1$
form a Markov chain. Clearly, $\mathbb{E}(U_1|X^G,U'_2)\rightarrow
(X^G,U'_2)\rightarrow U_1$ also form a Markov chain because
$\mathbb{E}(U_1|X^G,U'_2)$ is determined by $(X^G,U'_2)$. Thus we
have
\begin{eqnarray*}
I(X^G,U'_2;U_1)=I(\mathbb{E}(U_1|X^G,U'_2);U_1),
\end{eqnarray*}
and this gives
\begin{eqnarray*}
R^G_1&=&I(X^G,U'_2;U_1)\\
&=&I(\mathbb{E}(U_1|X^G,U'_2);U_1)\\
&=&\frac{1}{2}\log\frac{\mathbb{E}U^2_1}{\mathbb{E}{\widetilde{B}_3}^2}\\
&=&\frac{1}{2}\log\frac{(\sigma^2_X+\sigma^2_{T_0}+\sigma^2_{T_1})(\sigma^2_{T_0}+\sigma^2_{T_2}+\sigma^2_{T_3})}{\sigma^2_{T_0}(\sigma_{T_1}+\sigma_{T_2})^2+\sigma^2_{T_3}(\sigma^2_{T_0}+\sigma^2_{T_1})}.
\end{eqnarray*}
Hence $\sigma^2_{T_3}$ is uniquely determined by
\begin{eqnarray}
\sigma^2_{T_3}=\frac{\sigma^2_{T_0}(\sigma_{T_1}+\sigma_{T_2})^22^{2R_1}-(\sigma^2_{T_0}+\sigma^2_{T_2})(\sigma^2_X+\sigma^2_{T_0}+\sigma^2_{T_1})}{\sigma^2_X+\sigma^2_{T_0}+\sigma^2_{T_1}-2^{2R_1}(\sigma^2_{T_0}+\sigma^2_{T_1})}.\label{T_3}
\end{eqnarray}
We also can readily compute
\begin{eqnarray*}
R^G_2&=&I(X^G;U_1)+I(X^G;U_2)+I(U_1;U_2|X^G)-R^G_1\\
&=&\frac{1}{2}\log\frac{[\sigma^2_{T_0}(\sigma_{T_1}+\sigma_{T_2})^2+\sigma^2_{T_3}(\sigma^2_{T_0}+\sigma^2_{T_1})](\sigma^2_X+\sigma^2_{T_0}+\sigma^2_{T_2})}{\sigma^2_{T_0}(\sigma^2_{T_0}+\sigma^2_{T_2}+\sigma^2_{T_3})(\sigma_{T_1}+\sigma_{T_2})^2},
\end{eqnarray*}
\begin{eqnarray*}
R^G_1|_{\sigma^2_{T_3}=0}&=&\frac{1}{2}\log\frac{(\sigma^2_X+\sigma^2_{T_0}+\sigma^2_{T_1})(\sigma^2_{T_0}+\sigma^2_{T_2})}{\sigma^2_{T_0}(\sigma_{T_1}+\sigma_{T_2})^2}=R_1(V^G_2),\\
R^G_2|_{\sigma^2_{T_3}=0}&=&\frac{1}{2}\log\frac{\sigma^2_X+\sigma^2_{T_0}+\sigma^2_{T_2}}{\sigma^2_{T_0}+\sigma^2_{T_2}}=R_2(V^G_2),\\
R^G_1|_{\sigma^2_{T_3}=\infty}&=&\frac{1}{2}\log\frac{\sigma^2_X+\sigma^2_{T_0}+\sigma^2_{T_1}}{\sigma^2_{T_0}+\sigma^2_{T_1}}=R_1(V^G_1),
\end{eqnarray*}
and
\begin{eqnarray*}
R^G_2|_{\sigma^2_{T_3}=\infty}&=&\frac{1}{2}\log\frac{(\sigma^2_X+\sigma^2_{T_0}+\sigma^2_{T_2})(\sigma^2_{T_0}+\sigma^2_{T_1})}{\sigma^2_{T_0}(\sigma_{T_1}+\sigma_{T_2})^2}=R_2(V^G_1).
\end{eqnarray*}
Hence, as $\sigma^2_{T_3}$ varies from 0 to $\infty$, all the rate
pairs on the dominant face of $\mathcal{R}^G(U_1,U_2)$ are
achieved.

For rate pair $(R^G_1,R^G_2)$, we have
\begin{eqnarray}
R^G_1&=&I(X^G,U'_2;U_1)=I(\mathbb{E}(U_1|X^G,U'_2);U_1)\nonumber\\
&=&I(b_1X^G+b_2U'_2;b_1X^G+b_2U'_2+\widetilde{B}_3),\label{GR_1}\\
R^G_2&=&I(X^G;U'_2)+I(X^G,U_1;U_2|U'_2)\nonumber\\
&=&I(X^G;X^G+\widetilde{B}_2)+I(X^G,U_1;U_2|U'_2).\nonumber
\end{eqnarray}
To remove the conditioning term $U'_2$ in $I(X^G,U_1;U_2|U'_2)$,
we apply the Gram-Schmidt procedure to $(U'_2,X^G,U_1,U_2)$. It
yields
\begin{eqnarray*}
\overline{B}_1&=&U'_2,\\
\overline{B}_2&=&X-\mathbb{E}(X^G|\overline{B}_1)=X^G-b_3\overline{B}_1,\\
\overline{B}_3&=&U_1-\mathbb{E}(U_1|\overline{B}_1)-\mathbb{E}(U_1|\overline{B}_2)=U_1-b_4\overline{B}_1-b_5\overline{B}_2,\\
\overline{B}_4&=&U_2-\sum\limits_{i=1}^3\mathbb{E}(U_2|\overline{B}_i)=U_2-b_6\overline{B}_1-b_7\overline{B}_2-b_8\overline{B}_3,
\end{eqnarray*}
where
\begin{eqnarray}
b_3&=&\frac{\sigma^2_X}{\sigma^2_X+\sigma^2_{T_0}+\sigma^2_{T_2}+\sigma^2_{T_3}},\label{b_3}\\
b_4&=&\frac{\sigma^2_X+\sigma^2_{T_0}-\sigma_{T_1}\sigma_{T_2}}{\sigma^2_X+\sigma^2_{T_0}+\sigma^2_{T_2}+\sigma^2_{T_3}},\label{b_4}\\
b_5&=&\frac{\sigma^2_{T_2}+\sigma^2_{T_3}+\sigma_{T_1}\sigma_{T_2}}{\sigma^2_{T_0}+\sigma^2_{T_2}+\sigma^2_{T_3}},\label{b_5}\\
b_6&=&\frac{\sigma^2_X+\sigma^2_{T_0}+\sigma^2_{T_2}}{\sigma^2_X+\sigma^2_{T_0}+\sigma^2_{T_2}+\sigma^2_{T_3}},\label{b_6}\\
b_7&=&\frac{\sigma^2_{T_3}}{\sigma^2_{T_0}+\sigma^2_{T_2}+\sigma^2_{T_3}},\label{b_7}\\
b_8&=&\frac{\sigma^2_{T_3}(\sigma^2_{T_0}-\sigma_{T_1}\sigma_{T_2})}{\sigma^2_{T_0}(\sigma_{T_1}+\sigma_{T_2})^2+\sigma^2_{T_3}(\sigma^2_{T_0}+\sigma^2_{T_1})}.\label{b_8}
\end{eqnarray}
The following quantities are also needed
\begin{eqnarray}
\mathbb{E}\overline{B}^2_2&=&\frac{\sigma^2_X(\sigma^2_{T_0}+\sigma^2_{T_2}+\sigma^2_{T_3})}{\sigma^2_X+\sigma^2_{T_0}+\sigma^2_{T_2}+\sigma^2_{T_3}},\nonumber\\
\mathbb{E}\overline{B}^2_3&=&\frac{\sigma^2_{T_0}(\sigma_{T_1}+\sigma_{T_2})^2+\sigma^2_{T_3}(\sigma^2_{T_0}+\sigma^2_{T_1})}{\sigma^2_{T_0}+\sigma^2_{T_2}+\sigma^2_{T_3}},\nonumber\\
\mathbb{E}\overline{B}^2_4&=&\frac{\sigma^2_{T_3}(\sigma^2_X+\sigma^2_{T_0}+\sigma^2_{T_2})}{\sigma^2_X+\sigma^2_{T_0}+\sigma^2_{T_2}+\sigma^2_{T_3}}-b^2_7\mathbb{E}\overline{B}^2_2-b^2_8\mathbb{E}\overline{B}^2_3\nonumber\\
&=&\frac{\sigma^2_{T_3}(\sigma^2_X+\sigma^2_{T_0}+\sigma^2_{T_2})}{\sigma^2_X+\sigma^2_{T_0}+\sigma^2_{T_2}+\sigma^2_{T_3}}-\frac{\sigma^2_X\sigma^4_{T_3}}{(\sigma^2_X+\sigma^2_{T_0}+\sigma^2_{T_2}+\sigma^2_{T_3})(\sigma^2_{T_0}+\sigma^2_{T_2}+\sigma^2_{T_3})}\nonumber\\
&&-\frac{\sigma^4_{T_3}(\sigma^2_{T_0}-\sigma_{T_1}\sigma_{T_2})^2}{[\sigma^2_{T_0}(\sigma_{T_1}+\sigma_{T_2})^2+\sigma^2_{T_3}(\sigma^2_{T_0}+\sigma^2_{T_1})](\sigma^2_{T_0}+\sigma^2_{T_2}+\sigma^2_{T_3})}.\label{EB_4}
\end{eqnarray}
Now write
\begin{eqnarray*}
I(X^G,U_1;U_2|U'_2)&=&I(b_3\overline{B}_1+\overline{B}_2,b_4\overline{B}_1+b_5\overline{B}_2+\overline{B}_3;b_6\overline{B}_1+b_7\overline{B}_2+b_8\overline{B}_3+\overline{B}_4|\overline{B}_1)\\
&=&I(\overline{B}_2,b_5\overline{B}_2+\overline{B}_3;b_7\overline{B}_2+b_8\overline{B}_3+\overline{B}_4|\overline{B}_1).
\end{eqnarray*}
Since $\overline{B}_1$ is independent of
$(\overline{B}_2,\overline{B}_3,\overline{B}_4)$, it follows that
\begin{eqnarray*}
I(\overline{B}_2,b_5\overline{B}_2+\overline{B}_3;b_7\overline{B}_2+b_8\overline{B}_3+\overline{B}_4|\overline{B}_1)=I(\overline{B}_2,b_5\overline{B}_2+\overline{B}_3;b_7\overline{B}_2+b_8\overline{B}_3+\overline{B}_4).
\end{eqnarray*}
The fact that $\overline{B}_4$ is independent of
$(\overline{B}_2,\overline{B}_3)$ implies that
$(\overline{B}_2,b_5\overline{B}_2+\overline{B}_3)\rightarrow
b_7\overline{B}_2+b_8\overline{B}_3\rightarrow
b_7\overline{B}_2+b_8\overline{B}_3+\overline{B}_4$ form a Markov
chain. This observation, along with the fact that
$b_7\overline{B}_2+b_8\overline{B}_3$ is a deterministic function
of $(\overline{B}_2,b_5\overline{B}_2+\overline{B}_3)$, yields
\begin{eqnarray*}
I(\overline{B}_2,b_5\overline{B}_2+\overline{B}_3;b_7\overline{B}_2+b_8\overline{B}_3+\overline{B}_4)=I(b_7\overline{B}_2+b_8\overline{B}_3;b_7\overline{B}_2+b_8\overline{B}_3+\overline{B}_4).
\end{eqnarray*}
Hence
\begin{eqnarray}
R^G_2=I(X^G;X^G+\widetilde{B}_2)+I(b_7\overline{B}_2+b_8\overline{B}_3;b_7\overline{B}_2+b_8\overline{B}_3+\overline{B}_4).
\end{eqnarray}
Moreover, since
\begin{eqnarray}
b_7{\overline{B}_2}+b_8{\overline{B}_3}&=&(b_7-b_5b_8)X^G+b_8U_1+(b_3b_5b_8-b_3b_7-b_4b_8){U'_2},
\label{relation1}\\
b_7{\overline{B}_2}+b_8{\overline{B}_3}+{\overline{B}_4}&=&U_2-b_6{U'_2}.\label{relation2}
\end{eqnarray}
it follows that
\begin{eqnarray}
R^G_2&=&I(X^G;X^G+\widetilde{B}_2)+I\left((b_7-b_5b_8)X^G+b_8U_1+(b_3b_5b_8-b_3b_7-b_4b_8)U'_2;U_2-b_6{U'_2}\right).\label{GR_2}
\end{eqnarray}
Let $b^*_1=b_1$, $b^*_2=b_2$, $b^*_3=b_7-b_5b_8$, $b^*_4=b_8$,
$b^*_5=b_3b_5b_8-b_3b_7-b_4b_8$ and $b^*_6=b_6$. Then (\ref{GR_1}) and (\ref{GR_2}) can be simplified to
\begin{eqnarray}
R^G_1&=&I(b_1X^G+b_2U'_2;U_1)=I(b^*_1X^G+b^*_2U'_2;b^*_1X^G+b^*_2U'_2+\widetilde{B}_3),\label{GR1}\\
R^G_2&=&I(X^G;U'_2)+I\left(b^*_3X^G+b^*_4U_1+b^*_5U'_2;U_2-b^*_6{U'_2}\right)\nonumber\\
&=&I(X^G;X^G+\widetilde{B}_2)+I\left(b^*_3X^G+b^*_4U_1+b^*_5U'_2;b^*_3X^G+b^*_4U_1+b^*_5U'_2+\overline{B}_4\right)\label{GR2}.
\end{eqnarray}

\begin{figure}[thb]
\centering
\begin{psfrags}
\psfrag{x}[c]{$\mathbf{X}^G$}%
\psfrag{xhat1}[l]{$\mathbf{\hat{X}}^G_1$}%
\psfrag{xhat2}[l]{$\mathbf{\hat{X}}^G_2$}%
\psfrag{xhat3}[l]{$\mathbf{\hat{X}}^G_3$}%
\psfrag{delta}[c]{$ $}%
\psfrag{w1}[c]{$\mathbf{U}_1$}%
\psfrag{w2}[c]{$\mathbf{U}_2$}%
\psfrag{w2'}[c]{${\mathbf{U}'_2}$}%
\psfrag{t0}[c]{$\mathbf{\widetilde{B}}_2$}%
\psfrag{t1}[c]{$\mathbf{\widetilde{B}}_3$}%
\psfrag{t2}[c]{$\mathbf{\overline{B}}_4$}%
\psfrag{b1}[c]{$b^*_1$}%
\psfrag{b2}[r]{$b^*_2$}%
\psfrag{b3}[c]{$b^*_3$}%
\psfrag{b4}[l]{$b^*_4$}%
\psfrag{b5}[c]{$b^*_5$}%
\psfrag{b6}[c]{$b^*_6$}%
\psfrag{times}[c]{$\times$}%
\psfrag{plus}[c]{$+$}%
\psfrag{alpha1}[c]{$\alpha_1$}%
\psfrag{alpha2}[c]{$\alpha_2$}%
\psfrag{beta1}[c]{$\beta_1$}%
\psfrag{beta2}[c]{$\beta_2$}%
\includegraphics[scale=1]{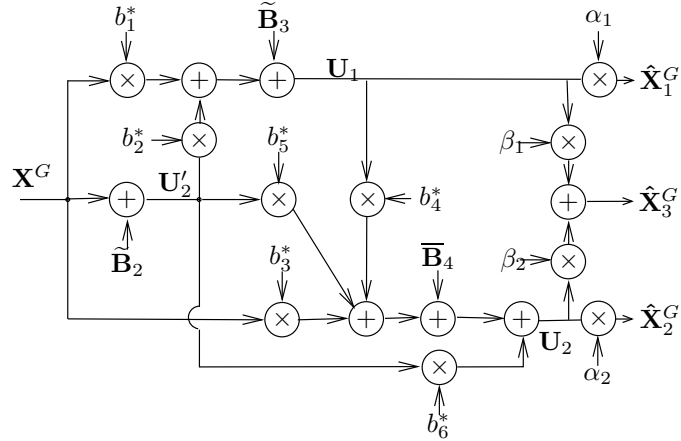}
\end{psfrags}
\caption{MD quantization scheme for $(R^G_1,R^G_2)$.\label{GGQ}}
\end{figure}

Equations (\ref{GR1}) and (\ref{GR2}) suggest the following
optimal MD quantization system (also see Fig. \ref{GGQ}):
\begin{enumerate}
\item Encoder 1 is a quantizer of rate $R^G_1$ with input $b^*_1\mathbf{X}^G+b^*_2{\mathbf{U}'_2}$ and output $\mathbf{U}_1$. The quantization error $\mathbf{\widetilde{B}}_3=\mathbf{U}_1-b^*_1\mathbf{X}-b^*_2{\mathbf{U}'_2}$ is Gaussian with covariance matrix $\mathbb{E}{\widetilde{B}_3}^2I_n$.
\item Encoder 2 consists of two quantizers. The rate of the first quantizer is $R^G_{2,1}$. Its input and output are $\mathbf{X}^G$ and ${\mathbf{U}'_2}$ respectively.
Its quantization error
$\mathbf{\widetilde{B}}_2={\mathbf{U}'_2}-\mathbf{X}^G$ is
Gaussian with covariance matrix
$\mathbb{E}{\widetilde{B}_2}^2I_n$. The second quantizer is of
rate $R^G_{2,2}$. It has input
$b^*_3\mathbf{X}^G+b^*_4\mathbf{U}_1+b^*_5\mathbf{U}'_2$ and
output $\mathbf{U}_2-b^*_6{\mathbf{U}'_2}$. Its quantization error
${\mathbf{\overline{B}}_4}$ is Gaussian with covariance matrix
$\mathbb{E}{\overline{B}^2_4}I_n$. The sum-rate of these two
quantizers is the rate of encoder 2, which is $R^G_2$. Here
$R^G_{2,1}=I(X^G;U'_2)$, and $R^G_{2,2}=R^G_2-R^G_{2,1}$.
\end{enumerate}
Remarks:
\begin{enumerate}
\item $\mathbf{U}_1$ is revealed to decoder 1 and decoder 3. ${\mathbf{U}'_2}$ and
$\mathbf{U}_2-b^*_6{\mathbf{U}'_2}$ are revealed to decoder 2 and
decoder 3. Decoder 1 constructs $\mathbf{\hat X}^G_1=\alpha_1
\mathbf{U}_1$. Decoder 2 first constructs $\mathbf{U}_2$ using
${\mathbf{U}'_2}$ and $\mathbf{U}_2-b^*_6{\mathbf{U}'_2}$, and
then constructs $\mathbf{\hat X}^G_2=\alpha_2 \mathbf{U}_2$.
Decoder 3 also first constructs $\mathbf{U}_2$, then constructs
$\mathbf{\hat X}_3=\beta_1 \mathbf{U}_1+\beta_2 \mathbf{U}_2$. It
is clear what decoder 2 and decoder 3 want is $\mathbf{U}_2$, not
${\mathbf{U}'_2}$ or $\mathbf{U}_2-b^*_6{\mathbf{U}'_2}$.
Furthermore, the construction of $\mathbf{U}_2$ can be moved to
the encoder part. That is, encoder 2 can directly construct
$\mathbf{U}_2$ with ${\mathbf{U}'_2}$ and
$\mathbf{U}_2-b^*_6{\mathbf{U}'_2}$; then, only $\mathbf{U}_2$
needs to be revealed to decoder 2 and decoder 3.
\item That ${\mathbf{\overline{B}}_4}$ is independent of
$(\mathbf{X}^G,\mathbf{U}_1,{\mathbf{U}'_2})$ and
$({\mathbf{\widetilde{B}}_2},{\mathbf{\widetilde{B}}_3})$ is a
deterministic function of
$(\mathbf{X}^G,\mathbf{U}_1,{\mathbf{U}'_2})$ implies that
${\mathbf{\overline{B}}_4}$ is independent of
$({\mathbf{\widetilde{B}}_2},{\mathbf{\widetilde{B}}_3})$.
\item The MD quantization scheme for $(R^G_1,R^G_2)$ essentially
consists of two Gram-Schmidt procedures, one operating on
$(\mathbf{X}^G,\mathbf{U}'_2,\mathbf{U}_1)$ and the other on
$(\mathbf{U}'_2,\mathbf{X}^G,\mathbf{U}_1,\mathbf{U}_2)$. The
formal description and analysis of this scheme from the
perspective of dithered quantization is left to Section V.
\end{enumerate}

\subsection{Discussion of Special Cases}
Next we consider three cases for which the MD quantizers have some
special properties.

\subsubsection{The case $D_3=(1/D_1+1/D_2-1/\sigma^2_X)^{-1}$}

\begin{figure}[hbt]
\centering
\begin{psfrags}
\psfrag{x}[c]{$\mathbf{X}^G$}%
\psfrag{xhat1}[l]{$\mathbf{\hat{X}}^G_1$}%
\psfrag{xhat2}[l]{$\mathbf{\hat{X}}^G_2$}%
\psfrag{xhat3}[l]{$\mathbf{\hat{X}}^G_3$}%
\psfrag{w1}[l]{$\mathbf{U}_1$}%
\psfrag{w2}[l]{$\mathbf{U}_2$}%
\psfrag{t0t1}[c]{$\mathbf{B}_2$}%
\psfrag{t0t2}[c]{$\mathbf{B}_3$}%
\psfrag{alpha1}[c]{$\alpha_1$}%
\psfrag{alpha2}[c]{$\alpha_2$}%
\psfrag{beta1}[r]{$\beta_1$}%
\psfrag{beta2}[r]{$\beta_2$}%
\psfrag{plus}[c]{$+$}%
\psfrag{times}[c]{$\times$}%
\includegraphics[scale=1]{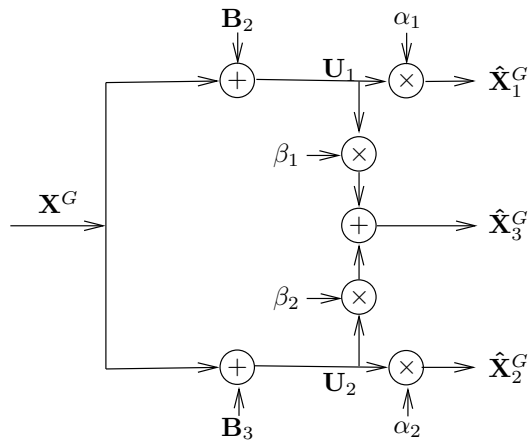}
\end{psfrags}
\caption{Special case:
$D_3=(1/D_1+1/D_2-1/\sigma^2_X)^{-1}$.\label{GE1Q}}
\end{figure}

For this case, we have
\begin{eqnarray*}
\left(R_1(V^G_1),R_2(V^G_1)\right)=\left(R_1(V^G_2),R_2(V^G_2)\right)=\left(\frac{1}{2}\log\frac{\sigma^2_X}{D_1},\frac{1}{2}\log\frac{\sigma^2_X}{D_2}\right),
\end{eqnarray*}
which is referred to as the case of no excess marginal rate. Since
the dominant face of $\mathcal{R}^G(U_1,U_2)$ degenerates to a
single point, the quantization splitting becomes unnecessary.
Moreover, (\ref{errorcorrelaton}) gives
$\mathbb{E}(U_1-X^G)(U_2-X^G)=0$, i.e., two quantization errors
are uncorrelated (and thus independent since $(X^G, U_1, U_2)$ are
jointly Gaussian) in this case. This
further implies that $U_1\rightarrow X^G\rightarrow U_2$ form a
Markov chain. Due to this fact, the Gram-Schmidt othogonalization
for $(X^G,U_1,U_2)$ becomes particularly simple:
\begin{eqnarray*}
B_1&=&X^G,\\
B_2&=&U_1-\mathbb{E}(U_1|X)=U_1-X^G,\\
B_3&=&U_2-\mathbb{E}(U_2|X,U_1)=\mathbb{E}(U_2|X^G)=U_2-X^G,
\end{eqnarray*}
and
\begin{eqnarray}
\mathbb{E}B^2_2&=&\sigma^2_{T_0}+\sigma^2_{T_1},\\
\mathbb{E}B^2_3&=&\sigma^2_{T_0}+\sigma^2_{T_2}.
\end{eqnarray}
The resulting MD quantization system is
\begin{enumerate}
\item Encoder 1 is a quantizer of rate $I(X^G;U_1)$ whose input is $\mathbf{X}^G$ and output is
$\mathbf{U}_1$. The quantization error $\mathbf{B}_2$ is
(approximately) Gaussian with covariance matrix
$\mathbb{E}B^2_2I_n$.
\item Encoder 2 is a quantizer of rate $I(X^G;U_2)$ with input $\mathbf{X}^G$ and
output $\mathbf{U}_2$. The quantization error $\mathbf{B}_3$ is
(approximately) Gaussian with covariance matrix
$\mathbb{E}B^2_3I_n$.
\end{enumerate}
So for this case, the conventional separate quantization scheme
\cite{Zamir2} suffices. See Fig. \ref{GE1Q}.

\subsubsection{The case $D_3=D_1+D_2-\sigma^2_X$}

For this case, we have
\begin{eqnarray*}
R_1(V^G_1)+R_2(V^G_1)=R_1(V^G_2)+R_2(V^G_2)=\frac{1}{2}\log\frac{\sigma^2_X}{D_3},
\end{eqnarray*}
which corresponds to the case of no excess sum-rate. Since
$D_3=D_1+D_2-\sigma^2_X$ implies $\sigma^2_X-D_1=D_2-D_3$ and
$\sigma^2_X-D_2=D_1-D_3$, it follows that
\begin{eqnarray}
\mathbb{E}(U_1U_2)&=&\sigma^2_X+\sigma^2_{T_0}-\sigma_{T_1}\sigma_{T_2}\nonumber\\
&=&\sigma^2_X+\frac{D_3\sigma^2_X}{\sigma^2_X-D_3}-\sqrt{\left(\frac{D_1\sigma^2_X}{\sigma^2_X-D_1}-\frac{D_3\sigma^2_X}{\sigma^2_X-D_3}\right)\left(\frac{D_2\sigma^2_X}{\sigma^2_X-D_2}-\frac{D_3\sigma^2_X}{\sigma^2_X-D_3}\right)}\nonumber\\
&=&\sigma^2_X+\frac{D_3\sigma^2_X}{\sigma^2_X-D_3}-\sqrt{\frac{\sigma^8_X(D_1-D_3)(D_2-D_3)}{(\sigma^2_X-D_1)(\sigma^2_X-D_2)(\sigma^2_X-D_3)^2}}\nonumber\\
&=&\sigma^2_X+\frac{D_3\sigma^2_X}{\sigma^2_X-D_3}-\frac{\sigma^4_X}{\sigma^2_X-D_3}\nonumber\\
&=&0. \label{independent}
\end{eqnarray}
Since $U_1$ and $U_2$ are jointly Gaussian, (\ref{independent})
implies $U_1$ and $U_2$ are independent. This is consistent with
the result in \cite{Ahlswede} although only discrete memoryless
sources were addressed there due to technical reasons. The
interpretation of (\ref{independent}) is that the outputs of the
two encoders (/quantizers) should be independent. This is
intuitively clear because otherwise these two outputs can be
further compressed to
 reduce the sum-rate but still achieve distortion $D_3$ for
the joint description. But that would violate the rate distortion
theorem, since $\frac{1}{2}\log\frac{\sigma^2_X}{D_3}$ is the
minimum $D_3$-admissible rate for the quadratic Gaussian case.

\begin{figure}[hbt]
\centering
\begin{psfrags}
\psfrag{r1}[c]{$R_1$}%
\psfrag{r2}[r]{$R_2$}%
\psfrag{d1}[c]{$\frac{1}{2}\log \frac{\sigma^2_X}{D_1}$}%
\psfrag{d2}[r]{$\frac{1}{2}\log \frac{\sigma^2_X}{D_2}$}%
\psfrag{d3}[c]{$\frac{1}{2}\log \frac{\sigma^2_X}{D_3}$}%
\psfrag{text}[l]{achievable with timesharing}%
\includegraphics[scale=1]{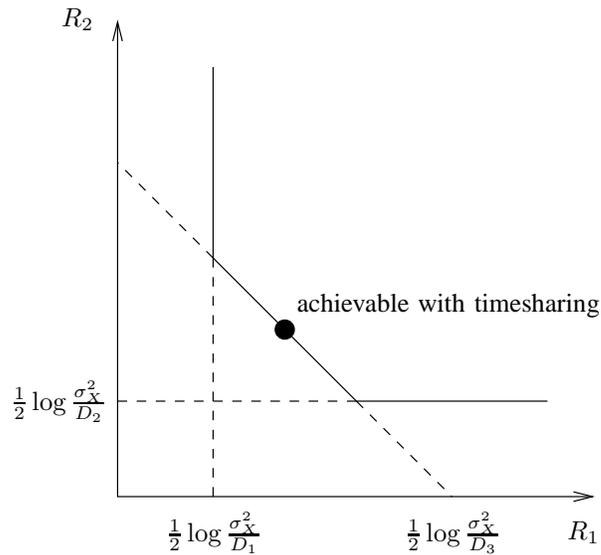}
\end{psfrags}
\caption{Special case: $D_3=D_1+D_2-\sigma^2_X$.\label{GE2R}}
\end{figure}

Now consider the following timesharing scheme: Construct an
optimal rate-distortion codebook of rate
$\frac{1}{2}\log\frac{\sigma^2_X}{D_3}$ that can achieve
distortion $D_3$. Encoder 1 uses this codebook a fraction $\gamma$
($0\leq\gamma\leq 1$) of the time and encoder 2 uses this codebook
the remaining $1-\gamma$ fraction of the time. For this scheme,
the resulting rates and distortions are given by
$R_1=\frac{\gamma}{2}\log\frac{\sigma^2_X}{D_3}$,
$R_2=\frac{1-\gamma}{2}\log\frac{\sigma^2_X}{D_3}$, $D_1=\gamma
D_3+(1-\gamma)\sigma^2_X$, $D_2=(1-\gamma)D_3+\gamma\sigma^2_X$,
and $D_3$. Conversely, for any fixed $D^*_1$ and $D^*_2$ with
$D^*_1+D^*_2=\sigma^2_X+D_3$, there exists a $\gamma^*\in[0,1]$
such that $D^*_1=\gamma^* D_3+(1-\gamma^*)\sigma^2_X$,
$D^*_2=(1-\gamma^*)D_3+\gamma^*\sigma^2_X$. The associated rates
are $R_1=\frac{\gamma^*}{2}\log\frac{\sigma^2_X}{D_3}$ and
$R_2=\frac{1-\gamma^*}{2}\log\frac{\sigma^2_X}{D_3}$.
 So, the timesharing scheme can achieve any point on the dominant face of
the rate region for the special case $D_3=D_1+D_2-\sigma^2_X$ (See
Fig. \ref{GE2R}). Specifically,  for the symmetric case where
$D^*_1=D^*_2=\frac{1}{2}(D_3+\sigma^2_X)$, we have
$\gamma^*=\frac{1}{2}$ and
$R_1=R_2=\frac{1}{4}\log\frac{\sigma^2_X}{D_3}$.

\subsubsection{The symmetric case $D_1=D_2\triangleq D_{12}$} The
symmetric case is of particular practical importance.  Moreover,
many previously derived expressions take simpler forms if
$D_1=D_2$. Specifically, we have
\begin{eqnarray*}
\sigma^2_{T_1}=\sigma^2_{T_2}=\frac{D_{12}\sigma^2_X}{\sigma^2_X-D_{12}}-\frac{D_3\sigma^2_X}{\sigma^2_X-D_3}\triangleq\sigma^2_{T_{12}},
\end{eqnarray*}
and
\begin{eqnarray*}
\alpha_1&=&\alpha_2=\frac{\sigma^2_X}{\sigma^2_X+\sigma^2_{T_0}+\sigma^2_{T_{12}}}\triangleq\alpha,\\
\beta_1&=&\beta_2=\frac{\sigma^2_X}{2(\sigma^2_X+\sigma^2_{T_0})}\triangleq\beta.
\end{eqnarray*}
The coordinates of $V^G_1$ and $V^G_2$ become
\begin{eqnarray*}
R_1(V^G_1)&=&R_2(V^G_2)=\frac{1}{2}\log\frac{\sigma^2_X}{D_{12}},\\
R_2(V^G_1)&=&R_1(V^G_2)=\frac{1}{2}\log\frac{D_{12}(\sigma^2_X-D_3)^2}{4D_3(\sigma^2_X-D_{12})(D_{12}-D_3)}.
\end{eqnarray*}
The expressions for $R^G_1$ and $R^G_2$ can be simplified to
\begin{eqnarray*}
R^G_1&=&\frac{1}{2}\log\frac{(\sigma^2_X+\sigma^2_{T_0}+\sigma^2_{T_{12}})(\sigma^2_{T_0}+\sigma^2_{T_{12}}+\sigma^2_{T_3})}{4\sigma^2_{T_0}\sigma^2_{T_{12}}+\sigma^2_{T_3}(\sigma^2_{T_0}+\sigma^2_{T_{12}})},\\
R^G_2&=&\frac{1}{2}\log\frac{[4\sigma^2_{T_0}\sigma^2_{T_{12}}+\sigma^2_{T_3}(\sigma^2_{T_0}+\sigma^2_{T_{12}})](\sigma^2_X+\sigma^2_{T_0}+\sigma^2_{T_{12}})}{4\sigma^2_{T_0}\sigma^2_{T_{12}}(\sigma^2_{T_0}+\sigma^2_{T_{12}}+\sigma^2_{T_3})}.
\end{eqnarray*}
To keep the rates equal, i.e., $R^G_1=R^G_2$, it must be true that
\begin{eqnarray*}
&&4\sigma^2_{T_0}\sigma^2_{T_{12}}+\sigma^2_{T_3}(\sigma^2_{T_0}+\sigma^2_{T_{12}})=2\sigma_{T_0}\sigma_{T_{12}}(\sigma^2_{T_0}+\sigma^2_{T_{12}}+\sigma^2_{T_3})\\
&\Leftrightarrow&(\sigma_{T_0}-\sigma_{T_{12}})^2(\sigma^2_{T_3}-2\sigma_{T_0}\sigma_{T_{12}})=0.
\end{eqnarray*}
If $\sigma_{T_0}\neq\sigma_{T_{12}}$, then
\begin{eqnarray}
\sigma^2_{T_3}&=&2\sigma_{T_0}\sigma_{T_{12}}\\
&=&2\sqrt{\frac{D_3\sigma^2_X}{\sigma^2_X-D_3}\left(\frac{D_{12}\sigma^2_X}{\sigma^2_X-D_{12}}-\frac{D_3\sigma^2_X}{\sigma^2_X-D_3}\right)}.
\end{eqnarray}
If $\sigma_{T_0}=\sigma_{T_{12}}$, then
\begin{eqnarray*}
R_1(V^G_1)=R_2(V^G_1)=R^G_1=R^G_2=R_1(V^G_2)=R_2(V^G_2),\quad\forall
\sigma^2_{T_3}\in\mathcal{R}^+,
\end{eqnarray*}
i.e., $(R^G_1,R^G_2)$ is not a function of $\sigma^2_{T_3}$.

\section{Optimal Multiple Description Quantization System}
In the MD quantization scheme for the quadratic Gaussian case
outlined in the preceding section, only the second order
statistics are needed and the resulting quantization system
naturally consists mainly of linear operations. In this section we
develop this system in the context of the Entropy Coded Dithered
(lattice) Quantization (ECDQ) for general sources with the squared
distortion measure. The proposed system may not be optimal for
general sources; however, if all the underlying second order
statistics are kept identical with those of the quadratic Gaussian
case, then the resulting distortions will also be the same.
Furthermore, since among all the i.i.d. sources with the same
variance, the Gaussian source has the highest differential
entropy, the rates of the quantizers can be upper-bounded by the
rates in the quadratic Gaussian case. At high resolution, we prove
a stronger result that the proposed MD quantization system is
asymptotically optimal for all i.i.d. sources that have finite
differential entropy.

In the sequel we discuss the MD quantization schemes in an order
that parallels the development in the preceding section. The
source $\{X(t)\}_{t=1}^\infty$ is assumed to be an i.i.d. random
process (not necessarily Gaussian) with $\mathbb{E}X(t)=0$ and
$\mathbb{E}X^2(t)=\sigma^2_X$ for all $t$.

\subsection{Successive Quantization Using ECDQ}
\begin{figure}[hbt]
\centering
\begin{psfrags}
\psfrag{x}[c]{$\mathbf{X}$}%
\psfrag{xhat1}[l]{$\mathbf{\hat{X}}_1$}%
\psfrag{xhat2}[l]{$\mathbf{\hat{X}}_2$}%
\psfrag{xhat3}[l]{$\mathbf{\hat{X}}_3$}%
\psfrag{w1}[c]{$\mathbf{W}_1$}%
\psfrag{w2}[l]{$\mathbf{W}_2$}%
\psfrag{q1}[c]{ECDQ1}%
\psfrag{q2}[c]{ECDQ2}%
\psfrag{a1}[c]{$a_1$}%
\psfrag{a2}[l]{$a_2$}%
\psfrag{alpha1}[c]{$\alpha_1$}%
\psfrag{alpha2}[c]{$\alpha_2$}%
\psfrag{beta1}[r]{$\beta_1$}%
\psfrag{beta2}[r]{$\beta_2$}%
\psfrag{plus}[c]{$+$}%
\psfrag{times}[c]{$\times$}%
\includegraphics[scale=1]{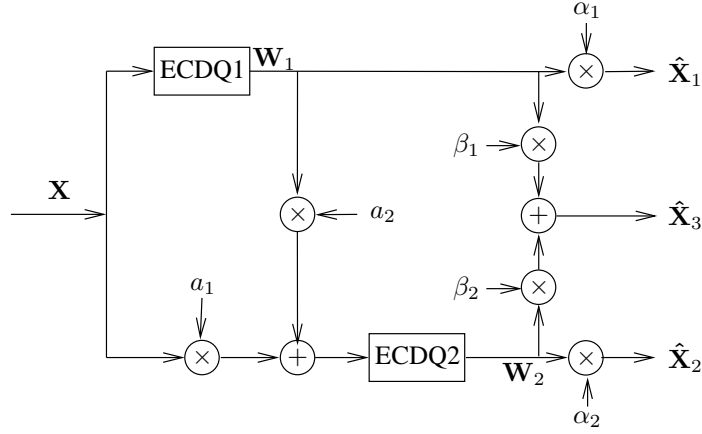}
\end{psfrags}
\caption{Successive quantization.\label{SuccQ}}
\end{figure}

Consider the MD quantization system depicted in Fig. \ref{SuccQ},
which corresponds to the Gaussian MD coding scheme for $V^G_1$.
Let $Q_{1,n}(\cdot)$ and $Q_{2,n}(\cdot)$ denote optimal
$n$-dimensional lattice quantizers. Let $\mathbf{Z}_1$ and
$\mathbf{Z}_2$ be $n$-dimensional random vectors which are
statistically independent and each is uniformly distributed over
the basic cell of the associated lattice quantizer. The lattices
have a ``white" quantization noise covariance matrix of the form
$\sigma_i^2I_n=\mathbb{E}\mathbf{Z}_i\mathbf{Z}^T_i$, where
$\sigma^2_i$ is the second moment of the lattice quantizer
$Q_{i,n}(\cdot)$, $i=1,2$; more specifically, let
$\sigma^2_1=\mathbb{E}B^2_2$, $\sigma^2_2=\mathbb{E}B^2_3$, where
$\mathbb{E}B^2_2$ and $\mathbb{E}B^2_3$ are given by (\ref{B_2})
and (\ref{B_3}), respectively. Furthermore, let
\begin{eqnarray*}
\mathbf{W}_1&=&Q_{1,n}(\mathbf{X}+\mathbf{Z}_1)-\mathbf{Z}_1,\\
\mathbf{W}_2&=&Q_{2,n}(a_1\mathbf{X}+a_2\mathbf{W}_1+\mathbf{Z}_2)-\mathbf{Z}_2,
\end{eqnarray*}
where $a_1$ and $a_2$ are given by (\ref{a_1}) and (\ref{a_2}),
respectively.

\begin{theorem}\label{thm51}
The first and second order statistics of
$(\mathbf{X},\mathbf{W}_1,\mathbf{W}_2)$ are the same as the first
and second order statistics of
$(\mathbf{X}^G,\mathbf{U}_1,\mathbf{U}_2)$ in Section IV.
\end{theorem}

Remark: The first order statistics of
$(\mathbf{X},\mathbf{W}_1,\mathbf{W}_2)$ and
$(\mathbf{X}^G,\mathbf{U}_1,\mathbf{U}_2)$ are all zero. Actually
all the random variables and random vectors in this paper (except
those Sections I and III) are of zero mean, so we focus on the
second order statistics.

\begin{proof}
The theorem follows directly from the correspondence between the
Gram-Schmidt orthogonalization and the sequential (dithered)
quantization established in Section II, and it is straightforward
by comparing Fig. \ref{GVQ} and Fig. \ref{SuccQ}. Essentially,
$\mathbf{X}$, $\mathbf{Z}_1$ and $\mathbf{Z}_2$ serve as the
innovations that generate the first and second order statistics of
the whole system.
\end{proof}
By Theorem \ref{thm51},
\begin{eqnarray*}
&&\frac{1}{n}\mathbb{E}\left|\left|\mathbf{X}-\alpha_i\mathbf{W}_i\right|\right|^2=\frac{1}{n}\mathbb{E}\left|\left|\mathbf{X}^G-\alpha_i\mathbf{U}_i\right|\right|^2= D_i,\quad i=1,2,\\
&&\frac{1}{n}\mathbb{E}\left|\left|
\mathbf{X}-\sum\limits_{i=1}^2\beta_i\mathbf{W}_i\right|\right|^2=\frac{1}{n}\mathbb{E}\left|\left|\mathbf{X}^G-\sum\limits_{i=1}^2\beta_i\mathbf{U}_i\right|\right|^2=D_3.
\end{eqnarray*}

Let $\mathbf{N}_i$ be an $n$-dimensional random vector distributed
as $-\mathbf{Z}_i$, $i=1,2,3$. By property 2) of the ECDQ, we have
\begin{eqnarray*}
H(Q_{1,n}(\mathbf{X}+\mathbf{Z}_1)|\mathbf{Z}_1)&=&I(\mathbf{X};\mathbf{X}+\mathbf{N}_1)\\
&=&h(\mathbf{X}+\mathbf{N}_1)-h(\mathbf{N}_1), \\
H(Q_{2,n}(a_1\mathbf{X}+a_2\mathbf{W}_1+\mathbf{Z}_2)|\mathbf{Z}_2)&=&I(a_1\mathbf{X}+a_2\mathbf{W}_1;a_1\mathbf{X}+a_2\mathbf{W}_1+\mathbf{N}_2)\\
&=&h(a_1\mathbf{X}+a_2\mathbf{W}_1+\mathbf{N}_2)-h(\mathbf{N}_2).
\end{eqnarray*}
Thus, we can upper-bound the rate of $Q_{1,n}(\cdot)$ (conditioned
on $\mathbf{Z}_1$) as follows.
\begin{eqnarray*}
R_1&=&\frac{1}{n}H(Q_{1,n}(\mathbf{X}+\mathbf{Z}_1)|\mathbf{Z}_1)\\
&=&\frac{1}{n}h(\mathbf{X}+\mathbf{N}_1)-\frac{1}{n}h(\mathbf{N}_1)\\
&=&\frac{1}{n}h(\mathbf{W}_1)-\frac{1}{n}h(\mathbf{N}_1)\\
&\leq&\frac{1}{n}h(\mathbf{U}_1)-\frac{1}{n}h(\mathbf{N}_1)\\
&=&\frac{1}{2}\log\left[2\pi
e(\sigma^2_X+\mathbb{E}B^2_2)\right]-\frac{1}{2}\log\frac{\mathbb{E}B^2_2}{G^{opt}_{n}}\\
&=&R_1(V^G_1)+\frac{1}{2}\log\left(2\pi eG^{opt}_{n}\right),
\end{eqnarray*}
where the inequality follows from Theorem \ref{thm51} and the fact
that for a given covariance matrix, the joint Gaussian
distribution maximizes the differential entropy. Similarly, the
rate of $Q_{2,n}(\cdot)$ (conditioned on $\mathbf{Z}_2$) can be
upper-bounded as follows.
\begin{eqnarray*}
R_2&=&\frac{1}{n}H(Q_2(a_1\mathbf{X}+a_2\mathbf{W}_1+\mathbf{Z}_2)|\mathbf{Z}_2)\\
&=&\frac{1}{n}h(a_1\mathbf{X}+a_2\mathbf{W}_1+\mathbf{N}_2)-\frac{1}{n}h(\mathbf{N}_2)\\
&=&\frac{1}{n}h(\mathbf{W}_2)-\frac{1}{n}h(\mathbf{N}_1)\\
&\leq&\frac{1}{n}h(\mathbf{U}_2)-\frac{1}{n}h(\mathbf{N}_1)\\
&=&\frac{1}{2}\log\left[2\pi
e(\sigma^2_X+\sigma^2_{T_0}+\sigma^2_{T_2})\right]-\frac{1}{2}\log\frac{\mathbb{E}B^2_3}{G^{opt}_{n}}\\
&=&R_2(V^G_1)+\frac{1}{2}\log\left(2\pi eG^{opt}_{n}\right).
\end{eqnarray*}
Since $G^{opt}_{n}\rightarrow\frac{1}{2\pi e}$ as
$n\rightarrow\infty$, we have $R_1\leq R_1(V^G_1)$ and $R_2\leq
R_2(V^G_1)$ as $n\rightarrow\infty$.

\subsection{Successive Quantization With Quantization Splitting Using ECDQ}
Now we proceed to construct the MD quantization system using ECDQ
in a manner which corresponds to that for the Gaussian MD
quantization scheme for an arbitrary rate pair $(R^G_1,R^G_2)$.

\begin{figure}[hbt]
\centering
\begin{psfrags}
\psfrag{x}[c]{$\mathbf{X}$}%
\psfrag{xhat1}[l]{$\mathbf{\hat{X}}_1$}%
\psfrag{xhat2}[l]{$\mathbf{\hat{X}}_2$}%
\psfrag{xhat3}[l]{$\mathbf{\hat{X}}_3$}%
\psfrag{delta}[c]{$\mathbf{\Delta}$}%
\psfrag{w1}[c]{$\mathbf{\widetilde{W}}_1$}%
\psfrag{w2}[c]{$\mathbf{\widetilde{W}}_2$}%
\psfrag{w2'}[c]{${\mathbf{\widetilde{W}}'_2}$}%
\psfrag{q1}[l]{\small ECDQ1}%
\psfrag{q2}[l]{\small ECDQ2}%
\psfrag{q3}[l]{\small ECDQ3}%
\psfrag{b1}[c]{$b^*_1$}%
\psfrag{b2}[r]{$b^*_2$}%
\psfrag{b3}[c]{$b^*_3$}%
\psfrag{b4}[l]{$b^*_4$}%
\psfrag{b5}[c]{$b^*_5$}%
\psfrag{b6}[c]{$b^*_6$}%
\psfrag{times}[c]{$\times$}%
\psfrag{plus}[c]{$+$}%
\psfrag{alpha1}[c]{$\alpha_1$}%
\psfrag{alpha2}[c]{$\alpha_2$}%
\psfrag{beta1}[c]{$\beta_1$}%
\psfrag{beta2}[c]{$\beta_2$}%
\includegraphics[scale=1.1]{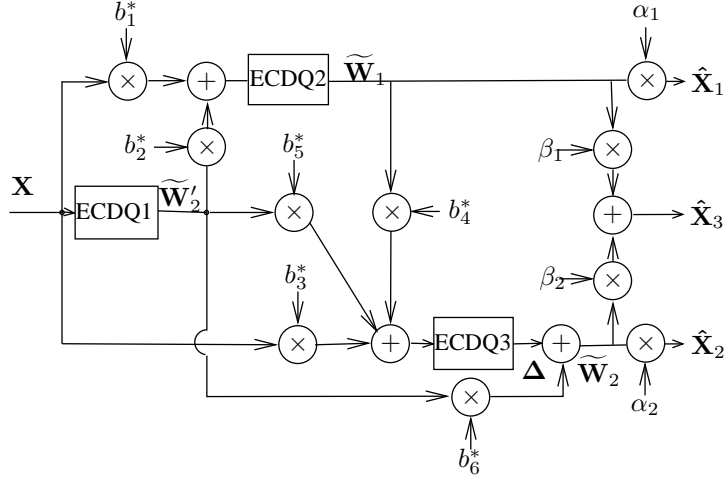}
\end{psfrags}
\caption{Successive quantization with quantization
splitting.\label{QuanS}}
\end{figure}

Let $Q^*_{1,n}(\cdot), Q^*_{2,n}(\cdot)$, and $Q^*_{3,n}(\cdot)$
denote optimal $n$-dimensional lattice quantizers. Let
${\mathbf{Z}^*_1}, {\mathbf{Z}^*_2}$, and ${\mathbf{Z}^*_3}$ be
$n$-dimensional random vectors which are statistically independent
and each is uniformly distributed over the basic cell of the
associated lattice quantizer. The lattices have a ``white"
quantization noise covariance matrix of the form
${\sigma^*_i}^2I_n=\mathbb{E}{\mathbf{Z}^*_i}{\mathbf{Z}^*_i}^T$,
where ${\sigma^*_i}^2$ is the second moment of the lattice
quantizer $Q^*_{i,n}(\cdot)$, $i=1,2,3$; more specifically, let
${\sigma^*_1}^2=\mathbb{E}{\widetilde{B}_2}^2$,
${\sigma^*_2}^2=\mathbb{E}{\widetilde{B}_3}^2$, and
${\sigma^*_3}^2=\mathbb{E}{\overline{B}^2_4}$, where
$\mathbb{E}{\widetilde{B}^2_2}$, $\mathbb{E}{\widetilde{B}_3}^2$
and $\mathbb{E}{\overline{B}^2_4}$ are given by (\ref{EB_2}),
(\ref{EB_3}) and (\ref{EB_4}) respectively. Define
\begin{eqnarray*}
{{\mathbf{\widetilde{W}}}'_2}&=&Q^*_{1,n}(\mathbf{X}+\mathbf{Z}^*_1)-{\mathbf{Z}^*_1},\\
\mathbf{\widetilde{W}}_1^n&=&Q^*_{2,n}(b^*_1\mathbf{X}+b^*_2{\mathbf{\widetilde{W}}'_2}+{\mathbf{Z}^*_2})-{\mathbf{Z}^*_2},\\
\mathbf{\Delta}&=&Q^*_{3,n}(b^*_3\mathbf{X}+b^*_4\mathbf{\widetilde{W}}_1+b^*_5{\mathbf{\widetilde{W}}'_2}+{\mathbf{Z}^*_3})-{\mathbf{Z}^*_3}\\
\mathbf{\widetilde{W}}_2&=&\mathbf{\Delta}+b^*_6{\mathbf{\widetilde{W}}'_2}.
\end{eqnarray*}
The system diagram is shown in Fig. \ref{QuanS}.

\begin{theorem}\label{thm52}
The first and second order statistics of
$(\mathbf{X},\mathbf{\widetilde{W}}_1,\mathbf{\widetilde{W}}_2,\mathbf{\widetilde{W}}'_2,\mathbf{\Delta})$
equal the first and second order statistics of
$(\mathbf{X}^G,\mathbf{U}_1,\mathbf{U}_2,\mathbf{U}'_2,\mathbf{U}_2-b_6\mathbf{U}'_2)$
in Section IV.
\end{theorem}
\begin{proof}
By comparing Fig. \ref{GGQ} and Fig. \ref{QuanS}, it is clear that
the theorem follows from the correspondence between the
Gram-Schmidt orthogonalization and the sequential (dithered)
quantization. The following 1-1 correspondences should be
emphasized: ${\mathbf{\widetilde{B}}_2}$ and $-{\mathbf{Z}^*_1}$,
${\mathbf{\widetilde{B}}_3}$ and $-{\mathbf{Z}^*_2}$,
${\mathbf{\overline{B}}_4}$ and $-{\mathbf{Z}^*_3}$. $\mathbf{X}$,
${\mathbf{Z}^*_1}$, ${\mathbf{Z}^*_2}$ and ${\mathbf{Z}^*_3}$ are
the innovations that generate the first and second order
statistics of the whole system.
\end{proof}
It follows from Theorem \ref{thm52} that
\begin{eqnarray*}
&&\frac{1}{n}\mathbb{E}\left|\left|\mathbf{X}-\alpha_i\mathbf{\widetilde{W}}_i\right|\right|^2=\frac{1}{n}\mathbb{E}\left|\left|\mathbf{X}^G-\alpha_i\mathbf{U}_i\right|\right|^2= D_i,\quad i=1,2,\\
&&\frac{1}{n}\mathbb{E}\left|\left|
\mathbf{X}-\sum\limits_{i=1}^2\beta_i\mathbf{\widetilde{W}}_i\right|\right|^2=\frac{1}{n}\mathbb{E}\left|\left|\mathbf{X}^G-\sum\limits_{i=1}^2\beta_i\mathbf{U}_i\right|\right|^2=D_3.
\end{eqnarray*}

Let ${\mathbf{N}^*_i}$ be an $n$-dimensional random vector
distributed as $-{\mathbf{Z}^*_i}$, $i=1,2,3$. By property 2) of
the ECDQ, we have
\begin{eqnarray*}
H(Q^*_{1,n}(\mathbf{X}+{\mathbf{Z}^*_1})|{\mathbf{Z}^*_1})&=&I(\mathbf{X};\mathbf{X}+{\mathbf{N}^*_1})\\
&=&h(\mathbf{X}+{\mathbf{N}^*_1})-h({\mathbf{N}^*_1}), \\
H(Q^*_{2,n}(b^*_1\mathbf{X}+b^*_2{\mathbf{\widetilde{W}}'_2}+{\mathbf{Z}^*_2})|{\mathbf{Z}^*_2})&=&I(b^*_1\mathbf{X}+b^*_2{\mathbf{\widetilde{W}}'_2};b^*_1\mathbf{X}+b^*_2{\mathbf{\widetilde{W}}'_2}+{\mathbf{N}^*_2})\\
&=&h(b^*_1\mathbf{X}+b^*_2{\mathbf{\widetilde{W}}'_2}+{\mathbf{N}^*_2})-h({\mathbf{N}^*_2}),\\
H(Q^*_{3,n}(b^*_3\mathbf{X}+b^*_4\mathbf{\widetilde{W}}_1+b^*_5{\mathbf{\widetilde{W}}'_2}+{\mathbf{Z}^*_3})|{\mathbf{Z}^*_3})&=&I(b^*_3\mathbf{X}+b^*_4\mathbf{\widetilde{W}}_1+b^*_5{\mathbf{\widetilde{W}}'_2};b^*_3\mathbf{X}+b^*_4\mathbf{\widetilde{W}}_1+b^*_5{\mathbf{\widetilde{W}}'_2}+{\mathbf{N}^*_3})\\
&=&h(b^*_3\mathbf{X}+b^*_4\mathbf{\widetilde{W}}_1+b^*_5{\mathbf{\widetilde{W}}'_2}+{\mathbf{N}^*_3})-h({\mathbf{N}^*_3}).
\end{eqnarray*}

Thus we can upper-bound the rate of $Q^*_{2,n}(\cdot)$
(conditioned on ${\mathbf{Z}^*_2}$) as follows.
\begin{eqnarray}
R_1&=&\frac{1}{n}H(Q^*_{2,n}(b^*_1\mathbf{X}+b^*_2{\mathbf{\widetilde{W}}'_2}+{\mathbf{Z}^*_2}^n)|{\mathbf{Z}^*_2})\nonumber\\
&=&\frac{1}{n}h(b^*_1\mathbf{X}+b^*_2{\mathbf{\widetilde{W}}'_2}+{\mathbf{N}^*_2})-\frac{1}{n}h({\mathbf{N}^*_2})\nonumber\\
&=&\frac{1}{n}h({\mathbf{\widetilde{W}}_1})-\frac{1}{n}h({\mathbf{N}^*_2})\nonumber\\
&\leq&\frac{1}{n}h({\mathbf{U}_1})-\frac{1}{n}h({\mathbf{N}^*_2})\nonumber\\
&=&\frac{1}{2}\log\left[2\pi
e(\sigma^2_X+\sigma^2_{T_0}+\sigma^2_{T_1})\right]-\frac{1}{2}\log\frac{\mathbb{E}{\widetilde{B}_3}^2}{G^{opt}_{n}}\nonumber\\
&=&R^G_1+\frac{1}{2}\log\left(2\pi
eG^{opt}_{n}\right)\label{nbound1},
\end{eqnarray}
where the inequality follows from Theorem \ref{thm52} and the fact
that for a given covariance matrix, the joint Gaussian
distribution maximizes the differential entropy.

Similarly, the sum-rate of $Q^*_{1,n}(\cdot)$ (conditioned on
${\mathbf{Z}^*_1}$) and $Q^*_{3,n}(\cdot)$ (conditioned on
${\mathbf{Z}^*_3}$) can be upper-bounded as follows.
\begin{eqnarray}
R_2&=&\frac{1}{n}H(Q^*_{1,n}(\mathbf{X}+{\mathbf{Z}^*_1})|{\mathbf{Z}^*_1})+\frac{1}{n}H(Q^*_{3,n}(b^*_3\mathbf{X}+b^*_4\mathbf{\widetilde{W}}_1+b^*_5{\mathbf{\widetilde{W}}'_2}+{\mathbf{Z}^*_3})|{\mathbf{Z}^*_3})\nonumber\\
&=&\frac{1}{n}h(\mathbf{X}+{\mathbf{N}^*_1})-\frac{1}{n}h({\mathbf{N}^*_1})+\frac{1}{n}h(b^*_3\mathbf{X}+b^*_4\mathbf{\widetilde{W}}_1^n+b^*_5{\mathbf{\widetilde{W}}'_2}+{\mathbf{N}^*_3})-\frac{1}{n}h({\mathbf{N}^*_3})\nonumber\\
&=&\frac{1}{n}h({\mathbf{\widetilde{W}}'_2})-\frac{1}{n}h({\mathbf{N}^*_1})+\frac{1}{n}h(\mathbf{\Delta})-\frac{1}{n}h({\mathbf{N}^*_3})\nonumber\\
&\leq&\frac{1}{n}h({\mathbf{U}'_2})-\frac{1}{n}h({\mathbf{N}^*_1})+\frac{1}{n}h(\mathbf{U}_2-b_6{\mathbf{U}'_2})-\frac{1}{n}h({\mathbf{N}^*_3})\nonumber\\
&\stackrel{(a)}{=}&\frac{1}{n}h({\mathbf{U}'_2})-\frac{1}{n}h({\mathbf{N}^*_1})+\frac{1}{n}h(b_7{\mathbf{\overline{B}}_2}+b_8{\mathbf{\overline{B}}_3}+{\mathbf{\overline{B}}_4})-\frac{1}{n}h({\mathbf{N}^*_3})\nonumber\\
&=&\frac{1}{2}\log\left[2\pi
e(\sigma^2_X+\sigma^2_{T_0}+\sigma^2_{T_2}+\sigma^2_{T_3})\right]-\frac{1}{2}\log\frac{\mathbb{E}{\widetilde{B}_2}^2}{G^{opt}_{n}}+\frac{1}{2}\log\left[2\pi
e(b^2_7\mathbb{E}{\overline{B}^2_2}+b^2_8\mathbb{E}{\overline{B}^2_3}+\mathbb{E}{\overline{B}^2_4})\right]\nonumber\\
&&-\frac{1}{2}\log\frac{\mathbb{E}{\overline{B}_4}^2}{G^{opt}_{n}}\nonumber\\
&=&R^G_2+\log\left(2\pi eG^{opt}_{n}\right),\label{nbound2}
\end{eqnarray}
where (a) follows from (\ref{relation2}). Remark: Since the
decoders only need to know
$\mathbf{\widetilde{W}}_2=\mathbf{\Delta}+b^*_6{\mathbf{\widetilde{W}}'_2}$
instead of $\mathbf{\widetilde{W}}'_2$ and $\mathbf{\Delta}$
separately, we can actually further reduce $R_2$ to
$\frac{1}{n}H(\mathbf{\widetilde{W}}_2|{\mathbf{Z}^*_1},{\mathbf{Z}^*_2},{\mathbf{Z}^*_3})$.
Since $G^{opt}_{n}\rightarrow\frac{1}{2\pi e}$ as
$n\rightarrow\infty$, it follows from (\ref{nbound1}) and
(\ref{nbound2}) that $R_1\leq R^G_1$, $R_2\leq R^G_2$ as
$n\rightarrow\infty$.

\begin{figure}[hbt]
\centering
\begin{psfrags}
\psfrag{x}[c]{$\mathbf{X}$}%
\psfrag{xhat1}[l]{$\mathbf{\hat{X}}_1$}%
\psfrag{xhat2}[l]{$\mathbf{\hat{X}}_2$}%
\psfrag{xhat3}[l]{$\mathbf{\hat{X}}_3$}%
\psfrag{w1}[l]{$\mathbf{W}_1$}%
\psfrag{w2}[l]{$\mathbf{W}_2$}%
\psfrag{q1}[c]{ECDQ1}%
\psfrag{q2}[c]{ECDQ2}%
\psfrag{alpha1}[c]{$\alpha_1$}%
\psfrag{alpha2}[c]{$\alpha_2$}%
\psfrag{beta1}[r]{$\beta_1$}%
\psfrag{beta2}[r]{$\beta_2$}%
\psfrag{plus}[c]{$+$}%
\psfrag{times}[c]{$\times$}%
\includegraphics[scale=1]{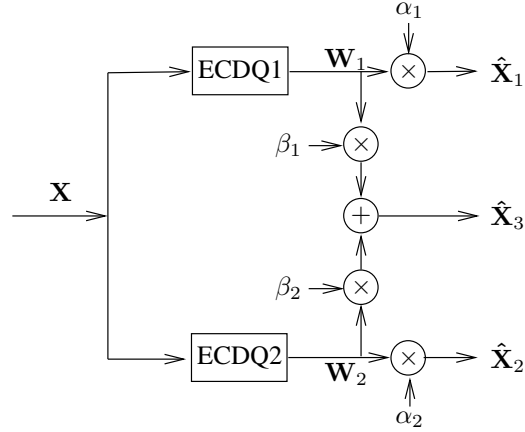}
\end{psfrags}
\caption{Separate quantization.\label{SeQ}}
\end{figure}

For the special case when $D_3=(1/D_1+1/D_2-1/\sigma^2_X)^{-1}$,
the MD quantization systems in Fig. \ref{SuccQ} and Fig.
\ref{QuanS} degenerate to two independent quantization operations
as shown in Fig. \ref{SeQ}. The connection between Fig. \ref{SeQ}
and Fig. \ref{GE1Q} is apparent.

The above results imply that for general i.i.d. sources, under the
same distortion constraints, the rates required by our scheme are
upper-bounded by the rates required for the quadratic Gaussian
case. This further implies our scheme can achieve the whole
Gaussian MD rate-distortion region as the dimension of the
(optimal) lattice quantizers becomes large.

\subsection{Optimality and An Upper Bound on the Coding Rates}
Define $\mathcal{Q}_{out}$ such that
$(R_1,R_2,D_1,D_2,D_3)\in\mathcal{Q}_{out}$ if and only if
\begin{eqnarray*}
R_i&\geq&\frac{1}{2}\log\frac{P_X}{D_i},\quad i=1,2,\\
R_1+R_2&\geq&\frac{1}{2}\log\frac{P_X}{D_3}+\frac{1}{2}\log\phi(D_1,D_2,D_3),
\end{eqnarray*}
where
\begin{eqnarray*}
\phi(D_1,D_2,D_3)=\left\{\begin{array}{ll}
  1, \hspace{1.95in} D_3<D_1+D_2-P_X\\
  \frac{P_XD_3}{D_1D_2}, \hspace{1.7in}D_3> \left(\frac{1}{D_1}+\frac{1}{D_2}-\frac{1}{P_X}\right)^{-1}\\
  \frac{(P_X-D_3)^2}
{(P_X-D_3)^2-[\sqrt{(P_X-D_1)(P_X-D_2)}-\sqrt{(D_1-D_3)(D_2-D_3)}]^2},  \mbox{ o.w.} \\
\end{array}\right.&\\
\end{eqnarray*}
and $P_X=2^{2h(X)}/2\pi e$ is the entropy power of $X$.  It was
shown by Zamir \cite{Zamir} that for i.i.d. sources with finite
differential entropy, $\mathcal{Q}_{out}$ is an outer bound of the
MD rate-distortion region and is asymptotically tight at high
resolution (i.e., $D_1,D_2,D_3\rightarrow 0$). Again, we only need
consider the case $D_1+D_2-P_X\leq D_3\leq
\left(\frac{1}{D_1}+\frac{1}{D_2}-\frac{1}{P_X}\right)^{-1}$. At
high resolution, we can write
\begin{eqnarray*}
\frac{1}{2}\log\phi(D_1,D_2,D_3)=\frac{1}{2}\log\frac{P_X}
{(\sqrt{D_1-D_3}+\sqrt{D_2-D_3})^2}+o(1),
\end{eqnarray*}
where $o(1)\rightarrow 0$ as $D_1,D_2,D_3\rightarrow 0$.

The following theorem says our scheme is asymptotically optimal at
high resolution for general smooth i.i.d. sources.
\begin{theorem}\label{asyoptimal}
The region
\begin{eqnarray*}
R_i&\geq&\frac{1}{2}\log\frac{P_X}{D_i}+\frac{1}{2}\log(2\pi eG^{opt}_n)+o(1),\quad i=1,2,\\
R_1+R_2&\geq&\frac{1}{2}\log\frac{P_X}{D_3}+\frac{1}{2}\log\frac{P_X}
{(\sqrt{D_1-D_3}+\sqrt{D_2-D_3})^2}+\frac{3}{2}\log(2\pi
eG^{opt}_n)+o(1)
\end{eqnarray*}
is achievable using optimal $n$-dimensional lattice quantizers via
successive quantization with quantization splitting.
\end{theorem}
\begin{proof}
See Appendix II.
\end{proof}
Remark:
\begin{enumerate}
\item As $D_1,D_2,D_3\rightarrow 0$ and $n\rightarrow\infty$, the
above region converges to the outer bound and thus is
asymptotically tight.
\item The sum-rate redundancy
of our MD quantization scheme (i.e., successive quantization with
quantization splitting) is at most three times the redundancy of
an optimal $n$-dimensional lattice quantizer in the high
resolution regime. It is easy to see from (\ref{nbound1}) and
(\ref{nbound2}) that for the Gaussian source, this is true at all
resolutions. Specifically, for scalar quantizers, we have
$G^{opt}_1=\frac{1}{12}$, and thus the redundancy is
$\frac{3}{2}\log\frac{\pi e}{6}$. This actually overestimates the
sum-rate redundancy of our scheme in certain cases. It will be
shown in the next section that for the scalar case, the redundancy
is approximately twice the redundancy of a scalar quantizer at
high resolution.
\item The successive quantization with quantization splitting can be
replaced by timesharing the quantization schemes for two vertices.
Since for vertices it only requires two quantization operations,
one can show that the redundancy of the timesharing approach is at
most twice the redundancy of an optimal $n$-dimensional lattice
quantizer.
\item The reason that our MD quantization scheme is asymptotically
optimal for all smooth sources is that the universal lossless
entropy encoder incorporated in ECDQ can, to some extent,
automatically exploit the real distribution of the source.
\end{enumerate}

The following theorem gives a single letter upper bound on the
rates of our scheme at all resolutions as the dimension of the
optimal lattices becomes large.
\begin{theorem}\label{highdim}
There exists a sequence of lattice dimensions $n_1,n_2,\cdots,$
such that
\begin{eqnarray*}
&&\limsup\limits_{m\rightarrow\infty}\frac{1}{n_m}\left[h(\mathbf{X}+{\mathbf{N}^*_1})-\frac{1}{n}h({\mathbf{N}^*_1})\right]\leq h\left(X+N^G_1\right)-h\left(N^G_1\right),\\
&&\limsup\limits_{m\rightarrow\infty}\frac{1}{n_m}\left[h(b^*_1\mathbf{X}+b^*_2{\mathbf{\widetilde{W}}'_2}+{\mathbf{N}^*_2})-\frac{1}{n}h({\mathbf{N}^*_2})\right]\leq h\left(X+b^*_2N^G_1+N^G_2\right)-h\left(N^G_2\right),\\
\end{eqnarray*}
and
\begin{eqnarray*}
&&\limsup\limits_{m\rightarrow\infty}\frac{1}{n_m}\left[h(b^*_3\mathbf{X}+b^*_4\mathbf{\widetilde{W}}_1^n+b^*_5{\mathbf{\widetilde{W}}'_2}+{\mathbf{N}^*_3})-\frac{1}{n}h({\mathbf{N}^*_3})\right]\\
&\leq&
h\left((b^*_3+b^*_1b^*_4+b^*_2b^*_4+b^*_5)X+(b^*_2b^*_4+b^*_5)N^G_1+b^*_4N^G_2+N^G_3\right)-h\left(N^G_3\right),
\end{eqnarray*}
where $N^G_1\sim\mathcal{N}(0,\mathbb{E}\widetilde{B}^2_2)$,
$N^G_1\sim\mathcal{N}(0,\mathbb{E}\widetilde{B}^2_3)$,
$N^G_1\sim\mathcal{N}(0,\mathbb{E}\overline{B}^2_4)$, and the
generic source variable $X$ are all independent.
\end{theorem}
Remark: This theorem implies that as the dimension of the optimal
lattices goes to infinity, the rates required by our scheme can be
upper-bounded as
\begin{eqnarray*}
R_1&\leq& h\left(X+b^*_2N^G_1+N^G_2\right)-h\left(N^G_2\right)\\
R_2&\leq&
h\left(X+N^G_1\right)-h\left(N^G_1\right)+h\left((b^*_3+b^*_1b^*_4+b^*_2b^*_4+b^*_5)X+(b^*_2b^*_4+b^*_5)N^G_1+b^*_4N^G_2+N^G_3\right)-h\left(N^G_3\right).
\end{eqnarray*}
By comparing the above two expressions with (\ref{GR1}) and
(\ref{GR2}), we can see that if $X$ is not Gaussian, then $R_1<
R^G_1$, $R_2< R^G_2$.

\begin{proof}
See Appendix III.
\end{proof}

\begin{figure}[hbt]
\centering
\begin{psfrags}
\psfrag{x}[c]{$\mathbf{X}$}%
\psfrag{xhat1}[l]{$\mathbf{\hat{X}}_1$}%
\psfrag{xhat2}[l]{$\mathbf{\hat{X}}_2$}%
\psfrag{xhat3}[l]{$\mathbf{\hat{X}}_3$}%
\psfrag{delta}[c]{\tiny $ $}%
\psfrag{w1}[c]{\tiny $ $}%
\psfrag{w2}[c]{\tiny $ $}%
\psfrag{w2'}[c]{\tiny $ $}%
\psfrag{q1}[l]{\hspace{0.02in}ECDQ}%
\psfrag{q2}[l]{\hspace{0.015in}ECDQ}%
\psfrag{q3}[l]{\hspace{0.015in}ECDQ}%
\psfrag{b1}[c]{$b'_1$}%
\psfrag{b2}[r]{$b'_2$}%
\psfrag{b3}[c]{$b'_3$}%
\psfrag{b4}[l]{$b'_4$}%
\psfrag{b5}[c]{$b'_5$}%
\psfrag{b6}[c]{$b'_6$}%
\psfrag{times}[c]{$\times$}%
\psfrag{plus}[c]{$+$}%
\psfrag{alpha1}[c]{$\alpha'_1$}%
\psfrag{alpha2}[c]{$\alpha'_2$}%
\psfrag{beta1}[c]{$\beta'_1$}%
\psfrag{beta2}[c]{$\beta'_2$}%
\includegraphics[scale=1.1]{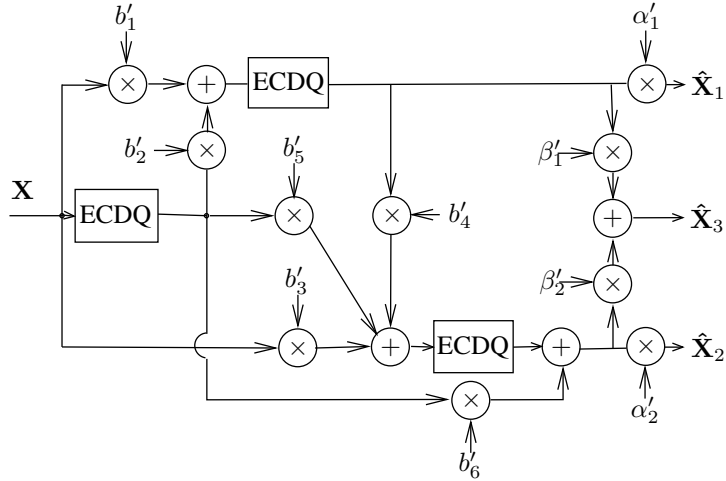}
\end{psfrags}
\caption{MD lattice quantization with quantizer
reuse.\label{Qreuse}}
\end{figure}

As mentioned in Section II, by incorporating pre- and postfilters,
a single quantizer can be used to sequentially perform three
quantization operations instead of using three different
quantizers. Let $Q^*_{n}(\cdot)$ be an optimal $n$-dimensional
lattice quantizers. The lattices have a ``white" quantization
noise covariance matrix of the form ${\sigma^*}^2I_n$, where
${\sigma^*}^2$ is the second moment of the lattice quantizer
$Q^*_{n}(\cdot)$. Without loss of generality, we assume
${\sigma^*}^2={\sigma^*_1}^2$, i.e.,
$Q^*_{n}(\cdot)=Q^*_{1,n}(\cdot)$. We can convert $Q^*_n(\cdot)$
to $Q^*_{i,n}(\cdot)$ by introducing the prefilter
$\frac{1}{a^*_i}$ and postfilter $a^*_i$, where
$a^*_i=\frac{\sigma^*_i}{\sigma}$, $i=2,3$. Incorporating the
filters into the coefficients of the system gives the system
diagram shown in Fig. \ref{Qreuse}. Here
$b'_1=\frac{b^*_1}{a^*_2}$, $b'_2=\frac{b^*_2}{a^*_2}$,
$b'_3=\frac{b^*_3}{a^*_3}$, $b'_4=\frac{b^*_4a^*_2}{a^*_3}$,
$b'_5=\frac{b^*_5}{a^*_3}$, $b'_6=\frac{b^*_6}{a^*_3}$,
$\alpha'_1=\alpha_1a^*_2$, $\alpha'_2=\alpha_2a^*_3$,
$\beta'_1=\beta_1a^*_2$, and $\beta'_2=\beta_2a^*_3$. Although the
quantizer $Q^*_n(\cdot)$ can be reused, the dither introduced in
each quantization operation should be independent.

\section{The Geometric Interpretation of the Scalar Quantization Scheme}
In this section we give a geometric interpretation of our MD
quantization scheme when undithered scalar quantization is used in
the proposed framework. This interpretation serves as a bridge
between the information theoretic description of the coding scheme
\footnote{Although the ECDQ-based MD scheme considered in the
preceding section is certainly of practical value, we mainly use
it as an analytical tool to establish the optimality of our
scheme. In practice, it is more desirable to have a MD scheme
based on low-complexity undithered quantization.} and the
practical quantization operation. Furthermore, it facilitates a
high-resolution analysis, which offers a performance comparison
between the proposed quantization scheme and existing multiple
description quantization techniques. Though only scalar
quantization is considered here, the interpretation can also be
extended to the vector quantization case.

\subsection{The Geometric Interpretation}
It is beneficial to clarify the definition of the encoder and
decoder functions of a classical scalar quantizer. The overall
quantization can be modeled to be composed of three components
\cite{Gra:98}:
\begin{enumerate}
\item The {\em lossy encoder} is a mapping $q: \mathcal{R}\rightarrow \mathcal{I}$, where the index set $\mathcal{I}$ is usually taken as a collection of consecutive integers. Commonly, this lossy encoder is alternatively specified by a partition of $\mathcal{R}$, i.e., the boundary points of the partition segments.
\item The {\em lossy decoder} is a mapping $q^{-1}: \mathcal{I} \rightarrow \mathcal{R'}$, where $\mathcal{R'}\subset\mathcal{R}$ is the reproduction codebook.
\item The {\em lossless encoder} $\gamma: \mathcal{I}\rightarrow\mathcal{C}$ is an invertible mapping into a collection $\mathcal{C}$ of variable-length binary vectors. This is essentially the entropy coding of the quantization indices.
\end{enumerate}

\begin{figure}[tb]
  \centering
    \includegraphics[width=8cm]{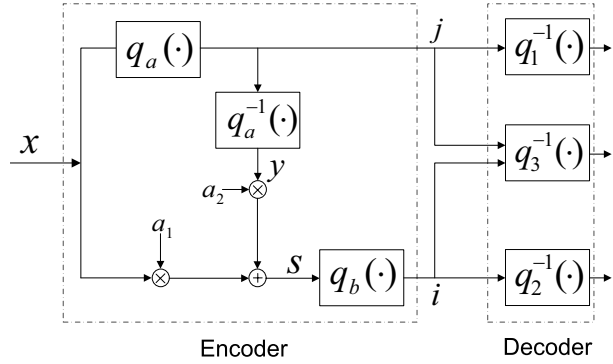}
\caption{Coding scheme using successive quantization in terms of
quantization encoder and decoder. \label{fig:V1_diagram}}
\end{figure}

The successive quantization coding scheme in Fig. \ref{SuccQ} is
redrawn in terms of quantization encoder and decoder in Fig.
\ref{fig:V1_diagram}. The scaling factors $\alpha_1$, $\alpha_2$,
$\beta_1$ and $\beta_2$ are absorbed into the lossy decoders. The
lossless encoder $\gamma$, though important, is not essential in
this interpretation and is thus omitted in Fig.
\ref{fig:V1_diagram}. The lossy decoders in the receiver are
mappings $q_1^{-1}: \mathcal{I}_1 \rightarrow \mathcal{R'}_1$,
$q_2^{-1}: \mathcal{I}_2 \rightarrow \mathcal{R'}_2$, and
$q_3^{-1}: \mathcal{I}_1\times\mathcal{I}_2 \rightarrow
\mathcal{R'}_3$, respectively; notice that the corresponding lossy
encoders do not necessarily exist in the system.

For simplicity, we assume the lossy encoder $q_a$ and $q_b$
generate uniform partitions of $\mathcal{R}$, respectively, while
the lossy decoder $q_a^{-1}$ takes the center points of the
partition cells of $q_a$ as the reproduction codebook. Notice that
function $y=q_a(q_a^{-1}(x))$ is piecewise constant. A linear
combination of $x$ and $y$ is then formed as $s=a_1x+a_2y$, which
is then mapped by the lossy encoder $q_b$ to an quantization index
$q_b(s)$. The task is to find the partition formed by these
operations, and it can be done by considering a partition cell
$i$, given by $(s_i,s_{i+1}]$, in the lossy encoder $q_b$.

\begin{figure}[tb]
  \centering
    \includegraphics[width=8cm]{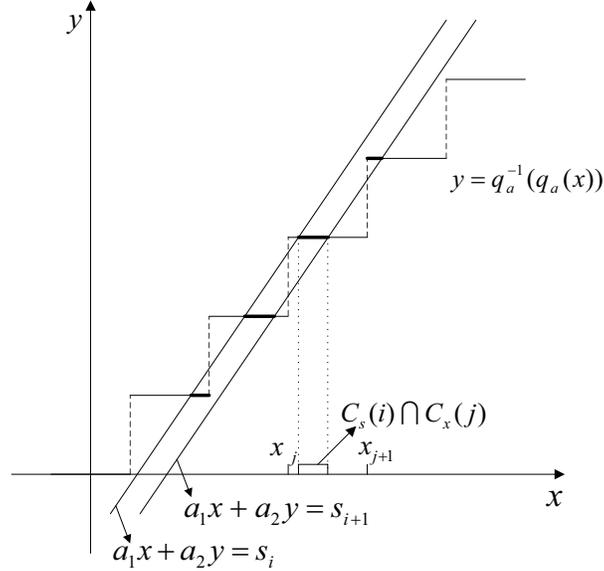}
\caption{The geometric interpretation of the partitions using
successive quantization. \label{fig:V1_stair}}
\end{figure}

In Fig. \ref{fig:V1_stair}, this partition cell is represented on
the $(x,y)$ plane. For operating points on the dominant face of
the SEGC region, it is always true that
$\sigma_{T_0}^2\leq\sigma_{T_1}\sigma_{T_2}$, which implies
$a_2\leq0$ [from (\ref{eqn:gaussiana1a2})], and thus the slope of
the line $a_1x+a_2y=s_i$ is always positive. It is clear that,
given $q_b(s)=i$, $x$ can fall only into the several segments
highlighted by the thicker lines in Fig. \ref{fig:V1_stair}, i.e.,
into the set
$C_s(i)=\{x:a_1x+a_2q_a(q_a^{-1}(x))\in(s_i,s_{i+1}]\}$. The
information regarding $x$ is thus revealed to the lossy decoder
$q_2^{-1}$. In the lossy encoder $q_a$, the information is
revealed to the lossy decoder $q_1^{-1}$ in the traditional manner
that, when index $j$ is specified, $x$ is in the $j$-th cell,
which is $(x_j,x_{j+1}]$; denote it as $C_x(j)=(x_j,x_{j+1}]$.
Jointly, the lossy decoder $q_3^{-1}$ has the information that $x$
is in the intersection of the two sets as $C_s(i)\cap C_x(j)$.

\begin{figure}[tb]
  \centering
    \includegraphics[width=10cm]{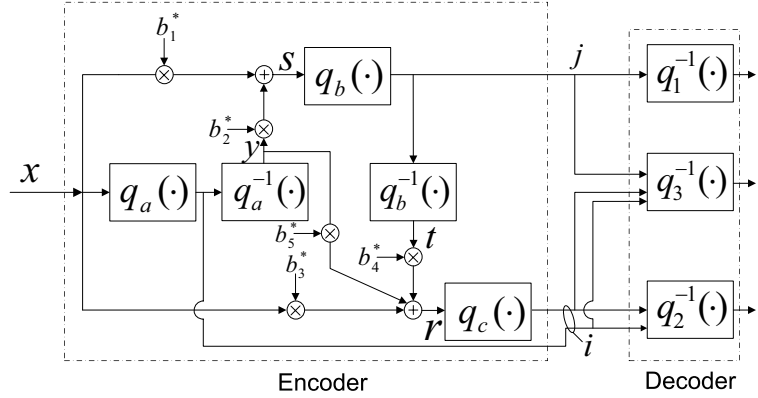}
\caption{Coding scheme using quantization splitting in terms of
quantization encoder and decoder. \label{fig:SP_diagram}}
\end{figure}

Now we briefly discuss the extension of this interpretation to the
case of quantization splitting. The coding scheme in Fig.
\ref{QuanS} is redrawn in Fig. \ref{fig:SP_diagram}. Some of the
operations in Fig. \ref{QuanS} are absorbed into the lossy
decoders. It can be observed that $q_a$, $q_a^{-1}$ and $q_b$ play
roles similar to those in Fig. \ref{fig:V1_diagram}; thus, the
geometric interpretation for successive quantization can still be
utilized. Let $s=b_1^*x+b_2^*q_a^{-1}(q_a(x))$ and define
$C_{s}(j)=\{x:b_1^*x+b_2^*q_a^{-1}(q_a(x))\in(s_j,s_{j+1}]\}$,
where $(s_j,s_{j+1}]$ is the $j$-th partition cell in the lossy
encoder $q_b$. The variable $s$ is defined differently from that
in successive quantization, but this slight abuse of the notation
does not cause any ambiguity.

Notice the index $i=(i_a,i_c)$ has two components, one is the
output of $q_a$, and the other is that of $q_c$. In a sense, $q_a$
and $q_c$ are formed in a refinement manner. Thus, the lossy
encoder $q_c$ and the lossy decoders $q_2^{-1}$ and $q_3^{-1}$
always have the exact output from $q_a$, which in effect confines
the source to a finite range. Thus, we need to consider only the
case for a fixed $q_a(x)$ value. It is obvious that when
$q_a(x)=i_0$ is fixed, $q_a^{-1}(i_0)=y_{i_0}$. Consider the
linear combination of $r=b_3^*x+b_4^*t+b_5^*q_a^{-1}(q_a(x))$,
where $t=q_b^{-1}(q_b(s))$. It is similar to the linear
combination of $s=b_1^*x+b_2^*y$, but with the additional constant
term $b_5^*y_{i_0}$, when  $i_0$ is given. It can be shown that
this constant term in fact removes the conditional mean such that
$E(r|q_a(x)=i_0)\approx 0$, and the lossy encoder $q_c$ is merely
a partition of an interval near zero. Thus with $q_a(x)=i_0$
given, $q_b$, $q_b^{-1}$ and $q_c$ essentially adopt the same
roles as $q_a$, $q_a^{-1}$ and $q_b$, respectively, in Fig.
\ref{fig:V1_diagram}. This implies a similar geometric
interpretation again holds for the additional components in Fig.
\ref{fig:SP_diagram}, since $b_4^*=b_8\leq 0$ and
$b_3^*=b_7-b_5b_8>0$. Define
$C_{xr}(i_a,i_c)=\{x:x\in(x_{i_a},x_{i_a+1}],
r\in(r_{i_c},r_{i_c+1}]\}$, where $(x_{i_a},x_{i_a+1}]$ is the
$i_a$-th partition cell in the lossy encoder $q_a$ and
$(r_{i_c},r_{i_c+1}]$ is the $i_c$-th partition cell in the lossy
encoder $q_c$. Given the index pair $(i,j)=(i_a,i_c,j)$, the joint
lossy decoder $q_3^{-1}$ is provided with information that $x\in
C_s(j)\cap C_{xr}(i_a,i_c)$.

\subsection{High-Resolution Analysis of Several Special Cases}
Below, the high-resolution performance of the proposed coding
scheme using scalar quantization is analyzed under several special
conditions. Of particular interest is the balanced case, where
$R_1=R_2=R$ and two side distortions are equal, $D_1=D_2$;
significant research effort has been devoted to this case. In the
analysis that follows, simplicity is often given priority over
rigor; this corresponds to the motivation to introduce this
section, which is to provide an intuitive interpretation of the
coding schemes.

For the balanced case, it can be shown \cite{Vai:98} that at
high-resolution if the side distortion is of the form
$D_1=b\sigma_x^22^{-2(1-\eta) R}$, where $0\leq\eta<1$ and
$b\geq1$, the central distortion of an MD system can
asymptotically achieve
\begin{equation}
D_3\geq\left\{\begin{array}{lr}
              \textstyle{\sigma_x^22^{-2R}/2(b+\sqrt{b^2-1})}& \eta=0;\\
              \textstyle{\sigma_x^22^{-2R(1+\eta)}/4b}  & 0<\eta<1.\\
              \end{array}\right.
\label{equ:prod2}
\end{equation}
Notice the condition $0<\eta<1$ in fact corresponds to the
condition that $\sigma_x^2\gg{D_1}$ and $D_1\gg{D_3}$ at high
rate. In this case, the central and side distortion product
remains bounded by a constant at fixed rate, which is $D_3
D_1\geq\frac{\sigma_x^42^{-4R}}{4}$, independent of the tradeoff
between them. This product has been used as the information
theoretical bound to measure the efficiency of quantization
methods \cite{Cha:041, Vai:01, Cha:044, Vaishampayan1,
Vaishampayan2, Cha:042}. For the sake of simplicity, we focus on
the zero-mean Gaussian source, however, because of the tightness
of the Shannon lower bound at high-resolution \cite{Zamir}, the
results of the analysis are applicable with minor changes for
other continuous sources with smooth probability density function.

\subsubsection{High-resolution analysis for successive quantization}
Consider using the quantization method depicted in Fig.
\ref{fig:V1_diagram} to construct two descriptions, such that
$D_1=D_2$, though the rates of the two descriptions are not
necessarily equal. For the case $\sigma_x^2\gg{D_1}$ and
$D_1\gg{D_3}$ at high rate, it is clear that
$\sigma_{x}^2\gg\sigma_{T_1}^2=\sigma_{T_2}^2\gg\sigma_{T_0}^2$.
Thus $a_1\approx 2$ and $a_2\approx -1$ [from
(\ref{eqn:gaussiana1a2})], which suggests that the slope of the
line $a_1x+a_2y=s_i$ should be approximately $2$ in this case.

\begin{figure}[tb]
  \centering
    \includegraphics[width=16.5cm]{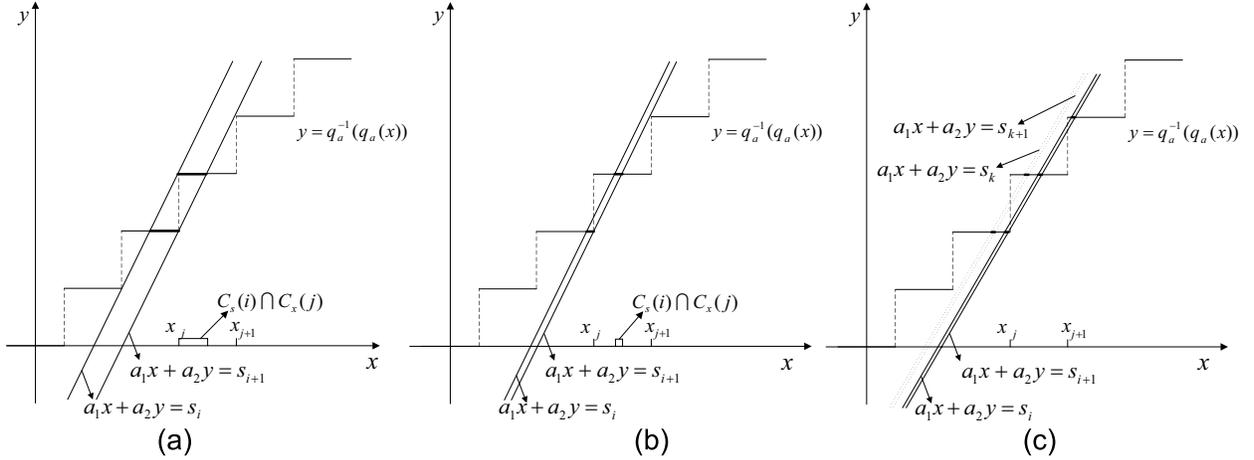}
\caption{Several special cases of the partition formed using
successive quantization. (a) $a_1=2$, $a_2=-1$, and the stepsize
of $q_a$ is the same as that of $q_b$. (b) $a_1=2$, $a_2=-1$, but
the stepsize of $q_a$ is much larger than that of $q_b$. (c) When
the stepsize of $q_a$ is much larger than that of $q_b$, by
slightly varying $a_1$ and $a_2$, the two side distortions can be
made equal.\label{fig:V1_slope2}}

\end{figure}

Next we consider the three cases depicted in Fig.
\ref{fig:V1_slope2}. In Fig. \ref{fig:V1_slope2} (a), $a_1=2$,
$a_2=-1$ are chosen. By properly choosing the thresholds and the
stepsize, a symmetric (between the two descriptions) partition can
be formed. In this partition, cells $C_x(\cdot)$ and cells
$C_s(\cdot)$ both are intervals. Furthermore, they form two
uniform scalar quantizers with their bins staggered by half the
stepsize. This in effect gives the staggered index assignment of
\cite{Ser:00,Cha:041}. By using this partition, the central
distortion is reduced to $1/4$ of the side distortions. Notice
that in this case the condition $D_1\gg{D_3}$ does not hold, but
choosing $a_1=2$, $a_2=-1$ indeed generates two balanced
descriptions; this suggests that certain discrepancy occurs when
applying the information theoretic results directly to the scalar
quantization case. The high-resolution performance of the
partition in Fig. \ref{fig:V1_slope2} (a) is straightforward,
being given by
$D_1=D_2\approx\frac{1}{12}\Delta_a^2\approx\frac{2\pi
e}{12}2^{-2R_1}\sigma_x^2$, where the second equality is true when
entropy coding is assumed, and $D_3\approx\frac{1}{4}D_1$ (also
see \cite{Cha:044}).

In Fig. \ref{fig:V1_slope2} (b), the stepsize in $q_b$, which is
denoted by $\Delta_b$, is chosen to be much smaller than that of
$q_a$, which is denoted as $\Delta_a$; however, $a_1=2$ and
$a_2=-1$ are kept unchanged. In this case, the partition by $q_a$
is still uniform, and the performance of $q_1^{-1}$ is given by
$D_1\approx\frac{1}{12}\Delta_a^2\approx\frac{2\pi
e}{12}2^{-2R_1}\sigma_x^2$. This differs from the previous case in
that most of the cells $C_s(i)$ are no longer intervals, but
rather the union of two non-contiguous intervals, when
$\Delta_a\gg\Delta_b$; for a small portion of the $C_s$ cells,
each of them can consist of three non-contiguous intervals, but
when $\Delta_a\gg\Delta_b$, this portion is negligible and will be
omitted in the discussion which follows. Furthermore, cell
$C_s(i)$ approximately consists of two length $\Delta_b/2$
intervals whose midpoints are $\frac{1}{2}\Delta_a$ apart. The
distortion achieved by using this partition in the lossy decoder
$q_2$ is
\begin{eqnarray}
D_2\approx(\frac{1}{4}\Delta_a)^2=\frac{3}{4}D_1
\end{eqnarray}
Intuitively, this says that the average distance of the points in
the cell $C_s(i)$ from its reproduction codeword is approximately
$\frac{1}{4}\Delta_a$, which is obviously true given the geometric
structure of the cell $C_s(i)$. Note that $D_1$ and $D_2$ are not
of equal value.

The rate of the second description is less straightforward, but
consider the joint partition revealed to $q_3^{-1}$. This
partition is almost uniform, while the rate of the output of $q_b$
after entropy coding is one bit less than that when the same
partition is used in a classical quantizer, because each cell
$C_s(i)$ consists of two local intervals instead of one as in the
classical quantizer. Thus,
\begin{eqnarray}
D_3\approx\frac{1}{12}\Delta_b^2\approx\frac{2\pi
e}{12}2^{-2(R_2+1)}\sigma_x^2=\frac{2\pi e}{48}2^{-2R_2}\sigma_x^2
\end{eqnarray}
It follows that an achievable high-resolution operating point
using scalar quantization is given by $(R_1,R_2,D_1,D_2,D_3)$,
where $D_1=\frac{2\pi e}{12}2^{-2R_1}\sigma_x^2$,
$D_2=\frac{3}{4}D_1$, $D_3=\frac{2\pi e}{48}2^{-2R_2}\sigma_x^2$;
by symmetry, the operating point $(R_2,R_1,D_2,D_1,D_3)$ is also
achievable. By time-sharing, an achievable balanced point is
$(\frac{R_1+R_2}{2},\frac{R_1+R_2}{2},\frac{7}{8}D_1,\frac{7}{8}D_1,D_3)$.
Obviously the central and side distortion product is
$\frac{7}{8}(\frac{2\pi e}{12})^22^{-2(R_1+R_2)}\sigma_x^2$, which
is only $2.5$ dB away from the information theoretic distortion
product. However, time-sharing is not strictly scalar
quantization, and later we discuss a method to avoid the
time-sharing argument.

In order to make $D_1=D_2$ when $\Delta_a\gg\Delta_b$, the values
of $a_2$ can be varied slightly. First, let $\Delta_a$ be fixed
such that $D_1(\approx\frac{1}{12}\Delta_a^2\approx\frac{2\pi
e}{12}2^{-2R_1}\sigma_x^2)$ and $R_1$ are then both fixed. It is
clear with stepsize $\Delta_b$ fixed, as $a_2$ decreases from
$-1$, the distortion $D_2$ increases. A simple calculation shows
that when $a_2=-4/3$, $D_2>D_1$; thus, the desired value of $a_2$
is in $(-4/3,-1)$, and we find this value to be $a_2=-1.0445$. The
detailed calculation is relegated to Appendix IV, where the
computation of the distortions and rates of this particular
quantizer also is given. By using such a value, it can be shown
that an achievable high-resolution operating point is
$(R_1,R_2,D_1,D_2,D_3)$, where $D_1=D_2\approx\frac{2\pi
e}{12}2^{-2R_1}\sigma_x^2$ and $D_3 \approx 0.8974\cdot\frac{2\pi
e}{48}2^{-2R_2}\sigma_x^2$. The rates $R_1$ and $R_2$ usually are
not equal, but the results derived here will be used to construct
two balanced descriptions next.

\subsubsection{Balanced descriptions using quantization splitting}
As previously pointed out, in the quantization splitting coding
scheme $\sigma_{T_3}^2$ should be chosen to be
$2\sigma_{T_0}\sigma_{T_1}$ when balanced descriptions are
required; then $\sigma_{T_1}\gg\sigma_{T_0}$ implies
$\sigma_{T_1}^2\gg\sigma_{T_3}^2\gg\sigma_{T_0}^2$. It follows
that $b_1^*\approx2$, $b_2^*=-1$, $b_3^*\approx2$,
$b_4^*\approx-1$ and $b_5^*\approx3$. We make the following
remarks assuming these values.
\begin{itemize}
\item The conditional expectation $E(r|q_a(x)=i)$ is approximately zero, which implies only the case in which $q_a^{-1}((q(x)))=0$ needs to be considered. This is obvious from the geometric structure given in Fig. \ref{fig:V1_slope2} (b) and the values of $b^*$s.
\item The partition formed by $q_c$ does not improve the distortion $D_1$ over $q_a$. This is because the slope of the line $b_3^*x+b_4^*t+b_5^*y_{i_0}=r_i$ on the $(x,t)$ plane is given in such a way that it almost aligns with the function $t=f(x)$. In such a case, the cell $C_{xr}(i_a,i_b)$ consists of segments from almost every cell $C_s(j)$ for which $C_s(j)\cap \{x:q_a(x)=i_a\}\neq\emptyset$. Intuitively, it is similar to letting the slope of $a_1x+a_2y=s_i$ have a slope of $1$ in Fig. \ref{fig:V1_slope2} (b), such that the distortion $D_2$ does not improve much over $\sigma_x^2$ in the successive quantization case.
\end{itemize}
With these two remarks, consider constructing balanced
descriptions using scalar quantization for $R_1=R_2$ as follows.
Chose $b_1^*=2$  and $b_2^*=-1.0445$ such that, without the lossy
encoder $q_c$, the distortions $D_1$ and $D_2$ are made equal.
Denote the entropy rate of $q_a$ as $R_{1a}$ and that of $q_b$ as
$R_2$. Let $b_3^*=2$, $b_4^*=-1$ but $b_5^*=2.9555$ such that
$E(r|q_a(x)=i)$ is approximately zero. By doing this,
$b_3^*x+b_4^*t+b_5^*y_{i_0}=r_i$ on the $(x,t)$-plane aligns with
the function $t=f(x)$, and thus the remaining rate $R_1-R_{1a}$ is
used by $q_c$ to improve $D_3$, but $D_1$ and $D_2$ are not
further improved. Since $q_a$ and $q_b$ are both operating on high
resolution, assuming $R_1-R_{1a}$ is also high, then $q_c$
partitions each $x\in C_s(j)\cap C_{xr}(i_a,i_c)$ into
$2^{R_1-R_{1a}}$ uniform segments, thus improve $D_0$ by a factor
of $2^{-2(R_1-R_{1a})}$.

Using this construction, we can achieve a balanced high-resolution
operating point of $(R_1,R_1,D_1,D_1,D_0)$ without time-sharing,
where $D_1=\frac{2\pi e}{12}2^{-2R_{1a}}\sigma_x^2$ and
$D_0\approx0.8974\cdot\frac{2\pi
e\sigma_x^2}{48}2^{-2(2R_1-R_{1a})}$. Thus, when $\sigma_x^2\gg
D_1=D_2\gg D_3$, the central and side distortion product is
$2.596$ dB away from the information theoretic distortion product.
This is a better upper bound than the best known upper bound of
the granular distortion using scalar quantization, which is $2.67$
dB away from the information theoretic distortion product
\cite{Cha:041}; this previous bound was derived in \cite{Cha:041}
using the multiple description scalar quantization scheme proposed
by Vaishampayan \cite{Vaishampayan1, Vaishampayan2} with
systematic optimization of quantization thresholds. It should be
pointed out that the results regarding the granular distortion
also apply to other continuous source as in the approach taken in
\cite{Cha:041}. Thus for any sources with smooth pdf, this
granular distortion can be $2.596$ away from the Shannon outer
bound which is tight at high resolution.

\subsection{Optimization of Scalar Quantization Scheme}
The analysis in the previous subsection reveals that for the
scalar case the proposed coding scheme can potentially achieve
better performance than the previous techniques based on scalar
quantization \cite{Vaishampayan1, Vaishampayan2,Cha:041}. However,
for the proposed coding scheme to perform competitively at low
rate with scalar quantization, better methods to optimize the
quantizer should be used. Specifically, the following improvements
are immediate:
\begin{itemize}
\item Given the partition formed by the lossy encoders, the lossy decoder $q_1^{-1}$, $q_2^{-1}$ and $q_3^{-1}$ should optimize the reproduction codebook to be the conditional mean of the codecells.
\item The index $i_a$ and $i_b$ should be jointly entropy-coded instead of being separately coded, and such a joint codebook should be designed.
\item The lossy encoder $q_c$ can be designed for each output index of $q_a$, and thus operates adaptively.
\item The encoder partition should be better optimized; the design method for multi-stage vector quantization offers a possible approach \cite{Cha:92}.
\end{itemize}
These improvements currently are under investigation; a systematic
comparison of these improvements is beyond the scope of this
article and thus will not be included.

\section{Conclusion}
We proposed a lattice quantization scheme which can achieve the
whole Gaussian MD rate-distortion region. Our scheme is universal
in the sense that it only needs the information of the first and
second order statistics of the source. Our scheme is optimal for
Gaussian sources at any resolution and asymptotically optimal for
all smooth sources at high resolution.

Our results, along with a recent work by Erez and Zamir
\cite{Erez}, consolidate the link between MMSE estimation and
lattice coding (/quantization), or in a more general sense, the
connection between Wiener and Shannon theories as illuminated by
Forney \cite{Forney1, Forney2}.

Although the linear MMSE structure is optimal in achieving the
Gaussian MD rate-distortion region as the dimension of the
(optimal) lattice quantizers goes to infinity, it is not optimal
for finite dimensional lattice quantizers since the distribution
of quantization errors is no longer Gaussian. Using nonlinear
structure to exploit the higher order statistics may result in
better performance.

We also want to point out that our derivation does not rely on the
fact that the source is i.i.d. in time. The proposed MD
quantization system is directly applicable for a general
stationary source, although it may be more desirable to whiten the
process first.

\appendices
\section{Gram-Schmidt Orthogonalization for Random Vectors}


Let $\mathcal{H}_v$ denote the set of all
$n$-dimensional\footnote{This condition is introduced just for the
purpose of simplifying the notations.}, finite-covariance-matrix,
zero-mean, real random (column) vectors. $\mathcal{H}_v$ becomes a
Hilbert space under the inner product mapping
\begin{eqnarray*}
\langle
\mathbf{X},\mathbf{Y}\rangle=\mathbb{E}(\mathbf{X}\mathbf{Y}^T):
\mathcal{H}_v\times\mathcal{H}_v\rightarrow\mathcal{R}^{n\times
n}.
\end{eqnarray*}
For
$\mathbf{X}^M_1=(\mathbf{X}_1,\mathbf{X}_2,\cdots,\mathbf{X}_M)^T$
with $\mathbf{X}_i\in\mathcal{H}_v$, $i=1,\cdots,M$, the
Gram-Schmidt orthogonalization proceeds as follows:
\begin{eqnarray*}
\mathbf{B}_1&=&\mathbf{X}_1,\\
\mathbf{B}_i&=&\mathbf{X}_i-\sum\limits_{j=1}^{i-1}\frac{\mathbb{E}(\mathbf{X}_i\mathbf{B}_j^T)}{\mathbb{E}(\mathbf{B}_j\mathbf{B}_j^T)}\mathbf{B}_j,\quad
i=2,\cdots,M.
\end{eqnarray*}
Note:
$\frac{\mathbb{E}(\mathbf{X}_i\mathbf{B}_j^T)}{\mathbb{E}(\mathbf{B}_j\mathbf{B}_j^T)}$
can be any matrix in $\mathcal{R}^{n\times n}$ if
$\mathbf{B}_j=\mathbf{0}$.

We can also write
\begin{eqnarray*}
\mathbf{B}_1&=&\mathbf{X}_1,\\
\mathbf{B}_i&=&\mathbf{X}_i-\widetilde{K}_{i-1}\mathbf{X}^{i-1}_{1},\quad
i=2,\cdots,M,
\end{eqnarray*}
where $\widetilde{K}_{i-1}\in\mathcal{R}^{n\times (i-1)n}$ is a
matrix satisfying
$\widetilde{K}_{i-1}K_{\mathbf{X}^{i-1}_1}=K_{\mathbf{X}_i\mathbf{X}^{i-1}_1}$.
When $K_{\mathbf{X}^{i-1}_1}$ is invertible, we have
$\widetilde{K}_{i-1}=K_{\mathbf{X}_i\mathbf{X}^{i-1}_1}K^{-1}_{\mathbf{X}^{i-1}_1}$.
Here $K_{\mathbf{X}^{i-1}_1}$ is the covariance matrix of
$(\mathbf{X}_1,\cdots,\mathbf{X}_i)^T$ and
$K_{\mathbf{X}_i\mathbf{X}_1^{i-1}}=\mathbb{E}[\mathbf{X}_i(\mathbf{X}_1,\cdots,\mathbf{X}_{i-1})^T]$.

Again, a sequential quantization system can be constructed with
$\mathbf{X}_1$ as the input to generate a zero-mean random vector
$\mathbf{\widetilde{X}}^M_1=(\mathbf{\widetilde{X}}_1,\cdots,\mathbf{\widetilde{X}}_M)^T$
whose covariance matrix is also $K_{\mathbf{X}^M_1}$. Assume
$K_{\textbf{B}_{i}}=\mathbb{E}\mathbf{B}_{i}\mathbf{B}_{i}^T$ is
nonsingular for $i=2,\cdots,M$. Let $Q_{i,n}(\cdot)$ be an
$n$-dimensional lattice quantizer, $i=1,2,\cdots,L-1$. The dither
$\mathbf{Z}_i$ is an $n$-dimensional random vector, uniformly
distributed over the basic cell of $Q_{i,n}$, $i=1,2,\cdots,M-1$.
Suppose $(\mathbf{X}_1,\mathbf{Z}_1,\cdots,\mathbf{Z}_{M-1})$ are
independent, and
$\mathbb{E}\mathbf{Z}_i\mathbf{Z}_i^T=K_{\mathbf{B}_{i}}$,
$i=1,2,\cdots,M$. Define
\begin{eqnarray}
\mathbf{\widetilde{X}}_1&=&\mathbf{X}_1, \label{vector1}\\
\mathbf{\widetilde{X}}_i&=&Q_{i-1,n}\left(\widetilde{K}_{i-1}\mathbf{\widetilde{X}}^{i-1}_{1}+\mathbf{Z}_{i-1}\right)-\mathbf{Z}_{i-1},\quad
i=2,\cdots,M. \label{vector2}
\end{eqnarray}
It is easy to show that $\mathbf{X}^M_1$ and
$\mathbf{\widetilde{X}}^M_1$ have the same covariance matrix.

As in the scalar case, a single quantizer can be reused if pre-
and post-filters are incorporated. Specifically, given an
$n$-dimensional lattice quantizer $Q_n(\cdot)$, let the dither
$\mathbf{Z}'_i$ be an $n$-dimensional random vector, uniformly
distributed over the basic cell of $Q_n$ with nonsingular
covariance matrix
$K_{\mathbf{Z}'}=\mathbb{E}\textbf{Z}'_i{\textbf{Z}'}^T_i$. Let
$A_i$ be an $n\times n$ nonsingular matrix\footnote{$A_i$ is in
general not unique even if we view $A_i$ and $-A_i$ as the same
matrix. For example, let $K_{\mathbf{Z}'}=U_1U^T_1$ be the
Cholesky decomposition of $K_{\mathbf{Z}'}$ and
$K_{\mathbf{B}_{i+1}}=U_2U^T_2$ be the Cholesky decomposition of
$K_{\mathbf{B}_{i+1}}$, where $U_1$ and $U_2$ are lower triangular
matrices. We can set $A_i=U_2U^{-1}_1$. Let
$K_{\mathbf{Z}'}=V_1\Lambda_1 V^T_1$ and
$K_{\mathbf{B}_{i+1}}=V_2\Lambda_2 V^T_2$ be the eigenvalue
decompositions of $K_{\mathbf{Z}'}$ and $K_{\mathbf{B}_{i+1}}$
respectively. We can also set
$A=V_2\Lambda^{\frac{1}{2}}_2\Lambda^{-\frac{1}{2}}_1V^T_1$} such
that $A_iK_{\mathbf{Z}'}A^T_i=K_{\mathbf{B}_{i+1}}$,
$i=1,2,\cdots,M-1$. Suppose
$(\mathbf{X}_1,\mathbf{Z}'_1,\cdots,\mathbf{Z}'_{M-1})$ are
independent.  Define
\begin{eqnarray*}
\mathbf{\overline{X}}_1&=&\mathbf{X}_1,\\
\mathbf{\overline{X}}_i&=&A_{i-1}\left[Q_n\left(A^{-1}_{i-1}\widetilde{K}_{i-1}\mathbf{\overline{X}}^{i-1}_{1}+\mathbf{Z}'_{i-1}\right)-\mathbf{Z}'_{i-1}\right],\quad
i=2,\cdots,M.
\end{eqnarray*}
It is easy to verify that $\mathbf{X}^M_1$ and
$\mathbf{\widetilde{X}}^M_1$ have the same covariance matrix by
invoking property 2) of the ECDQ. Here introducing the prefilter
$A^{-1}_{i}$ and the postfilter $A_{i}$ is equivalent to shaping
$Q(\cdot)$ by $A_i$, which induces a new quantizer
$Q_{i,n}(\cdot)$ given by
$Q_{i,n}(\mathbf{x})=A_iQ_n(A^{-1}_i\mathbf{}x)$.

Suppose $K_{\mathbf{B}_i}$ is singular for some $i$, say
$K_{\mathbf{B}_i}$ is of rank $k$ with $k<n$. For this type of
degenerate case, the quantization operation should be carried out
in the nonsingular subspace of $K_{\mathbf{B}_i}$. Let
$K_{\mathbf{B}_i}=U\Lambda U^T$ be the eigenvalue decomposition of
$K_{\mathbf{B}_i}$. Without loss of generality, assume
$\Lambda=\mbox{diag}\{\lambda_1,\cdots,\lambda_k,0,\cdots,0\}$,
where $\lambda_i>0$ for all $i=1,2,\cdots,k$. Define
$\Lambda_k=\mbox{diag}\{\lambda_1,\cdots,\lambda_k\}$. Now replace
the $n$-dimensional quantizer $Q_{i-1,n}(\cdot)$ in
(\ref{vector2}) by a $k$-dimensional quantizer $Q_{i-1,k}(\cdot)$
and replace the dither $\mathbf{Z}_{i-1}$ by a dither
$\mathbf{\widetilde{Z}}_{i-1}$ which is a $k$-dimensional random
vector, uniformly distributed over the basic cell of $Q_{i-1,k}$
with
$\mathbb{E}\mathbf{\widetilde{Z}}_{i-1}\mathbf{\widetilde{Z}}_{i-1}^T=\Lambda_k$.
Let
\begin{eqnarray*}
\left[\mathbf{\widetilde{X}}_i\right]_{1,k}=Q_{i-1,k}\left(\left[U^T\widetilde{K}_{i-1}\mathbf{\widetilde{X}}^{i-1}_{1}\right]_{1,k}+\mathbf{\widetilde{Z}}_{i-1}\right)-\mathbf{\widetilde{Z}}_{i-1}
\end{eqnarray*}
and we have
\begin{eqnarray*}
\mathbf{\widetilde{X}}_i=U\begin{pmatrix}
  \left[\mathbf{\widetilde{X}}_i\right]_{1,k} \\
  \left[U^T\widetilde{K}_{i-1}\mathbf{\widetilde{X}}^{i-1}_{1}\right]_{k+1,n}
\end{pmatrix},
\end{eqnarray*}
where
$\left[U^T\widetilde{K}_{i-1}\mathbf{\widetilde{X}}^{i-1}_{1}\right]_{1,k}$
is a column vector containing the first $k$ entries of
$U^T\widetilde{K}_{i-1}\mathbf{\widetilde{X}}^{i-1}_{1}$ and
$\left[U^T\widetilde{K}_{i-1}\mathbf{\widetilde{X}}^{i-1}_{1}\right]_{k+1,n}$
is a column vector that contains the remaining entries of
$U^T\widetilde{K}_{i-1}\mathbf{\widetilde{X}}^{i-1}_{1}$.

\section{Proof of Theorem \ref{asyoptimal}}

It is easy to verify that as $D_1,D_2,D_3\rightarrow 0$, we have
$\frac{\sigma^2_{T_0}}{D_3}\rightarrow 1$,
$\frac{\sigma^2_{T_0}+\sigma^2_{T_i}}{D_3}\rightarrow 1$, and
$\frac{\sigma_{T_i}}{\sqrt{D_i-D_3}}\rightarrow 1$, $i=1,2$. Let
$\sigma^2_{T_3}\in\left[0,M\left(\frac{D_3}{D_1}(\sqrt{D_1-D_3}+\sqrt{D_2-D_3})^2+D_2\right)\right]$,
where $M$ is a fixed large number. Clearly,
$\sigma^2_{T_3}\rightarrow 0$ as $D_1,D_2,D_3\rightarrow 0$.

For the MD quantization scheme shown in Fig. \ref{QuanS}, we have
\begin{eqnarray*}
R_1&=&\frac{1}{n}H(Q^*_{2,n}(b^*_1\mathbf{X}+b^*_2{\mathbf{\widetilde{W}}'_2}+{\mathbf{Z}^*_2}^n)|{\mathbf{Z}^*_2})\\
&=&\frac{1}{n}h(\mathbf{X}+b^*_2{\mathbf{N}^*_1}+{\mathbf{N}^*_2})-\frac{1}{n}h({\mathbf{N}^*_2})\\
&=&\frac{1}{n}h(\mathbf{X}+b^*_2{\mathbf{N}^*_1}+{\mathbf{N}^*_2})-\frac{1}{2}\log\frac{\mathbb{E}{\widetilde{B}_3}^2}{G^{opt}_{n}},\\
R_2&=&\frac{1}{n}H(Q^*_{1,n}(\mathbf{X}+{\mathbf{Z}^*_1})|{\mathbf{Z}^*_1})+\frac{1}{n}H(Q^*_{3,n}(b^*_3\mathbf{X}+b^*_4\mathbf{\widetilde{W}}_1+b^*_5{\mathbf{\widetilde{W}}'_2}+{\mathbf{Z}^*_3})|{\mathbf{Z}^*_3})\\
&=&\frac{1}{n}h(\mathbf{X}+{\mathbf{N}^*_1})-\frac{1}{n}h({\mathbf{N}^*_1})+\frac{1}{n}h(b^*_3\mathbf{X}+b^*_4\mathbf{\widetilde{W}}_1^n+b^*_5{\mathbf{\widetilde{W}}'_2}+{\mathbf{N}^*_3})-\frac{1}{n}h({\mathbf{N}^*_3})\\
&\leq&\frac{1}{n}h(\mathbf{X}+{\mathbf{N}^*_1})-\frac{1}{n}h({\mathbf{N}^*_1})+\frac{1}{n}h(b_7{\mathbf{\overline{B}}_2}+b_8{\mathbf{\overline{B}}_3}+{\mathbf{\overline{B}}_4})-\frac{1}{n}h({\mathbf{N}^*_3})\\
&=&\frac{1}{n}h(\mathbf{X}+{\mathbf{N}^*_1})-\frac{1}{2}\log\frac{\mathbb{E}{\widetilde{B}_2}^2}{G^{opt}_{n}}+\frac{1}{2}\log\left[2\pi
e(b^2_7\mathbb{E}{\overline{B}^2_2}+b^2_8\mathbb{E}{\overline{B}^2_3}+\mathbb{E}{\overline{B}^2_4})\right]-\frac{1}{2}\log\frac{\mathbb{E}{\overline{B}^2_4}}{G^{opt}_{n}}.
\end{eqnarray*}

Since $\frac{1}{n}h(\mathbf{X}+{\mathbf{N}^*_1})=h(X)+o(1)$ and
$\frac{1}{n}h(\mathbf{X}+b^*_2{\mathbf{N}^*_1}+{\mathbf{N}^*_2})=h(X)+o(1)$
as $D_1,D_2,D_3\rightarrow 0$, it follows that
\begin{eqnarray*}
R_1&=&h(X)-\frac{1}{2}\log\frac{\mathbb{E}{\widetilde{B}_3}^2}{G^{opt}_{n}}+o(1)\\
&=&\frac{1}{2}\log\frac{P_X(D_2+\sigma^2_{T_3})}{D_3(\sqrt{D_1-D_3}+\sqrt{D_2-D_3})^2+\sigma^2_{T_3}D_1}+\frac{1}{2}\log(2\pi
eG^{opt}_{n})+o(1),\\
R_2&\leq&h(X)-\frac{1}{2}\log\frac{\mathbb{E}{\widetilde{B}_2}^2}{G^{opt}_{n}}+\frac{1}{2}\log\left[2\pi
e(b^2_7\mathbb{E}{\overline{B}^2_2}+b^2_8\mathbb{E}{\overline{B}^2_3}+\mathbb{E}{\overline{B}^2_4})\right]-\frac{1}{2}\log\frac{\mathbb{E}{\overline{B}_4}^2}{G^{opt}_{n}}+o(1)\\
&=&\frac{1}{2}\log\frac{P_X}{D_2+\sigma^2_{T_3}}+\frac{1}{2}\log\frac{D_3(\sqrt{D_1-D_3}+\sqrt{D_2-D_3})^2+\sigma^2_{T_3}D_1}{D_3(\sqrt{D_1-D_3}+\sqrt{D_2-D_3})^2}+\log(2\pi
e G^{opt}_{n})+o(1).
\end{eqnarray*}
So we have
\begin{eqnarray*}
R_1+R_2\leq\frac{1}{2}\log\frac{P^2_X}{D_2(\sqrt{D_1-D_3}+\sqrt{D_2-D_3})^2}+\frac{3}{2}\log(2\pi
e G^{opt}_{n}).
\end{eqnarray*}

When $\sigma^3_{T_3}=0$, there is no quantization splitting and
the quantizer $Q^*_{3,n}(\cdot)$ can be removed. In this case, we
have
\begin{eqnarray*}
R_1&=&\frac{1}{2}\log\frac{P_XD_2}{D_3(\sqrt{D_1-D_3}+\sqrt{D_2-D_3})^2}+\frac{1}{2}\log(2\pi
eG^{opt}_{n})+o(1),\\
R_2&=&\frac{1}{2}\log\frac{P_X}{D_2}+\frac{1}{2}\log(2\pi e
G^{opt}_{n})+o(1)
\end{eqnarray*}
When
$\sigma^3_{T_3}=M\left[\frac{D_3}{D_1}(\sqrt{D_1-D_3}+\sqrt{D_2-D_3})^2+D_2\right]$,
we have
\begin{eqnarray*}
R_1&=&\frac{1}{2}\log\frac{P_X}{D_1}+\frac{1}{2}\log(2\pi e
G^*_{2,n})+\epsilon(M)+o(1),\\
R_2&=&\frac{1}{2}\log\frac{P_XD_1}{D_3(\sqrt{D_1-D_3}+\sqrt{D_2-D_3})^2}+\log(2\pi
eG^{opt}_{n})-\epsilon(M)+o(1),
\end{eqnarray*}
where $\epsilon(M)\rightarrow 0$ as $M\rightarrow\infty$.
Therefore, the region
\begin{eqnarray*}
R_1&=&\frac{1}{2}\log\frac{P_X}{D_1}+\frac{1}{2}\log(2\pi e
G^{opt}_{n})+\epsilon(M)+o(1),\\
R_2&=&\frac{1}{2}\log\frac{P_X}{D_2}+\frac{1}{2}\log(2\pi e
G^{opt}_{n})+o(1),\\
R_1+R_2&=&\frac{1}{2}\log\frac{P^2_X}{D_2(\sqrt{D_1-D_3}+\sqrt{D_2-D_3})^2}+\frac{3}{2}\log(2\pi
e G^{opt}_{n})+o(1)
\end{eqnarray*}
is achievable.

By symmetry, the region
\begin{eqnarray*}
R_1&=&\frac{1}{2}\log\frac{P_X}{D_1}+\frac{1}{2}\log(2\pi e
G^{opt}_{n})+o(1),\\
R_2&=&\frac{1}{2}\log\frac{P_X}{D_2}+\frac{1}{2}\log(2\pi e
G^{opt}_{n})+\epsilon(M)+o(1),\\
R_1+R_2&=&\frac{1}{2}\log\frac{P^2_X}{D_2(\sqrt{D_1-D_3}+\sqrt{D_2-D_3})^2}+\frac{3}{2}\log(2\pi
e G^{opt}_{n})+o(1)
\end{eqnarray*}
is achievable via the other form of quantization splitting. The
desired result follows by combining these two regions and choosing
$M$ large enough.

\section{Proof of Theorem \ref{highdim}}
We shall only give a heuristic argument here. The rigorous proof
is similar to that of Theorem 3 in \cite{Feder} and thus is
omitted.

It is well-known that the distribution of the quantization noise
converges to a white Gaussian distribution in the divergence sense
as the dimension of the optimal lattice becomes large
\cite{Feder}. So we can approximate $\mathbf{N}^*_i$ by
$\mathbf{N}^G_i$, where $\mathbf{N}^G_i$ is a zero-mean Gaussian
vector with the same covariance as that of $\mathbf{N}^*_i$,
$i=1,2,3$. Therefore, for large $n$, we have
\begin{eqnarray*}
\frac{1}{n}h(\mathbf{X}+{\mathbf{N}^*_1})-\frac{1}{n}h({\mathbf{N}^*_1})&\approx&\frac{1}{n}h(\mathbf{X}+{\mathbf{N}^G_1})-\frac{1}{n}h({\mathbf{N}^G_1})\\
&=&h\left(X+N^G_1\right)-h\left(N^G_1\right),
\end{eqnarray*}
\begin{eqnarray*}
h\left(b^*_1\mathbf{X}+b^*_2{\mathbf{\widetilde{W}}'_2}+{\mathbf{N}^*_2}\right)-\frac{1}{n}h\left({\mathbf{N}^*_2}\right)&=&\frac{1}{n}h\left(\mathbf{X}+b^*_2\mathbf{N}^*_1+\mathbf{N}^*_2\right)-\frac{1}{n}h\left(\mathbf{N}^*_2\right)\\
&\approx&\frac{1}{n}h\left(\mathbf{X}+b^*_2\mathbf{N}^G_1+\mathbf{N}^G_2\right)-\frac{1}{n}h\left(\mathbf{N}^G_2\right)\\
&=&h\left(X+b^*_2N^G_1+N^G_2\right)-h\left(N^G_2\right),
\end{eqnarray*}
and
\begin{eqnarray*}
&&\frac{1}{n}h\left(b^*_3\mathbf{X}+b^*_4\mathbf{\widetilde{W}}_1^n+b^*_5{\mathbf{\widetilde{W}}'_2}+{\mathbf{N}^*_3}\right)-\frac{1}{n}h\left({\mathbf{N}^*_3}\right)\\
&=&\frac{1}{n}h\left((b^*_3+b^*_1b^*_4+b^*_2b^*_4+b^*_5)\mathbf{X}+(b^*_2b^*_4+b^*_5)\mathbf{N}^*_1+b^*_4\mathbf{N}^*_2+\mathbf{N}^*_3\right)-\frac{1}{n}h\left(\mathbf{N}^*_3\right)\\
&\approx&\frac{1}{n}h\left((b^*_3+b^*_1b^*_4+b^*_2b^*_4+b^*_5)\mathbf{X}+(b^*_2b^*_4+b^*_5)\mathbf{N}^G_1+b^*_4\mathbf{N}^G_2+\mathbf{N}^G_3\right)-\frac{1}{n}h\left(\mathbf{N}^G_3\right)\\
&=&h\left((b^*_3+b^*_1b^*_4+b^*_2b^*_4+b^*_5)X+(b^*_2b^*_4+b^*_5)N^G_1+b^*_4N^G_2+N^G_3\right)-h\left(N^G_3\right).
\end{eqnarray*}

\section{The calculation of scalar operating point using successive quantization}

Observe in Fig. \ref{fig:V1_slope2} (c) that the value $a_2$ is
slightly different from $-1$, such that a portion of the $C_s$
cells consist of three length $\frac{1}{2}\Delta_b$ intervals
which are approximately $\frac{-a_2}{2}\Delta_a$ apart (denote the
set of this first class of cells as $C_s'$), while the other $C_s$
cells consist of only two length $\frac{1}{2}\Delta_b$ which are
also $\frac{-a_2}{2}\Delta_a$ apart (denote the set of this second
class of cells as $C_s''$); the ratio between the cardinalities of
these two sets is function of $a_2$, which is approximately
$\frac{-3-3a_2}{4+3a_2}$. Here we again ignore the cells $C_s$
whose constituent segments are at the border of $q_a$ partition
cells, which is a negligible portion when $\Delta_a\gg\Delta_b$.
The average distortion for each first class cell $C_s$ is
approximately $\frac{2}{3}(\frac{-a_2}{2}\Delta_a)^2$, while the
average distortion for each second class cell $C_s$ is
approximately $(\frac{1}{2}\cdot\frac{-a_2}{2}\Delta_a)^2$. Thus,
the distortion $D_2$ can be approximated as
\begin{eqnarray}
D_2&\approx&(-3-3a_2)\cdot\frac{2}{3}(\frac{-a_2}{2}\Delta_a)^2+(4+3a_2)(\frac{1}{2}\frac{-a_2}{2}\Delta_a)^2\nonumber\\
&=&\frac{-1}{16}(5a_2+4)a_2^2{\Delta_a}^2
\end{eqnarray}
Notice that $-3-3a_2+4+3a_2=1$; thus, $(-3-3a_2)$ is the
percentage of the first class cells in all the $C_s$ cells.
Letting $D_1=D_2=\frac{1}{12}{\Delta_a}^2$, we can solve for
$a_2$; the only real solution to this equation is $a_2=-1.0445$.
The distortion $D_3$ is approximately
$\frac{1}{12}(\frac{1}{2}\Delta_b)^2$, by using an almost uniform
partition of stepsize $\frac{1}{2}\Delta_b$. To approximate the
entropy rate for $q_b$, consider the rate contribution from the
first class $C_s$ cells, namely
\begin{eqnarray}
R_2'&=&-\sum_{C_s(i)\in C_s'}p(q_2^{-1}(i))\frac{3}{2}\Delta_b\log_2(p(q_2^{-1}(i))\frac{3}{2}\Delta_b)\nonumber\\
&\approx&(3+3a_2)\log_2(\frac{3}{2}\Delta_b)-\sum_{C_s(i)\in
C_s'}p(q_2^{-1}(i))\frac{3}{2}\Delta_b\log_2(p(q_2^{-1}(i)))
\end{eqnarray}
where $p(x)$ is the pdf of the source, and the second
approximation comes from taking the percentage of the first class
cells in all the $C_s$ cells as the probability that a random
$C_s$ is a first class cell. Similarly the rate contribution from
the second class $C_s$ cells is
\begin{eqnarray}
R_2''&=&-\sum_{C_s(i)\in C_s''}p(q_2^{-1}(i))\frac{2}{2}\Delta_b\log_2(p(q_2^{-1}(i))\frac{2}{2}\Delta_b)\nonumber\\
&\approx&-(4+3a_2)\log_2(\frac{2}{2}\Delta_b)-\sum_{C_s(i)\in
C_s''}p(q_2^{-1}(i))\frac{2}{2}\Delta_b\log_2(p(q_2^{-1}(i)))
\end{eqnarray}
Thus, the rate $R_2$ can be approximated as
\begin{eqnarray}
\label{eqn:rate2}
R_2&\approx&R_2'+R_2''\nonumber\\
&\approx&-\log_2(\Delta_b)+ (3+3a_2)\log_2(\frac{3}{2})-\sum_{C_s(i)\in C_s'}p(q_2^{-1}(i))\frac{3}{2}\Delta_b\log_2(p(q_2^{-1}(i)))\nonumber\\
&&-\sum_{C_s(i)\in
C_s''}p(q_2^{-1}(i))\frac{2}{2}\Delta_b\log_2(p(q_2^{-1}(i))).
\end{eqnarray}
When $q_a(\cdot)$ is high resolution, $p(q_2^{-1}(i))$ is
approximately equal to $p(x)$, for any $x\in C_s(i)$, and thus
equal to $p(q_3^{-1}(i,\cdot))$. Using this approximation and
taking $\frac{1}{2}\Delta_b$ as $\delta x$, the last two terms in
(\ref{eqn:rate2}) can be approximated by an integral, which is in
fact $h(p)$, the differential entropy of the source. It follows
that
\begin{eqnarray}
R_2&\approx&R_2'+R_2''\nonumber\\
&\approx&-\log_2(\Delta_b)+ (3+3a_2)\log_2(\frac{3}{2})+h(p)
\end{eqnarray}
where $h(p)=\frac{1}{2}\log(2\pi e\sigma_x^2)$ for the Gaussian
source. Thus,
$D_3\approx\frac{1}{12}(\frac{1}{2}\Delta_b)^2\approx0.8974\cdot\frac{2\pi
e\sigma_x^2}{48}2^{-2R_2}$.


\end{document}